\documentclass[hyper]{JHEP3}

\usepackage{epsfig}
\usepackage{latexsym}
\usepackage{amsfonts}
\usepackage{amsmath}
\usepackage{amsthm}
\usepackage{amssymb}
\usepackage{amsbsy}
\usepackage{multirow}

\newcommand{\Dp}[1]{\mathrm{D {#1}}}
\newcommand{\antiDp}[1]{\overline{\Dp{#1}}}

\def\IZ{{\mathbb Z}}
\def\ICP{{\mathbb CP}}
\def\CM{{\mathcal M}}

\def\IIA          {\mathrm{IIA}}

\def\IR{{\mathbb R}}

\def\cala         {{\cal A}}

\def\cald         {{\cal D}}

\def\caln         {{\cal N}}
\def\calo         {{\cal O}}

\def\calr         {{\cal R}}

\newsavebox{\uuunit}
\sbox{\uuunit}
    {\setlength{\unitlength}{0.825em}
     \begin{picture}(0.6,0.7)
        \thinlines
        \put(0,0){\line(1,0){0.5}}
        \put(0.15,0){\line(0,1){0.7}}
        \put(0.35,0){\line(0,1){0.8}}
       \multiput(0.3,0.8)(-0.04,-0.02){12}{\rule{0.5pt}{0.5pt}}
     \end {picture}}

\def\be{\begin{equation}}
\def\ee{\end{equation}}
\def\bea{\begin{eqnarray}}
\def\eea{\end{eqnarray}}

\def\ul{\underline}
\def\a{\alpha}

\def\G{\Gamma}
\def\d{\delta}

\def\l{\lambda}

\def\m{\mu}

\def\o{\omega}
\def\O{\Omega}
\def\p{\pi}
\def\r{\rho}

\def\S{\Sigma}

\title{Black hole bound states in $\mathbf{AdS_3 \times S^2}$}

\author{Jan de Boer$^1$, Frederik Denef$^{2,3}$, Sheer El-Showk$^1$, Ilies Messamah$^1$
and Dieter Van den Bleeken$^{3,2}$\\
\\
$^1$ Instituut voor Theoretische Fysica, Universiteit Amsterdam, \\
Valckenierstraat 65, 1018XE Amsterdam, The Netherlands\\
\\
$^2$ Jefferson Physical Laboratory, Harvard University, \\
Cambridge, MA 02138, USA \\
\\
$^3$ Instituut voor Theoretische Fysica, KU Leuven, \\
Celestijnenlaan 200D, B-3001 Leuven, Belgium \\

} \abstract{We systematically construct the geometries dual to the
1+1 dimensional (0,4) conformal field theories that arise in the
low-energy description of wrapped M5-branes in $S^1 \times {\rm
CY}_3$ compactifications of M-theory. This includes a large number
of multicentered black hole bound states asymptotic to AdS$_3
\times S^2$. In addition, we find many geometries that develop
multiple, mutually decoupled AdS$_3 \times S^2$ throats. We argue
there is a useful one to one correspondence between the connected
components of the space of solutions and particular limits of type
IIA attractor flow trees. We point out that there is a
thermodynamic instability of small supersymmetric BTZ black holes
to localization on the $S^2$, a supersymmetric and exactly
solvable analog of the well known AdS-Schwarzschild localization
instability, and identify this with the ``Entropy Enigma'' in four
dimensions. We discuss the phase transition this suggests, and
initiate the CFT interpretation of these results. }
\preprint{KUL-TF-08/05 \\ITFA-2008-03 }

\begin{document}

\section{Introduction and summary}

Consider M-theory compactified on the product of a Calabi-Yau $X$ of
volume $V_X$ and a circle of radius $R$ in the limit $R/l_{11} \to \infty$,
$V_X/l_{11}^6$ fixed. In this limit the
worldvolume excitations of M5-branes wrapped on
4-cycles in $X$ and the $S^1$ which are of finite energy in $1/R$ units
decouple from the bulk. Their dynamics is described by a (0,4) supersymmetric 1+1 dimensional nonlinear
sigma model with target space given by the classical M5 moduli space. This is the MSW CFT \cite{Maldacena:1997de,Minasian:1999qn}.

The holographic dual to this CFT is thought to be quantum M-theory
with AdS$_3 \times S^2 \times X$ boundary asymptotics. The scales
of AdS$_3$ and $S^2$ are set by the central charge $c = p^3$ of
the CFT, where $p^3 = D_{ABC} p^A p^B p^C$ denotes the triple
self-intersection product of the 4-cycle homology class $p$
wrapped by the M5 \cite{Maldacena:1997de}.\footnote{We will in
this paper drop subleading contributions linear in $p$ to the
central charge, so $c_L=c_R \equiv c$.} When this is large,
semiclassical supergravity becomes reliable.

In this paper we will study systematically the BPS sector on the gravity side, and uncover some surprises.

A first one is based on the following observations. A priori we
can choose to embed the M5 in a Calabi-Yau $X$ with arbitrary
moduli. In particular, we are free to pick any value for the
normalized\footnote{Normalized such that $\frac{Y^3}{6} \equiv 1$.
The overall scale (volume) of $X$ in 11d Planck units is in a
hypermultiplet, while the relative scales are in vector
multiplets.} K\"ahler moduli $Y^A$. On the other hand, the AdS$_3
\times S^2 \times X$ solution freezes $Y^A = p^A/U$,
$U:=(p^3/6)^{1/3}$. This presents a puzzle: As we will show in an
explicit example, the spectrum of BPS states depends in general on
the moduli $Y^A$; there are walls of marginal stability in
$Y$-space where certain M5 states split into two different M5
constituents. The decay always happens in the direction towards
the attractor point $Y^A = p^A/U$. Moreover, congruent with this,
when taking the decoupling limit on the gravity side with the
$Y^A$ on the side of a wall of marginal stability where the
constituents are bound together, we will see that we do not end up
with a single decoupled asymptotically AdS$_3 \times S^2 \times
X|_p$ space, but with several mutually decoupled AdS$_3 \times S^2
\times X|_{p_i}$ spaces (embedded in asymptotically ${\mathbb R}^5
\times X|_Y$ space), each with its own attractor point $Y_i^A =
p_i^A/U_i$. This implies that the MSW CFT is not capable of
capturing the entire moduli space of M5 bound states, as in the
latter case sectors corresponding to M5-M5 bound state
constituents decouple from each other in the IR, with each sector
flowing to its own fixed point CFT. Some puzzles and questions
related to this will be discussed in section~\ref{commquest}.


In the bulk of the paper, we focus on a single asymptotically AdS$_3 \times S^2$ sector.
Suitable density matrices of CFT states will be dual in the semi-classical sense to BPS black hole solutions. The simplest BPS black hole solution with the correct charges and asymptotics
is the extremal rotating BTZ black hole
\cite{Banados:1992gq} times $S^2$. Its entropy agrees with the Cardy formula for the asymptotic
degeneracy of BPS states in the $(0,4)$ CFT for $L_0 \gg \frac{c}{24}$: $S_{\rm BTZ} = 4 \pi
\sqrt{\frac{c}{24}(L_0-\frac{c}{24})}$.

However, for $h \equiv (L_0 - \frac{c}{24})/c$ sufficiently small (below to a
critical value $h_c$),
when the BTZ black hole radius drops below the AdS$_3$ ($\sim
S^2$) radius, a thermodynamic instability occurs: at the given
value of $L_0$ it becomes entropically favorable for the black
hole to \emph{localize} on the sphere; this more entropic solution
looks locally like a 5d BMPV black hole
\cite{Breckenridge:1996is}, with M2 charge and $S^3$ horizon,
sitting at some point of the sphere and the center of AdS$_3$, the
sphere being supported still by flux. Once $L_0$ drops below zero,
the BTZ black hole ceases to exist altogether as a regular black hole solution; instead one finds 
a conical defect singularity.

This instability can be viewed as a supersymmetric version of the
instability pointed out for Schwarzschild-AdS black holes by
Banks, Douglas, Horowitz and Martinec \cite{Banks:1998dd}. Related
thermodynamical as well as dynamical instabilities were studied
among others in
\cite{Gregory:1993vy,Peet:1998cr,Martinec:1999sa,Cvetic:1999ne,Chamblin:1999tk,Chamblin:1999hg,Gubser:2000ec,Gubser:2000mm,Yamada:2006rx,Balasubramanian:2007bs}.
The nonsupersymmetric nature of these systems makes them difficult
to study. In contrast, here, supersymmetry allows us to obtain
completely explicit solutions.

In fact, the single sphere localized black hole is but one of a very intricate set of supersymmetric multi black hole bound states in AdS$_3 \times S^2$ we will construct in general. We will do this by lifting four dimensional
type IIA multi black hole bound states
\cite{Denef:2000nb,LopesCardoso:2000qm,Bates:2003vx,Denef:2007vg} with D4, D2 and D0 total charge but no net D6 charge to
five dimensions using the 4d-5d correspondence
\cite{Gaiotto:2005gf,Gaiotto:2005xt,Behrndt:2005he,Cheng:2006yq,
Bena:2005ay,Bena:2005ni,Elvang:2005sa,Berglund:2005vb,Balasubramanian:2006gi},
and then carefully taking the decoupling limit. The black hole localization instability in AdS$_3 \times S^2$, it turns out, is then nothing but the uplift of the four dimensional ``Entropy Enigma'' of \cite{Denef:2007vg,Denef:2007yn}! The Entropy Enigma is the observation that in the regime in which the total D4-D2-D0 charge is scaled up uniformly, for sufficiently large background type IIA CY volume, multicentered black holes dominate the entropy. Since (for zero D2 charge) $h = (L_0 - \frac{c}{24})/c = -q_0/p^3$, with $q_0$ the D0 charge and $p$ the D4 charge, this regime indeed corresponds to $h \to 0$.

General two black hole configurations in AdS$_3 \times S^2$ can be
viewed as fat, backreacting versions of the M2 and anti-M2 probe
particles sitting at the north and south poles of the $S^2$ which
were considered in \cite{Gaiotto:2006ns} in a derivation of the
OSV conjecture.

Although many multicentered black hole configurations exist, it
appears that the entropically most dominant one (or at least the
most entropic one we have been able to find) is the configuration
which in four dimensions consist of one pure $\overline{D6}$
particle with zero entropy and one large D6-D4-D2-D0 black hole;
this lifts in AdS$_3 \times S^2$ to a single BMPV black hole
localized on the sphere --- the localized black hole referred to
above.

Unlike the BTZ black hole, the localized black holes in general
have macroscopic $S^2$ angular momentum, up to values of order
$p^3$. (The maximal angular momentum is reached for the $L_0 = 0$
ground state, rotating global AdS$_3 \times S^2$, obtained by
uplifting a $D6-\overline{D6}$ 2-particle state.) Assuming the BTZ
alone dominates the entropy for $h > h_c$ and the single localized
black hole alone for $h<h_c$, we thus find that in the $c=p^3 \to
\infty$ limit, a first order phase transition occurs at $h=h_c$,
with order parameter given by the $S^2$ angular momentum.

We also argue that in the canonical ensemble, trading fixed $L_0$
for fixed left-moving temperature $T$, this localization
transition is not visible; instead the small BTZ destabilizes due
to a supersymmetric version of the Hawking-Page phase transition,
going from BTZ to rotating AdS$_3 \times S^2$ at $T_c =
\frac{1}{2\pi}$. Such a phase transition was observed already in
\cite{Dijkgraaf:2000fq} for AdS$_3 \times S^3$ by studying the
Fareytail expansion of the elliptic genus. Interestingly, this
critical temperature can also be obtained as the temperature of
the smallest supersymmetric BTZ black hole that can be made
adiabatically as the limit of a certain class of ``scaling''
solutions. These zero entropy scaling solutions can be viewed as
part of the microstates (in the sense of the fuzzball proposal)
that make up the BTZ black hole above $T_c$. Whether there are other
microstates contribute which do contribute to the unstable BTZ black
hole below the critical temperature remains to be investigated.

Finally, we initiate a discussion of the CFT interpretation of all
this, with particular attention paid to the $h \to 0$ regime. To this
end we refine and improve the original analysis of
\cite{Maldacena:1997de}, pointing out, in particular, the importance
of including all $c=p^3$ winding modes, which freeze 4-cycle
deformation moduli at special supersymmetric points and contribute
dominantly to the entropy at small $h$. We explain some of the
qualitative features observed on the gravity side, including the
decrease in $SU(2)_R$ charge expectation value with increasing
$L_0$, but we leave a more complete analysis for future work.

The largest part of the paper is devoted to laying the necessary
groundwork: constructing multicentered solutions in AdS$_3 \times
S^2$, finding ways to establish their existence without having to
construct them explicitly, classifying them, and identifying their
dual CFT quantum numbers.

As mentioned above, our strategy for finding the configurations
surviving the decoupling limit will consist of lifting four
dimensional type IIA multi black hole bound states to five
dimensions, and then carefully taking the decoupling limit. This
turns out to \emph{not} be the same as naively dropping the
constant terms in the defining harmonic functions.

The decoupling procedure is not entirely straightforward, since
from the type IIA point of view, it sends the CY volume in string
units $V_X/l_s^6 \sim (R^3/l_{11}^3) (V_X/l_{11}^6)$ to infinity,
while keeping the 4d string coupling $g_{4d} \sim
(V_X/l_{11}^6)^{-1/2}$ finite. (The 10d string coupling goes to
infinity too, of course, as it should for eleven dimensional
supergravity to become the proper low energy effective theory.)
Since the volume in string units is in a vector multiplet (unlike
the 4d string coupling, which belongs to a hypermultiplet), a IIA
multicentered solution existing at some finite value of
$V_X/l_s^6$ could be destroyed when taking the limit, as for
instance a wall of marginal stability may be encountered at some
value of $V_X/l_s^6$, where the solution decays.

The asymptotically AdS$_3 \times S^2$ solutions thus obtained (after a suitable 
rescaling of coordinates\footnote{Namely $\vec x \to \ell_5^3 \vec x$, where
$\ell_5 := l_{11}/4 \pi \tilde{V}_X^{1/3}$, $\tilde{V}_X := V_X/l_{11}^6$.}) can be written
explicitly, and are completely determined by $2\,{\rm dim}\,
H^2(X)+2$ harmonic functions (we put $R \equiv 1$ in what follows):
\begin{equation}
H^0=\sum_a\frac{p^0_a}{|x-x_a|}\,,\quad\ \,
H^A=\sum_a\frac{p^A_a}{|x-x_a|}\,,\quad\ \,
H_A=\sum_a\frac{q_A^a}{\,|x-x_a|}\,,\quad\ \,
H_0=\sum_a\frac{q_0^a}{|x-x_a|}-\frac{1}{4} \,.\nonumber
\end{equation}
Here the coordinate vector $x_a$ gives the position in the spatial
$\mathbb{R}^3$ of the $a$th center with charge
$\G_a=(p^0_a,p^A_a,q^a_A,q_0^a)$. Furthermore $\sum_a p^0_a=0$ and
$\sum_a p^A_a$ is positive, i.e.\ lies within the K\"ahler
cone.\footnote{This is necessary because the asymptotic K\"ahler
moduli are $Y^A= p^A/U$. By far not all holomorphic 4-cycles have
positive charge $p$. On the other hand, wrapping an M5 on such
4-cycles does seem to give rise to a sensible decoupled MSW
sigma-model. This presents a puzzle similar to the one caused by
M5-M5 bound states discussed above. If, as seems plausible, these
charges are realized as M5-M5 2-centered bound states in gravity,
the resolution would be the same as there.} The IIA interpretation
of these charges is (D6,D4,D2,D0); the M-theory one is
(KK,M5,M2,P). The positions $x_a$ have to satisfy the
integrability constraints
\begin{equation} \label{integrabconstr}
\sum_b\frac{\langle\G_a,\G_b\rangle}{|x_a-x_b|}=-\frac{p^0_a}{4}\,,
\end{equation}
where we define the symplectic intersection product
\begin{equation}
 \langle \Gamma_1,\Gamma_2 \rangle := - p_1^0 q^2_0 + p_1^A q^2_A - q^1_A p_2^A
 +  q^1_0 p_2^0.
\end{equation}
Half this product equals the amount of angular momentum
(corresponding to rotations of the asymptotic $S^2$) stored in the
electromagnetic field produced by this pair of charges.

The metric, gauge field and K\"ahler scalars of the solution are
given by
\begin{eqnarray}
ds^2_{5d}&=&2^{-2/3}\,Q^{-2}\left[-(H^0)^2(dt+\omega)^2-2L(dt
+\omega)(d\psi+\omega_0)
+\Sigma^2(d\psi+\omega_0)^2\right]\nonumber\\&&+2^{-2/3}\,Q\,dx^idx^i\,,\label{multicentersolutio}\\
A^A_{5d}&=&\frac{-H^0\,y^A}{Q^{3/2}}(dt+\o)+\frac{1}{H^0}\left(H^A-\frac{Ly^A}{Q^{3/2}}\right)(d\psi+\o_0)+\cala^A_d\,,\nonumber\\
Y^A&=&\frac{2^{1/3}y^A}{\sqrt{Q}}\,,\nonumber
\end{eqnarray}
where $x^i\in \mathbb{R}^3$ and $\psi$ is an angular coordinate with
period $4\pi$, and the functions appearing are given by
\begin{eqnarray}
 d\omega_0 & = &\star dH^0\,,\nonumber\\
 d\cala_d^A & = &\star dH^A \,,\nonumber\\
 \star d\o & = & \langle dH,H\rangle \,,\nonumber\\
 \S&=&\sqrt{\frac{Q^3-L^2}{(H^0)^2}}\,,\label{conditiones}\\
 L&=&H_0(H^0)^2+\frac{1}{3}D_{ABC}H^AH^BH^C-H^AH_AH^0\,,\nonumber\\
 Q&=&(\frac{1}{3}D_{ABC}y^Ay^By^C)^{2/3}\,,\nonumber\\
 D_{ABC}y^Ay^B&=&-2H_CH^0+D_{ABC}\,H^AH^B\,.\nonumber
\end{eqnarray}
Here the Hodge star is with respect to the flat $\mathbb{R}^3$
spanned by the coordinates $x^i$ and $D_{ABC}$ are the triple
intersection numbers of the chosen basis of $H^2(X)$. Note that
the only equation which might not have an explicit closed form
solution is the last one (the first three can be solved explicitly
as was done e.g.\ in \cite{Bates:2003vx}). In some cases, for
example when $b_2=1$, a closed form solution is easily obtained.

Asymptotically, the geometry is not quite AdS$_3 \times S^2$, but an $S^2$ bundle over AdS$_3$:
\begin{eqnarray}
 ds^2 &\approx& d\eta^2 + e^{\eta/U} (-d\tau^2+d\sigma^2)
+U^2\left(d\theta^2+\sin^2\theta \, ( d\phi + \tilde{A} )^2 \right) \,, \\
\tilde{A} & = & \mbox{$\frac{J}{J_{\rm max}}$} d(\tau-\sigma) \\
A^A_{\rm 5d}& \approx &-p^A\cos\theta \, (d\phi+\tilde{A}) + 2 D^{AB}q_B \, d(\sigma+\tau) \,, \\
Y^A& \approx &\frac{p^A}{U} \,.
\end{eqnarray}
where $U:=(\frac{1}{6} p^3)^{1/3}$, $D^{AB}=(D_{ABC}p^C)^{-1}$,
and we made the change of coordinates\footnote{Here, the
coordinates $(r,\theta,\phi)$ are standard spherical coordinates
for $\vec{x}$.} $(r,t,\psi) \to (\eta,\tau,\sigma)$ to
leading order given by:
\begin{equation}
 \eta:=U \, \log \frac{r}{U}, \qquad \tau:=t,
  \qquad \sigma:=\frac{\psi}{2} - t \, .
\end{equation}
The flat connection $\tilde{A}$ determines the twisting of the $S^2$ over the AdS$_3$ base. $J$ is the $S^2$-angular momentum of the solution and $J_{\rm max} := \frac{p^3}{12}$ is its maximal value for given $p$.

A solution to the integrability constraints (\ref{integrabconstr})
does not automatically imply a well-behaved full solution --- the
formally obtained metric may still have various unacceptable
pathologies such as closed timelike curves. Determining when an
actual well behaved solution exists is in general a difficult
problem. In asymptotically flat space this can up to a certain
extent\footnote{Multicentered ``scaling'' or ``abyss'' solutions
\cite{Denef:2007vg,Denef:2002ru,Bena:2007qc,deBoer:2007}, for
which the centers' coordinates can approach each other arbitrarily
closely, are viewed as being continuously connected to single
centered solutions, hence the split flow conjecture unfortunately
does not say anything about the existence of such solutions for a
given charge partitioning.} be circumvented by making use of the
``split attractor flow'' conjecture \cite{Denef:2007vg}, which
states that there is a one to one correspondence between connected
components of solution spaces of physical multicentered solutions
and attractor flow trees. An attractor flow tree consists of
single center attractor flows which are allowed to split on walls
of marginal stability. The starting point of the tree is the
asymptotic value of the moduli, and the end points of its branches
are the attractor points of the constituent charges. While still
somewhat involved, it is in general much simpler to establish the
existence of attractor flow trees than the existence of full
solutions. The basic idea behind the (well supported) conjecture
is that in asymptotically flat space, one can tune the asymptotic
moduli to follow precisely the behavior of the moduli along a
particular flow tree. By doing so, one can adiabatically assemble
or disassemble multicentered solutions. In this way, we also get a
natural partitioning of the Hilbert space of BPS states at a given
point in IIA moduli space, according to their flow tree
association, and this was further used in \cite{Denef:2007vg} to
derive various wall crossing formulae for BPS indices.

In the decoupled asymptotically AdS$_3 \times S^2$ limit there are
no moduli to tune --- all relevant asymptotic moduli are frozen to
their attractor values. Nevertheless, our uplift and decoupling
procedure combined with the split attractor flow conjecture allows
to conclude that there is a one to one correspondence between
connected components of the solution space of multicentered
asymptotic AdS$_3 \times S^2$ solutions of total (M5,M2) charge
$(p^A,q_A)$ and IIA attractor flow trees which persist in the limit
in which we take the starting point of the flow to
\begin{equation} \label{llllll}
 B^A + i J^A = D^{AB} q_B + i \Lambda p^A, \quad \Lambda \to \infty.
\end{equation}
Here $B^A$ and $J^A$ are the components of the IIA B-field and
K\"ahler form. This limiting point is essentially the attractor
point associated to the AdS$_3 \times S^2 \times X$ geometry for
the given charges, translated to type IIA variables. Split flows
which disappear when going from $J^A = \infty Y^A$ to $J^A =
\infty p^A$ due to wall crossing encode the fragmentation into
multiple decoupled AdS$_3 \times S^2$ geometries dual to the
different CFT's that appear in the low energy limit of wrapped M5
branes in a background with K\"ahler moduli $Y^A$.

The flow tree picture is also useful to understand certain
possible degenerations of solution spaces in the decoupling limit.
(Degenerations are more an issue now than in the asymptotically
flat case, precisely because we no longer have asymptotic moduli
we can use to tune away accidental degenerations, and because in
the $\Lambda \to \infty$ limit, central charges are prone to line
up as they become dominated by their leading terms.) To this end,
we distinguish the notions of marginal and threshold stability
walls. Both are associated to central charges lining up, but in
the former case the charges have intersection product nonzero, in
the latter case zero\footnote{This is the criterion for threshold
stability for two charges, a precise definition for more than two
charges is more involved and will not be discussed in this paper.}.
When crossing the former, flow trees and solutions disappear and
BPS indices jump, whereas when crossing the latter, flow trees
merely change topology, solution spaces expand, hit infinite
extent and contract again, and BPS indices remain invariant. If
the limiting value of $B+iJ$ given in (\ref{llllll}) lies on a
threshold stability wall, the corresponding solution exists in the
decoupling limit, but the solution space will be noncompact in the
sense that some centers can reach the boundary of
AdS$_3$.\footnote{However, as will be shown in the companion paper
\cite{deBoer:2007}, the solution space, viewed as a BPS phase
space, still has finite symplectic volume.}

The outline of this paper is as follows. In section
\ref{sol_review}, we review multicentered black hole solutions in
four dimensions, their uplift to five dimensions, and the split
flow conjecture. In section \ref{sec3}, we take the decoupling
limit, study its asymptotics and determine the dual CFT quantum
numbers of the solutions. For a single 5d black ring, the
decoupling limit is closely related to the decoupling limits considered in
\cite{Elvang:2004ds,Bena:2004tk}. We are not aware though of any
systematic discussion in the literature of the decoupling limit in
the case of multicentered solutions. We end the section by
formulating the existence criterion based on the split flow
conjecture, discussing various possible behaviors of the solutions
in the decoupling limit in this picture. In section
\ref{examples}, we give some examples. This includes the uplift of
a 4d D6-anti-D6 dipole, which becomes global (twisted) AdS$_3
\times S^2$ in the decoupling limit, as well as configurations
giving rise to the 4d Entropy Enigma. Section \ref{entropyenigma}
is devoted to demystifying the enigma by identifying it as a
supersymmetric version of the Banks-Douglas-Horowitz-Martinec
localization instability. We discuss the corresponding phase
transitions in the microcanonical and canonical ensembles. In
section \ref{cftenigma}, we initiate interpretations of the
observed gravitational phenomena in the MSW CFT. We conclude in
section \ref{conclusions}. Appendix \ref{conventions} details our
conventions, appendix \ref{margtresh} details the distinction between
marginal and threshold stability walls, appendix \ref{app_rescaled} restates the supergravity
solutions in a rescaled form, convenient for taking the decoupling
limit, and appendix \ref{D4D2D0}
gives a nontrivial, explicit example of a D4-D4 (or M5-M5) two
centered bound state with a line of marginal stability extending
all the way into the large volume limit, for the 2-modulus
Calabi-Yau $X_8[1,1,2,2,2]$. These are the bound states that give
rise to the puzzle mentioned in the beginning. Finally, appendix
\ref{csterm} gives some details of the computation of the CFT
quantum numbers from the solution geometries.

In a companion paper \cite{deBoer:2007}, the (quantum) structure of
the solutions spaces will be analyzed.


\section{Black constellations in four and five dimensions}\label{sol_review}

We begin with a brief review of multicentered black hole solutions of $\mathcal{N}=2$
supergravity in 4 dimensions and their lift to 5 dimensions. The four dimensional theory is obtained by
compactifying IIA on a proper $SU(3)$ holonomy Calabi-Yau manifold $X$, the five dimensional theory from compactifying M-theory on the same Calabi-Yau manifold. In the regime of interest to us, we can restrict to the cubic part of the IIA prepotential.

The multicentered solutions are determined by specifying a number of
charges, $\Gamma_a$, and their locations, $\vec{x}_a$, in the spatial
$\mathbb{R}^3$. These charged centers correspond in the 10 dimensional picture
to branes wrapping even cycles in the CY$_3$. There are $2b_2+2$ independent
such cycles in homology, with $b_2$ the second Betti-number of $X$, each giving
rise to a charge in 4d sourcing one of the $2b_2+2$ vector fields of the
$\caln=2$ supergravity. We will often denote the charges by their coefficients
in a basis of cohomology, i.e.\  $\G=(p^0,p^A,q_A,q_0)=p^0+p^AD_A+q_A\tilde D^A+q_0
\, dV$, where the $D_A$ form a basis of $\mathrm{H}^2(X,\mathbb{Z})$, the $\tilde
D_A$ make up a dual basis and $dV$ is the unit volume element of X; $\int_X dV \equiv 1$.

The moduli of the Calabi-Yau appear as scalar fields in
the 4d/5d effective theories. In the solutions we will be considering the
hypermultiplet moduli will be constant (and will mostly be irrelevant)
while the moduli in the vector multiplets will vary dynamically in response to charged sources. An
important boundary condition in these solutions is then the value of these
vector multiplet moduli at infinity.

Our review of these solutions will be concise,
as they are discussed in great detail in e.g.\ the references \cite{Denef:2000nb,Bates:2003vx,Denef:2007vg}. We will recall the {\em split attractor flow conjecture}, which relates the existence of solutions
at particular values of the moduli at infinity to the existence of certain flow
trees in moduli space. A short discussion of the concept of marginal stability, distinguishing between proper marginal stability and what we call threshold stability, is given in appendix \ref{margtresh}.

\subsection{Four dimensional solutions}

Our starting point are the multicentered black hole solutions of \cite{Denef:2000nb,LopesCardoso:2000qm,Bates:2003vx}.
The solutions are entirely determined in terms of a single function $\Sigma$, which is obtained from
the charge $(p^0,p^A,q_A,q_0)$ single centered BPS black hole entropy $S(p^0,p^A,q_A,q_0)$ by
substituting
\begin{equation}
 \Sigma := \frac{1}{\pi} \, S(H^0,H^A,H_A,H_0)\,,
\end{equation}
where
\begin{equation}
H \equiv
(H^0,H^A,H_A,H_0):=\sum_a\frac{\G_a\,\sqrt{G_4}}{|{x}-{x}_a|}-2\mathrm{Im}(e^{-i\alpha}\O)|_{r=\infty}\,
.
\end{equation}
Here $G_4$ is the four dimensional Newton constant (i.e the
Einstein-Hilbert action is of the form
$S^{\mathrm{EH}}_{4}=\frac{1}{16\pi\,G_4}\int\sqrt{-g_4}\calr_4$).  We keep this
dependence on $G_4$ explicit for now as it will be important when we take
the decoupling limit. The $\G_a$ in the $2b_2+2$ harmonic functions take values in $H^{\rm ev}(X, \mathbb{Z})$, the
integral even cohomology of the Calabi-Yau $X$, $e^{i\a}$ is the
phase of the total central charge\footnote{More explicitly
$Z(\G)=\langle\sum_a\G_a,\O\rangle$ and
$e^{i\a}=\frac{Z}{|Z|}$.}
and $\Omega$ is the normalized period vector defining the special geometry. $\G_a$ is
the charge vector of the center at position $\vec{x}_a$. The constant term of the harmonic functions is such that
$\Sigma|_{r = \infty} = 1$.

The solutions are now given by the following four dimensional metric, gauge fields and
moduli\,\footnote{We will work for the moment in conventions where we take
$c=\hbar=1$ but keep dimensions of length explicit. The formulae here can be
compared with those of e.g.\   \cite{Bates:2003vx} by noting that there the
convention $G_4=1$ was used. For more information concerning the
conventions and different length scales used in this paper, see appendix
\ref{conventions}.}:
\begin{eqnarray}
 ds^2 & = & -\frac{1}{\S}(dt+\sqrt{G_4}\,\o)^2+\S\, dx^idx^i\,, \nonumber\\
 \cala^0 & = & \frac{\partial \log \S}{\partial H_0}\left(\frac{dt}{\sqrt{G_4}}+\o\right)+\omega_0\,,\label{multicenter}\\
 \cala^A & = & \frac{\partial \log \S}{\partial H_A}\left(\frac{dt}{\sqrt{G_4}}+\o\right)+\cala_d^A\,,\nonumber\\
 t^A&=&B^A+i\,J^A=\frac{H^A-i\frac{\partial \S}{\partial H_A}}{H^0+i\frac{\partial \S}{\partial H_0}},\nonumber
\end{eqnarray}
The off diagonal metric components can be found explicitly too \cite{Bates:2003vx} by solving
\begin{equation} \label{omegaeom}
 \star d\o = \frac{1}{\sqrt{G_4}}\langle dH,H\rangle \, ,
\end{equation}
where the Hodge $\star$ is on flat $\mathbb{R}^3$. The Dirac parts $\cala_d^A$, $\omega_0=\cala_d^0$ of the vector potentials are obtained by solving
\begin{eqnarray}
 d\omega_0 & = & \frac{1}{\sqrt{G_4}}\star dH^0 \,, \\
 d\cala_d^A & = & \frac{1}{\sqrt{G_4}}\star dH^A \,.
\end{eqnarray}
Again the Hodge star $\star$ is on flat
$\mathbb{R}^3$. Asymptotically for $r \to \infty$ we have\footnote{Here
$\mathcal{A}_d$ includes both $\omega_0$ and $\mathcal{A}_d^A$. }
\begin{equation} \label{asympt4d}
 ds^2 = -dt^2 + d\vec x^2, \qquad \cala = 2\, {\rm Re}\,(e^{-i \alpha}
 \Omega)|_{\infty} \, \frac{dt}{\sqrt{G_4}}
 + \mathcal{A}_d|_\infty
\end{equation}

The above form of the solution holds for any prepotential. However it still requires finding the entropy function $S(p,q)$
which in general cannot be obtained in closed form. If we take the prepotential to be cubic, which is
tantamount to taking the large volume limit in IIA, we can be more explicit. First, the period vector becomes $\Omega=-\frac{e^{B+iJ}}{\sqrt{\frac{4J^3}{3}}}$, considered as an element of $H^{\rm ev}(X,\mathbb{R})$. Furthermore \cite{Shmakova:1996nz,Gaiotto:2005xt},
\begin{eqnarray}
  \cala^0 & = & \frac{-L}{\S^2}\left(\frac{dt}{\sqrt{G_4}}+\o\right)+\omega_0\nonumber\\
  \cala^A & = &\frac{H^AL-Q^{3/2}y^A}{H^0\S^2}
  \left(\frac{dt}{\sqrt{G_4}}+\o\right)+\cala_d^A\,,\nonumber\\
 t^A&=&\frac{H^A}{H^0}+\frac{y^A}{Q^{\frac{3}{2}}}\left(i\S-\frac{L}{H^0}\right),\nonumber \\
 \S&=&\sqrt{\frac{Q^3-L^2}{(H^0)^2}}\,,\label{conditions}\\
 L&=&H_0(H^0)^2+\frac{1}{3}D_{ABC}H^AH^BH^C-H^AH_AH^0\,,\nonumber\\
 Q^3&=&(\frac{1}{3}D_{ABC}y^Ay^By^C)^2\,,\nonumber\\
 D_{ABC}y^Ay^B&=&-2H_CH^0+D_{ABC}H^AH^B\,.\nonumber
 \end{eqnarray}
The entropy function $\Sigma$ will play a central role in the
discussion that follows.  At the horizon of one of the bound black holes this
function will be proportional to the entropy, i.e.\
$\S(H)|_{(x \to x_a)}=\frac{G_4}{|x-x_a|^2}\S(\G_a)+\calo(\frac{\sqrt{G_4}}{|x-x_a|})$ where
$\pi\Sigma(\G_a)=S(\G_a)$ is the Bekenstein-Hawking entropy of the $a$th
center\footnote{Note that the entropy formula for black holes involving
$\Dp{6}$-charge is rather involved and might appear singular as $H^0$ (or $p^0$)
goes to zero, see (\ref{conditions}). This is however not the case and by
analysing the formula in an expansion around small $H^0$ one finds that the
leading term is the non-singular entropy function for a black hole without
$\Dp{6}$-charge, $\S=\sqrt{\frac{D_{ABC}H^AH^BH^C}{3}(D^{AB}H_AH_B-2H_0)}$, as
expected.}.

Finally there are $N-1$ independent consistency conditions on the relative
positions of the $N$ centers, reflecting the fact that these configurations really are bound states and one can't move the centers around freely. These conditions arise from requiring
integrability of (\ref{omegaeom}). They take the simple form
\begin{equation}
  \langle H,\G_s\rangle|_{x=x_s}=0\,,
 \label{consistency}
\end{equation}
or written out more explicitly\footnote{For brevity we use unconventional
notation here: by $\sum_{s\neq \ul r}$ we mean a sum over all $s$ different
from $r$ whereas $\sum_{s\neq r}$  denotes a doubles sum over all $s$ and $r$
such that $s$ and $r$ are different.}
\begin{equation}  \sqrt{G_4}\,\sum_{b \neq \ul a}
   \frac{\langle
    \G_a,\G_b\rangle}{r_{ab}}=\langle h,\G_a\rangle\,, \label{consistency2}
\end{equation}
where $r_{ab}=|x_{ab}|=|x_a-x_b|$ and $h=-2\mathrm{Im}(e^{-i\alpha}\O)|_\infty$
are the constant terms in the harmonic functions. Note that, as these depend on
the asymptotic values of the scalar fields, the equilibrium distances between the
different centers do so as well.

Since there are $N-1$ independent position constraints, the dimension of the moduli
space modulo the center of mass translations will generically be $2N-2$.

\subsection{Five dimensional solutions}

In \cite{Gaiotto:2005xt} (see also
\cite{Gaiotto:2005gf,Behrndt:2005he,Cheng:2006yq,
Bena:2005ay,Bena:2005ni,Elvang:2005sa,Berglund:2005vb,Balasubramanian:2006gi})
these solutions were lifted to five dimensions via the connection
between IIA and M-theory on a circle. The five dimensional
solution can be expressed in terms of the four dimensional one as (see appendix \ref{conventions} for more details about notations and conventions):
\begin{eqnarray}
ds^2_{5d}&=&\tilde{V}_{\IIA}^{2/3}\,\ell_5^2\left(d\psi+\cala^0\right)^2+\tilde{V}_{\IIA}^{-1/3}\,\frac{\hat{R}}{2}\,ds^2_{\mathrm{4d}}\,,\nonumber\\
A_{\mathrm{5d}}^A&=&\cala^A+B^A\left(d\psi+\cala^0\right)\,,\label{5Dsolution}\\
Y^A&=&\tilde{V}_{\IIA}^{-1/3}\,J^A\,,\qquad\tilde{V}_{\IIA}=\frac{D_{ABC}}{6}J^AJ^BJ^C=\frac{1}{2}\left(\frac{\Sigma}{Q}\right)^3\,.\nonumber
\end{eqnarray}
Here $\psi$ parametrizes the M-theory circle with periodicity
$4\pi$ and we define, in terms of the 11d Planck length $l_{11}$ and the physical asymptotic M-theory circle radius $R$,
\begin{equation}
 \ell_5 := \frac{l_{11}}{4 \pi \tilde{V}_M^{1/3}} \, , \qquad \hat{R} = \frac{R}{\ell_5} \, ,
\end{equation}
where $\tilde{V}_M = V_M/l_{11}^6$ is the M-theory volume of $X$ in 11d Planck units. The reduced 5d Planck length $\ell_5$ is related to the 4d Newton constant $G_4$ by
\begin{equation}
 \ell_5^3 = R \, G_4
\end{equation} 
and we have the relation $\hat{R}=2 \, \tilde{V}_{\IIA}^{1/3}|_\infty$. Note that unlike the M-theory volume in 11d Planck units, which is in a hypermultiplet and hence constant, the IIA volume in string units varies over space. Our normalizations are chosen such that asymptotically we have the
metric
\begin{eqnarray}
 ds^2_{5d}|_{\infty}&=&\frac{R^2}{4} \left(d\psi+\cala^0\right)^2 + d \vec x^2 - dt^2 \,,\\
  \cala^0& =& -2 \cos \alpha_\infty\, \frac{dt}{R}
 + p^0 \cos\theta \, d\phi\,,\nonumber
\end{eqnarray}
where $\cala^0$ was obtained from (\ref{asympt4d}), and we recall that $e^{i
\alpha}$ is the phase of the total central charge.  Recall that $p^0$ is the total
$D6$-charge of the solution and for most of this paper we will take this to be
zero.

The five dimensional vector multiplet scalars $Y^A$ are related to the M-theory K\"ahler moduli by
$J^A_M=\tilde{V}_M^{1/3}Y^A$. Here $\tilde{V}_M=\frac{V_M}{l_{11}^6}$ is the volume
of the internal Calabi-Yau as measured with the M-theory metric. This is
constant throughout the solution as it is in a hypermultiplet and hence
decoupled. For more details about all the different length scales and the
relation between M-theory and IIA variables in our conventions see appendix
\ref{conventions}.

For practical computations it is often useful to express the metric
(\ref{5Dsolution}) above more explicitly in terms of the functions
(\ref{conditions}):
\begin{eqnarray}
ds^2_{5d}&=&2^{-2/3}Q^{-2}\left[-\ell_5^2\,(H^0)^2\biggl(\sqrt{\frac{R}{\ell_5^3}}dt+\omega\biggr)^2-2\ell_5^2\,L\,\biggl(\sqrt{\frac{R}{\ell_5^3}}dt+\,\omega\biggr)(d\psi+\omega_0)
+\ell_5^2\,\Sigma^2(d\psi+\omega_0)^2\right]\nonumber\\&&+2^{-2/3}\frac{R}{\ell_5}Q\,dx^idx^i\,.
\end{eqnarray}

Finally, note that by construction, all these five dimensional solutions have a $U(1)$ isometry along the $\psi$ direction. They are therefore not the complete set of five dimensional BPS solutions.

\subsection{Properties}

Let us briefly
recall some relevant properties of these multicentered solutions.

The first new feature with respect to single black holes is that, as shown in \cite{Denef:2000nb}, they carry an angular momentum equal to
\begin{equation}
\vec J=\frac{1}{2}\sum_{ a<  b} \frac{\langle \G_a,\G_b\rangle\,\vec{x}_{ab}}{r_{ab}}\label{angularmomentum}\,.
\end{equation}
Note that Dirac quantization of the charges is equivalent to half integral quantization of the angular momentum of a two centered solution. This angular momentum is
associated to $SO(3)$ rotations in the three non-compact spacelike dimensions and
should not be confused by the momentum around the M-theory circle (which, in the
four dimensional picture, corresponds to the  $\Dp{0}$-charge $q_0$).

Another important property of a configuration with a sufficient number of centers is that although the centers bind to each other there is some freedom left to change their respective positions. These possible movements can be thought of as flat directions in the interaction potential. Equation (\ref{consistency2}) constrains the locations of the centers to the points where this potential is zero.  As for a system with
$N$ centers there are $N-1$ such equations for $3N-3$ coordinate
variables (neglecting the overall center of mass coordinate) there is, in general,
a $2N-2$ dimensional moduli space of solutions for fixed charges and
asymptotics.  This space may or may not be connected and it may even have
interesting topology.  We will refer to this as the moduli space of
solutions or solution space; the latter terminology will be preferred as it
is less likely to be confused with the moduli space of the Calabi-Yau, in which
the scalar fields $t^A$ take value.  The shape of this solution space does, in fact, depend quite
sensitively on where the moduli at infinity, $t^A|_\infty$, lie in the
Calabi-Yau moduli space (as the latter determine $h$ on the RHS of eqn.
(\ref{consistency2})). We will return in more detail to the geometry of the solution space and more specifically to its quantization in \cite{deBoer:2007}.

The space-time corresponding to a generic multicenter
configuration can be rather complicated as there can be many
centers of different kinds. Some properties of the 5 dimensional
geometry have been discussed in the literature, e.g.\
\cite{Bena:2005va,Cheng:2006yq,Berglund:2005vb,Gaiotto:2005xt} and
we won't repeat the details here. A basic understanding will be
useful when considering the decoupling limit so we shortly
summarize some points of interest.  The four dimensional solutions
are defined on a space that is topologically $\mathbb{R}^4$.  When
lifted to five dimensions, however, a Taub-NUT circle is fibred
over this space pinching at the location of any center with
$D6$-charge.  The resultant space typically has non-contractible
two-spheres extending between centers with $\Dp{6}$ charge and has
been referred to as a ``bubbling solution'' \cite{Bena:2005va}.
Generically a $\Dp{4}$ charged center will lift to a black string
unless it also caries $\Dp{6}$-charge in which case it lifts to
what locally looks like a BMPV black hole at the center of 5
dimensional Taub-NUT \cite{Gaiotto:2005gf}. The topology of the
horizon at a given center is that of an $S^1$-bundle over $S^2$ of
degree $p^0_a$, i.e. $S^1\times S^2$ for $p^0_a=0$ and
$S^3/\mathbb Z_{|p^0_a|}$ otherwise.

Finally let us mention a symmetry of the solutions (which is closely related to
the one observed in \cite{Bena:2005ni}, \cite{Cheng:2006yq}) given by the
following shift of the harmonic functions:
\begin{align}
 H^0 &\rightarrow H^0\,,\nonumber\\
 H^A &\rightarrow H^A + k^A H^0\,, \label{chargesym} \\
 H_A &\rightarrow H_A + D_{ABC}H^B k^C + \frac{1}{2} D_{ABC}k^B k^C H^0\,, \nonumber\\
 H_0 &\rightarrow H_0 + k^A H_A + \frac{1}{2} D_{ABC} H^A k^B k^C + \frac{1}{6} D_{ABC} k^A k^B k^C (H^0)\,.\nonumber
\end{align}
Under which the metric and the constraint equations are invariant and the gauge
field is transformed by a large gauge transformation
\be \label{large_gauge}
 A^A \rightarrow A^A + k^A d\psi\,.
\ee

\subsection{The split attractor flow conjecture}\label{afconjecture}

So far we have reviewed a class of 4 and 5 dimensional solutions. These
solutions are relatively complicated and it is non-trivial to determine if they
are well-behaved everywhere.  In particular one should be concerned about the
appearance of closed timelike curves or singularities. If the entropy function,
$\Sigma$, which involves a square root, becomes zero or takes imaginary values
in some regions the 4d solution is clearly ill behaved; this is equivalent to
closed timelike curves in the 5d metric as discussed in \cite{Cheng:2006yq} and
\cite{Bena:2005va}. One can on the other hand show that if $\S^2>\o_i\o^i$
everywhere then there can be no closed timelike curves \cite{Berglund:2005vb}. This
is a rather complicated condition to check for a generic multicenter solution
however and furthermore it is sufficient but not necessary; the condition could be violated
without closed timelike curves appearing.  In \cite{Denef:2000nb} and
\cite{Denef:2007vg} a simplified criteria was proposed for the existence of
(well-behaved) solutions which we will now relate.

In \cite{Denef:2000nb} a conjecture was motivated whereby the existence of a multicentered solution is equivalent to the existence of an {\em attractor flow
tree}.  The latter is a graph in the Calabi-Yau moduli space beginning at the
moduli at infinity, $t^A|_\infty$, and ending at the attractor points for each
center, see figure \ref{splitflowtree}.
\FIGURE{
\includegraphics[page=1,scale=0.7]{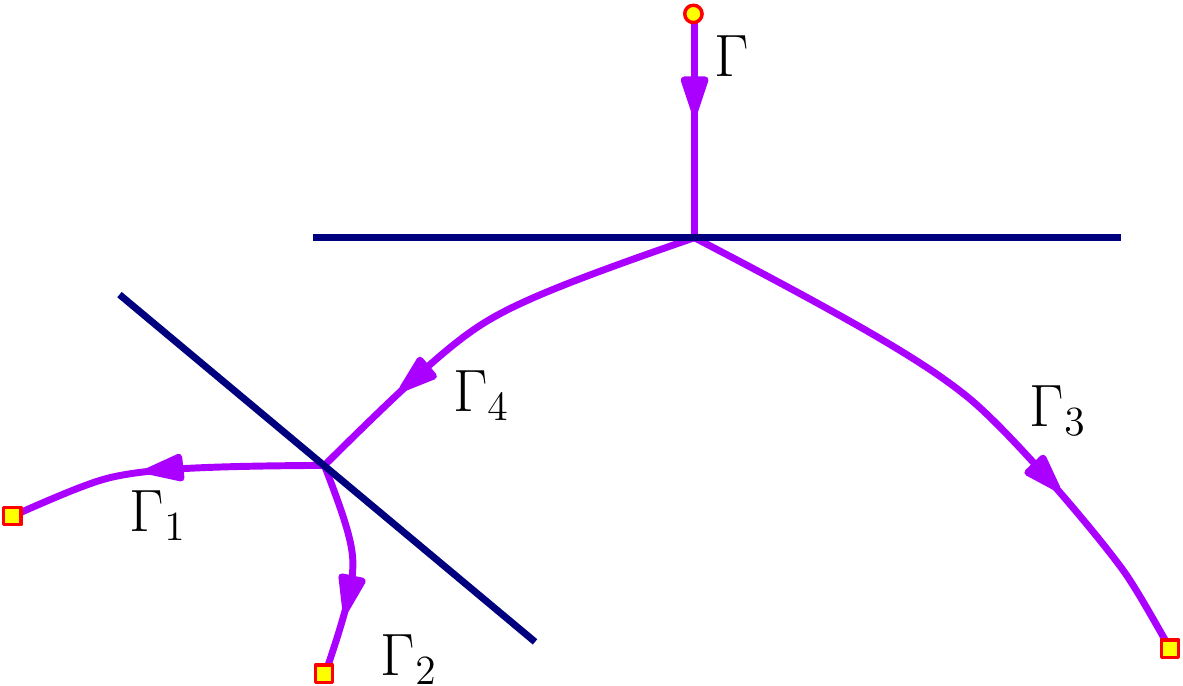}
\caption{Sketch of an attractor flow tree. The dark blue lines are lines of marginal stability, the purple lines are single center attractor flows. The tree starts at the yellow circle and flows towards the attractor points indicated by the yellow boxes.}\label{splitflowtree}
}
The edges correspond to single center flows towards the attractor point
for the sum of charges further down the tree.  Vertices can occur where single
center flows (for a charge $\Gamma = \Gamma_1 + \Gamma_2$) cross walls of
marginal stability where the central charges are all aligned ($|Z(\Gamma)| =
|Z(\Gamma_1)| + |Z(\Gamma_2)|$).  The actual (multi-parameter) flow of the
moduli $t^A$ for a multicentered solution will then be a thickening of this
graph as e.g.\ in figure \ref{splitflow} (see \cite{Denef:2000nb}, \cite{Denef:2007vg} for more details). For a given attractor flow there will be a single connected set of solutions to the equations (\ref{consistency2}) that all have a well-behaved space-time geometry.

When we consider the decoupling limit of the multicenter solutions in the
next section we will see that the attractor flow conjecture and its utility in
classifying solutions can be extended to AdS space.\\\\

\section{Decoupling limit}\label{sec3}
As outlined in the introduction, we want to study the geometries
dual to states of M5-branes wrapped on 4-cycles with total
homology class $p^A D_A$, in the decoupling limit $R/l_{11} \to
\infty$, $V_M/l_{11}^6$ fixed. A convenient way to take the limit
is to adapt units such that $R$ remains finite --- for example $R
\equiv 1$ --- while $\ell_5 \to 0$ (note that $l_{11}/\ell_5$ is
fixed because $V_M/l_{11}^6$ is fixed). Then the dynamics of
finite energy excitations of the M5 are described by a (0,4)
supersymmetric 1+1 dimensional nonlinear sigma model with target
space naively\footnote{As discussed in the introduction and
further in section~\ref{mswstring}, the precise M5-brane
interpretation of the decoupling limit is rather mysterious and
still poses various puzzles.} given by the classical M5 moduli
space, the MSW string \cite{Maldacena:1997de,Minasian:1999qn},
decoupled from bulk and KK modes. For example, Kaluza-Klein
excitations along the 4-cycle decouple as their mass is of order
$V_M^{-1/6}$, which scales to infinity.

We wish to find out how multicentered solutions with total charge
$(0,p^A,q_A,q_0)$ behave when we take this limit. The IIA K\"ahler
moduli $J^A$ are related to the normalized scalars $Y^A$ as $J^A
\sim \frac{R}{\ell_5} \, Y^A$, hence $J \to \infty$. For two
centered solutions involving D6-charges, the equilibrium
separation following from the integrability condition
(\ref{consistency}) asymptotes to
\begin{equation}
 |\vec{x}_1-\vec{x}_2| = \frac{\langle \Gamma_1,\Gamma_2 \rangle}{2 \, {\rm Im}(e^{-i \alpha} Z_1)|_{\infty}} \frac{\ell_5^{3/2}}{\sqrt{R}} \sim \frac{\langle \Gamma_1,\Gamma_2 \rangle}{R^2} \, \ell_5^3,
\end{equation}
where we used that for total D6-charge zero, $\alpha \to 0$ when $J \to \infty$, while $Z_1 \sim i  J^{3/2} \sim i (R/\ell_5)^{3/2}$.

To keep the coordinate separation finite in the limit $\ell_5 \to 0$, we should therefore rescale all coordinates as
\begin{equation} \label{rescalingcoords}
 \vec x = \ell_5^3 \, \vec{\mathsf{x}}.
\end{equation}
The finite $\vec{\mathsf{x}}$ region then has the expected properties for a decoupling limit. First, as we will see, at finite values of $\vec{\mathsf{x}}$, the metric converges to an expression of the form $ds^2=\ell_5^2 \,  \mathsf{ds^2}$ with $\mathsf{d s^2}$ finite. Finite fluctuations of $\mathsf{d s^2}$ thus give rise to finite action fluctuations --- the $\ell_5^2$ metric prefactor cancels the $\ell_5^{-3}$ in front of the Einstein-Hilbert action \cite{Maldacena:1997re}. Similarly, M2-branes wrapping the M-theory circle and stretched over finite $\vec{\mathsf{x}}$ intervals have finite energy. Finally, the geometry becomes asymptotically an $S^2$ bundle over AdS$_3$ at large $r=|\vec{\mathsf{x}}| \to \infty$:
\begin{eqnarray}
\mathsf{ds^2} &\approx& d\eta^2 + e^{\eta/U} (-d\tau^2+d\sigma^2)
+U^2\left(d\theta^2+\sin^2\theta \, ( d\phi + \tilde{A} )^2 \right) \,, \label{asn1} \\
\tilde{A} & = & \mbox{$\frac{J}{J_{\rm max}}$} d(\tau-\sigma) \\
A^A_{\rm 5d}& \approx &-p^A\cos\theta \, (d\phi+\tilde{A}) + 2 D^{AB}q_B \, d(\sigma+\tau) \,, \label{asn2} \\
Y^A& \approx &\frac{p^A}{U} \,. \label{asn3}
\end{eqnarray}
where $U:=(\frac{1}{6} D_{ABC} p^A p^B p^C)^{1/3}$ and we made the change of coordinates $(r,t,\psi) \to
(\eta,\tau,\sigma)$ to leading order given by:
\begin{equation}
 \eta:=U \, \log \frac{R^2 r}{U}, \qquad \tau:=\frac{t}{R},
  \qquad \sigma:=\frac{\psi}{2} - \frac{t}{R} \, .
\end{equation}
Notice that the normalized K\"ahler moduli $Y^A$ and the $U(1)$
vectors $A^A$ are fixed at attractor values determined by the M5
and M2 charges. The flat connection $\tilde{A}$ determines the
twisting of the $S^2$ over the AdS$_3$ base; $J$ is the
$S^2$-angular momentum of the solution and $J_{\rm max} :=
\frac{U^3}{2}$ is its maximal value for given $p$. Note that going
around the M-theory circle in the new coordinates corresponds to
\begin{equation} \label{periodicities}
 \sigma \to \sigma + 2\pi,
\end{equation}
with all other coordinates fixed. Parallel transport of the $S^2$ along this circle produces a rotation $\Delta \phi = \frac{J}{J_{\rm max}} 2 \pi$ around its $z$-axis (which is the axis determined by the direction of the four dimensional angular momentum). Because $\tilde A \sim d(\sigma-\tau)$, the sphere similarly gets rotated in time, resulting in angular momentum proportional to the amount of twisting around the $S^1$. Since the $S^2$ descends from the spatial sphere at infinity in four dimensions, this equals the 4d angular momentum of the 4d multicentered solution. In the dual CFT, it translates to $SU(2)_R$ charge.

The $\sigma \to \sigma+2\pi$ circle smoothly connects to the asymptotic M-theory circle in the original asymptotically flat geometry. Fermions must be periodic around this circle, as antiperiodic fermions would produce a nonzero vacuum energy. Therefore we have periodic boundary conditions for the fermions on the AdS$_3$ boundary circle, so the supersymmetric black hole configurations we are describing must correspond to supersymmetric states in the Ramond sector of the boundary CFT.

It is not true, however, that \emph{all} multicentered solutions with total charge $(0,p^A,q_A,q_0)$ give rise to  such asymptotic AdS$_3 \times S^2$ attractor geometries\footnote{Despite the nontrivial twist of the $S^2$, we will still loosely refer to the asymptotic geometry as AdS$_3 \times S^2$.} in the decoupling limit. For example D4-D4 2-centered solutions (i.e.\  $p^0_1=p^0_2=0$), of which explicit examples are given in appendix \ref{D4D2D0}, turn out to have equilibrium separations in the original coordinates scaling as
$|\vec x_1 - \vec x_2| \sim \langle \Gamma_1,\Gamma_2 \rangle \ell_5$. The different scaling compared to the case with nonzero D6-charges is due
to the fact that now $\arg Z_1 \to 0$ in the decoupling limit. In the rescaled coordinates (\ref{rescalingcoords})
the separation diverges, so these multicentered solutions therefore do not fit in the asymptotic AdS$_3 \times S^2 \times X$ attractor geometry associated to the total M5 charge $p^A$. Rather, they give rise to two mutually decoupled AdS$_3 \times S^2 \times X$ attractor geometries associated to the two individual centers. More elaborate configurations of this kind are possible too, for instance consisting of two clusters each with zero net D6-charge, but containing themselves more centers with nonzero D6-charge. The centers within each cluster will have rescaled coordinate separations of order 1, while the mutual separation between the clusters diverges like $\ell_5^{-2}$ in these coordinates.

These D4-D4 type BPS bound states exist in regions of K\"ahler moduli space separated from the overall M5 attractor point $Y^A=p^A/U$ by a wall of marginal stability. They correspond to ensembles of BPS states of the MSW string which exist at certain values of the $Y^A$ but not at the attractor point. Their interpretation in the AdS-CFT context is therefore less clear --- we will return to this in section \ref{commquest}.

In the following we wish to focus on solutions which do correspond to a single asymptotic AdS$_3 \times S^2$ in the decoupling limit, and in particular find practical criteria to determine when this will be the case. We will proceed
by rescaling coordinates as in (\ref{rescalingcoords}) and carefully studying the behavior of the solutions when $\ell_5 \to 0$. As the explicit form of the multicenter solutions is rather complicated we will first make the dependence on $\ell_5$ more clear by pulling it out through a rescaling of the variables in section \ref{rescaling}. After this rescaling the dependence on $\ell_5$ will simply be an overall factor in
the metric as described above and a dependence left in the equilibrium distance between the centers
and the constant terms of the harmonic functions. Once we have this simple form
we will take the decoupling limit by sending $\ell_5\rightarrow 0$. We calculate the asymptotics and some quantum numbers in sections \ref{asymptotics} and \ref{quantumnrs} and finally we
will discuss when the decoupling limit is well defined (in the sense that we do get a single asymptotically
AdS$_3 \times S^2$ geometry when $\ell_5 \to 0$) in section \ref{existence}.

\subsection{Rescaling}\label{rescaling}
As discussed above, to take the decoupling limit
we want to work with the rescaled coordinates, $\mathsf{x^i}$,
\begin{equation}
x^i=\ell_5^{3}\, \mathsf{x}^i\,.
\end{equation}
Furthermore we want to extract a factor of $\ell_5$ out of the 5d metric. As the multicenter solutions are rather complicated we will here first simplify the dependence on $\ell_5$ by redefining various quantities.
In the rescaled coordinates it is natural to define rescaled harmonic
functions, $\mathsf{H}$,
\begin{equation}
    \mathsf{H}= \ell_5^{3/2} H = \sum_a\frac{\G_a}{\sqrt{R}\,
     |\mathsf{x}-\mathsf{x}_a|}-2\ell_5^{3/2}\mathrm{Im}(e^{-i\alpha}\O)|_\infty\,.\label{rescaledharmonics}
\end{equation}
It is not difficult to verify that all functions appearing in
(\ref{conditions}) are actually homogenous under the rescaling of
the coordinates and harmonic functions given above. For instance
\begin{eqnarray}
y^A(H)&=&\ell_5^{-3/2}y^A(\mathsf{H})\,,\nonumber\\
Q(H)&=&\ell_5^{-3}\,Q(\mathsf{H})\,,\\
L(H)&=&\ell_5^{-9/2}L(\mathsf{H})\,,\nonumber\\
\Sigma(H)&=&\ell_5^{-3}\,\Sigma(\mathsf{H})\nonumber\,.
\end{eqnarray}
The scaling of $\omega$ is a little more subtle.  Here one has to
take into account that the $\star$ scales as well since the flat 3d
metric scales as $\ell_5^{-6}$ under the coordinate rescaling.
This implies
\begin{equation}
\star_x=\ell_5^{3(3-2p)}\star_\mathsf{x}\,,
\end{equation}
for the $\star$ acting on a $p$-form\,. So from its equation of
motion (\ref{conditions}) we see that
\begin{equation}
\omega(H,dx,\sqrt{G_4})=\ell_5^{-3/2}\omega(\mathsf{H},d\mathsf{x},R^{-1/2})\,,\label{rescalingomega}
\end{equation}
where the factor $\ell_5^{3/2}$ out of
$\sqrt{G_4}=\frac{\ell_5^{3/2}}{\sqrt{R}}$ is essential.


Note that the 4d metric from (\ref{multicenter}) scales as
\begin{equation}
ds^2_{\mathrm{4d}}(H,dx,\sqrt{G_4})=\ell_5^{-3}ds^2_{\mathrm{4d}}(\mathsf{H},d\mathsf{x},R^{-1/2})\,.
\end{equation}
Finally there are also some fields that remain invariant under the
rescaling:
\begin{eqnarray}
t^A(H)&=&t^A(\mathsf{H})\\
\o_0(H,dx,\sqrt{G_4})&=&\o_0(\mathsf{H},d\mathsf{x},R^{-1/2})\\
 \cala(H,dx,\sqrt{G_4})&=&\cala(\mathsf{H},d\mathsf{x},R^{-1/2})\,.
\end{eqnarray}
It is clear from the discussion above that the whole solution transforms
homogeneously under the rescaling of the coordinates and the redefinition of the
harmonic functions. In fact our solutions in rescaled coordinates take exactly
the same form as the original solutions in Section \ref{sol_review}, with the
only changes being the replacement of $\sqrt{G_4}$ with $R^{-1/2}$ and $H$ with
$\mathsf{H}$ everywhere.  For the readers convenience we provide the explicit
rescaled form of the solutions in Appendix
\ref{app_rescaled}.

The 5d metric in these coordinates now has a prefactor
$\ell^2_5$
\begin{eqnarray}
\frac{1}{\ell_5^2}\,ds^2_{5d}&=&2^{-2/3}\,Q^{-2}\left[-(\mathsf{H}^0)^2(\sqrt{R}dt+\omega)^2-2L(\sqrt{R}dt
+\omega)(d\psi+\omega_0)
+\Sigma^2(d\psi+\omega_0)^2\right]\nonumber\\&&+2^{-2/3}R\,Q\,d\mathsf{x}^id\mathsf{x}^i\,.\label{rescaledexplicit}
\end{eqnarray}
Otherwise, the only appearance of $\ell_5$ is
through the harmonic functions $\mathsf{H}$ in
(\ref{rescaledharmonics}). It enters there in two ways. First
through the constant terms
\begin{equation}
-2\ell_5^{3/2}\mathrm{Im}(e^{-i\alpha}\O)|_\infty\,,
\end{equation}
where it is important to recall that $\O|_\infty$ also depends on $\ell_5$ as
$J^A_\infty$ is related to $\ell_5$ by
$\frac{4J^3_\infty}{3}=\left(\frac{R}{l_5}\right)^3$.  Secondly, the equilibrium
positions $\mathsf{x}_i$ of the charged centers are determined by the
consistency condition
\begin{equation}
\langle\G_a,\mathsf{H}\rangle|_{\mathsf{x}_a}=0\,. \label{rescaledconsistency}
\end{equation}
By this equation they depend on the
constant part of the harmonics and thus $\ell_5$. We will elaborate in detail on
this dependence in the next subsection when we consider the $\ell_5\rightarrow
0$ limit.

From this point onwards we will always be working with rescaled
coordinates (unless we explicitly state otherwise).  Hence, for notational
simplicity {\bf we will revert to original notation} (e.g.\  $\S$, $ds^2_\mathrm{4d}$, $x$, $H$) though
we will be referring to the {\em rescaled} expressions (e.g.\  $\S(\mathsf{H})$,
$ds^2_{\mathrm{4d}}(\mathsf{H}, d\mathsf{t}, d\mathsf{x}, R^{-1/2})$, $\mathsf{x}$, $\mathsf{H}$).  Hopefully
this will not lead to excessive confusion.

\subsection{Decoupling}\label{decouplinglimit}

Having rewritten our solutions in a rescaled form where the
$\ell_5$ dependence is transparent (see e.g.\
(\ref{rescaled5Dsolutiona})) we can consistently take the
decoupling limit, $\ell_5\rightarrow 0$, while keeping $R,t,
{x}^i,\psi, \tilde V_M$ and $\G_i$ fixed.  As mentioned before, in
the rescaled variables $\ell_5$ only appears through the constants
in the harmonic functions so taking the limit $\ell_5\rightarrow
0$ will leave the whole structure of the solution invariant except
for replacing the harmonic functions by their limiting form.
Changing $\ell_5$ also effects the equilibrium distances of the
centers, $x_a$, in the solution due to the appearance of the
constant terms in the constraint equation (\ref{consistency2}). In
general the equilibrium distances will vary in a rather
complicated (an not unique) way, some interesting examples will be
discussed explicitly in section \ref{examples}.

Let us now examine the dependence on $\ell_5$ in the small $\ell_5$
regime.  The constant terms of the rescaled harmonic functions are
\begin{equation}
h=-2\ell_5^{3/2}\mathrm{Im}(e^{-i\alpha}\O)|_\infty\,,
\end{equation}
where $\Omega=-\frac{e^{B+iJ}}{\sqrt{\frac{4J^3}{3}}}$ and $J^A|_\infty=\frac{R}{2l_5}Y^A|_\infty$.
We can write those constant terms in an expansion for small $\ell_5$ as
\begin{eqnarray}
h^0&=&h^0_{(4)}\frac{\ell_5^4}{R^{5/2}}+\calo(\ell_5^{6})\,,\nonumber\\
h^A&=&h^A_{(2)}\frac{\ell_5^2}{\sqrt{R}}+h^A_{(4)}\frac{\ell_5^4}{R^{5/2}}+\calo(\ell_5^{6})\,,\label{formula3}\\
h_A&=&h_A^{(2)}\frac{\ell_5^2}{\sqrt{R}}+h_A^{(4)}\frac{\ell_5^4}{R^{5/2}}+\calo(\ell_5^{6})\,,\nonumber\\
h_0&=&-\frac{R^{3/2}}{4}+h_0^{(2)}\frac{\ell_5^2}{\sqrt{R}}+h_0^{(4)}\frac{\ell_5^4}{R^{5/2}}+\calo(\ell_5^{6})\,,\nonumber\,
\end{eqnarray}
where the leading terms are\footnote{To keep the formulas in (\ref{formula3}) readable we suppressed the various indices and contractions, these formulas should all be read as e.g.\  $
XYZ=D_{ABC}X^AY^BZ^C\,,\quad (XY)_A=D_{ABC}X^BY^C\,,\quad XY=X_AY^A\,.$}
\begin{eqnarray}
h_{(4)}^0&=&8\frac{pYB-qY}{pYY}|_\infty\nonumber\,,\\
h_{(2)}^A&=&Y^A_\infty\label{seriesconstants}\,,\\
h_A^{(2)}&=&(YB)_A|_\infty+\frac{Y^2_A}{pY^2}(qY-pYB)|_\infty\,,\nonumber\\
h_0^{(2)}&=&\frac{1}{2}YB^2|_\infty+\frac{BY^2}{pY^2}(qY-pYB)|_\infty+2\frac{(qY-pYB)^2}{(pY^2)^2}|_\infty\,.\nonumber
\end{eqnarray}
So in the limit $\ell_5\rightarrow 0$ all the constants in harmonics are sent to zero except for the one in the $\Dp{0}$ harmonic ${H}_0$ which reads
\begin{equation}
h_0\rightarrow-\frac{R^{3/2}}{4}\,.\label{limitconstants}
\end{equation}

The equilibrium distances also depend on the asymptotic moduli
through (\ref{consistency}). These constraints can be
written in the form
\begin{equation}
\sum_b\frac{\langle\G_a,\G_b\rangle}{\sqrt{R}\,|{x}_a-{x}_b|}=-\langle\G_a,h\rangle\,.\label{formula2}
\end{equation}
So from the behavior (\ref{limitconstants}) we see that in the
decoupling limit $\ell_5\rightarrow 0$ the consistency conditions
(\ref{consistency}) become
\begin{equation}
\sum_b\frac{\langle\G_a,\G_b\rangle}{|{x}_a-{x}_b|}=-\frac{p^0_a}{4}R^2\label{limitconsistency}\,.
\end{equation}

Summarized, the decoupling limit corresponds to replacing the harmonic
functions by
\begin{eqnarray}
{H}^0&=&\sum_a\frac{p^0_a}{\sqrt{R}\,|{x-x}_a|}\,,\nonumber\\
{H}^A&=&\sum_a\frac{p^A_a}{\sqrt{R}\,|{x-x}_a|}\,,\label{limitharmonics}\\
{H}_A&=&\sum_a\frac{q_A^a}{\sqrt{R}\,|{x-x}_a|}\,,\nonumber\\
{H}_0&=&\sum_a\frac{q_0^a}{\sqrt{R}\,|{x-x}_a|}-\frac{R^{3/2}}{4} \,.\nonumber
\end{eqnarray}
Furthermore the equilibrium distances are now determined by the
equations (\ref{limitconsistency}).

Note that this limit is similar to the usual near horizon limit, but
not quite the same, since we are not simply dropping all constant terms from the harmonic functions.
A similar situation was encountered for instance in \cite{Elvang:2004ds}, where a similar decoupling
limit is defined for the three charge super tubes.

It is useful to note that although under the decoupling limit the $\Dp{0}$
constant goes to a fixed non vanishing value, this constant can, however, be
removed by the following formal transformations
\begin{eqnarray}
     {H}_0 && \rightarrow {H}_0 + \frac{R^{3/2}}{4} \nonumber \\
     L && \rightarrow L + \frac{R^{3/2}}{4} \; ({H}^0)^2\label{zeroconstants} \\
     t && \rightarrow v=t - \frac{R}{4} \psi\nonumber
\end{eqnarray}
As this is the only effect of
the constant term in the $\Dp{0}$-brane harmonic function, we can set it to zero
while replacing
$t$ by $v = t - R/4 \; \psi$ and making a shift in $L$ at the same time. This is sometimes
technically convenient.

\subsection{Asymptotics}\label{asymptotics}

Now that we have implemented the decoupling limit we want to study the new
asymptotics of these solutions. This is completely determined by the asymptotics of the
harmonic functions. For ${r}\rightarrow\infty$ the harmonic functions
(\ref{limitharmonics}) can be expanded as
\begin{eqnarray}
{H}^0&\rightarrow&R^{-1/2}\frac{e\cdot d^0}{{r}^2}\,,\nonumber\\
{H}^A&\rightarrow&R^{-1/2}\left(\frac{p^A}{{r}}+\frac{e\cdot d^A}{{r}^2}\right)\,,\label{expandedharmonics}\\
{H}_A&\rightarrow&R^{-1/2}\left(\frac{q_A}{{r}}+\frac{e\cdot d_A}{{r}^2}\right)\,,\nonumber\\
{H}_0&\rightarrow&R^{-1/2}\left(\frac{q_0}{{r}}+\frac{e\cdot d_0}{{r}^2}\right) \nonumber\,,
\end{eqnarray}
where we have put the constant in ${H}_0$ to zero by the procedure
explained at the end of the last subsection. In our notation
\begin{equation}
 d:=\sum_a\,\G_a\,{x}_a
\end{equation}
is the dipole moment and
$\vec{e}=\frac{{\vec{x}}}{{r}}$, ${r}=|{x}|$, is the normalized
position vector that gives the direction on the S$^2$ at infinity.  Note that
for ${H}^0$ the dipole term is leading as we take the overall $\Dp{6}$
charge zero; the same is true for ${H}_A$ if the total $\Dp{2}$
charge is zero. As we will only consider cases of non-vanishing overall
$\Dp{4}$ charge here the dipole term is always subleading.

In studying the asymptotics of the physical fields it will
be most straightforward to work in a coordinate system where $d^0$ lies along the
$z$-axis. In this case
\begin{equation}
e\cdot d^0=\cos\theta |d^0|\,,
\end{equation}
with the standard spherical coordinates $(\theta,\phi)$. To simplify the
notation we will often write just $d^0$ for $|d^0|$; it should be clear from the
context when the vectorial quantity is intended and when the scalar.  Note that
the different dipole moments don't have to align so in general there is no simple
expression for e.g.\  $e\cdot d^A$ in this coordinate system.

In the decoupled geometry the $d^0$ plays a distinguished role as
it is proportional to the total angular momentum of the system. To
see this we start from the stability condition in the decoupled
theory, (\ref{limitconsistency}), multiply by $x_b$ and sum over
$b$ (note that this still is a vector identity):
\begin{equation}
 J = \frac{1}{2} \sum_{a\neq
    b}\frac{\langle\G_a,\G_b\rangle{x}_b}{|{x}_a-{x}_b|}=\frac{R^2}{8}\sum_a
    p^0_a{x}_a\, = \frac{R^2}{8} d^0 \,. \label{Jd0rel}
\end{equation}
From the above asymptotic expansion of the harmonics
(\ref{expandedharmonics}), we can determine the asymptotic behavior of all the
fields and functions appearing in our solution. First, let us determine the
large $r$ expansion of the functions $y^A$. These are given in the form of a
quadratic equation which can be solved in a $1/r$ expansion as
\begin{equation}
   y^A = H^A - H^0 D^{AB} H_B - \frac{1}{2} (H^0)^2 D^{FA} D_{FBC} D^{BD} H_D D^{CE} H_E + \calo(\frac{1}{r^4})\,,
\end{equation}
where we defined
\begin{equation}
    D^{AB} = (D_{ABC} H^{D})^{-1}\,.
\end{equation}
Armed with this expression for $y^A$ we compute
\begin{equation}
   D_{ABC} y^A y^B y^C = D_{ABC} H^A H^B H^C - 3 H^0 H^A H_A + \frac{3}{2} (H^0)^2 H_A D^{AB} H_B + \calo(\frac{1}{r^6})\,. \label{y3expansion}
\end{equation}
We can now evaluate the $1/r$ expansion of the coefficient $\frac{\S^2}{Q^2}$
appearing in front of $d \psi^2$ in the metric
\begin{equation}
    \frac{\S^2}{Q^2} = \left( H_A D^{AB} H_B - 2 H_0 \right) \left( \frac{D_{ABC} H^A H^B H^C}{3} \right)^{-1/3} + \calo(\frac{1}{r})\,.
\end{equation}
The expansion of $L$ is straightforward, and the the expansion for $Q$ follows
directly from (\ref{y3expansion}). The last non-trivial expansions to be calculated are those of $\omega$ and $\omega_0$. For those the following result is
convenient: for any vector $n^i\in\mathbb{R}^3$ one has
\begin{equation}
   d\left(\frac{\epsilon_{ijk} n^i r^j dr^k}{r^3}\right) = - \ast_3 d \left(\frac{n^i r^i}{r^3}\right)\,.
\end{equation}
In particular we find that
\begin{equation}
    \omega_0 = - \epsilon_{ijk} \frac{(d^0)^i r^j dr^k}{r^3} + \calo(\frac{1}{r^2}) = - \frac{\sin^2 \theta d^0}{r} d\phi + \calo(\frac{1}{r})\,,
\end{equation}
where in the last equality we used our choice to take the $z$ axis to be along
the D6 dipole moment $d^0$.  We will not need the explicit form of $\o$ because
its leading term goes like $\calo(r^{-2})$. This follows from the asymptotic
form of the equations of motion
\begin{equation}
d\o=\sqrt{R}\star\left(-h_0dH^0+\calo(\frac{1}{r^4})\right)\, ,
\end{equation}
where we have once more shifted the $\Dp{0}$ constant term to zero; see the end of section \ref{decouplinglimit} for the details.

We are now ready to spell out the asymptotic expansion of the metric.
We start from (\ref{rescaledexplicit}), use the expansions computed above and
replace $t$ by $v$ to compensate for shifting the $\Dp{0}$ constant $h_0$. The
result one gets up to terms of order\footnote{In this power counting we consider
$\calo(dr)=\calo(r)$.} $\calo(\frac{1}{r})$ is
\begin{eqnarray}
ds^2_{\mathrm{5d}}&=&-r\frac{R}{U}dvd\psi+\frac{U^{-4}}{4}\left[-R^2(d^0)^2dv^2+2R\left(\frac{e\cdot d^AD_{ABC}p^Bp^C}{3}
-\frac{p^Aq_Ad^0\cos\theta}{3}\right)dvd\psi+\cald d\psi^2\right]\nonumber\\
&&+U^2\frac{dr^2}{r^2}+U^2\left(d\theta^2+\sin^2\theta\,(d\phi+\tilde A)^2\right)+\calo(\frac{1}{r})\,.\label{intermetric}
\end{eqnarray}
Here we introduced the notation
\begin{equation}
v=t-R/4\psi\,, \quad U^3=\frac{p^3}{6}\,,\quad \cald=\frac{p^3}{3}\left(D^{AB}q_Aq_B-2q_0\right)\mbox{and }\quad \tilde A=\frac{J}{J_{\mathrm{max}}}
 \frac{2v}{R}\,.\label{ascoords}
\end{equation}
We used the relation between the D6-dipole moment $d^0$ and the
angular momentum $J$ given by (\ref{Jd0rel}) and the fact that
there is a maximal angular momentum
$J_{\mathrm{max}}=\frac{U^3}{2}$. Note that $\p^2\cald=S(\G_t)^2$,
so $\cald$ is the discriminant of the total charge. With a
coordinate transformation to a new radial variable $\rho$ one can
show that the angular dependent part in the second term of
\ref{intermetric} is really further subleading. The coordinate
$\rho$ is given by
\begin{eqnarray}
\frac{\r^2}{4U^2}=-\frac{U^{-4}}{2}R\left(\frac{e\cdot d^AD_{ABC}p^Bp^C}{3}
-\frac{p^Aq_Ad^0\cos\theta}{3}\right)+\frac{R}{U}r\,.
\end{eqnarray}
In this new radial coordinate the expansion in large $\r$ takes the following form
\begin{eqnarray}
ds^2_{\mathrm{5d}}&=&-\frac{\r^2}{4U^2}dvd\psi+\frac{U^{-4}}{4}\left[-R^2(d^0)^2dv^2+\cald d\psi^2\right]
+4U^2\frac{d\r^2}{\r^2}\\&&+U^2\left(d\theta^2+\sin^2\theta\, (d\phi+\tilde A)^2\right)+\calo(\frac{1}{\r^2})\,.\nonumber
\end{eqnarray}

Using the expansion formulas derived above it is straightforward to calculate
the asymptotics of the gauge field and the scalars. Putting everything together
we see that the solution asymptotes to
\begin{eqnarray}
ds^2_{\mathrm{5d}}&=&-\frac{\r^2}{4U^2}dvd\psi+\frac{U^{-4}}{4}\left[-R^2(d^0)^2dv^2+\cald d\psi^2\right]\nonumber\\
&&+4U^2\frac{d\r^2}{\r^2}+U^2\left(d\theta^2+\sin^2\theta\, (d\phi+\tilde A)^2\right)+\calo(\frac{1}{\r^2})\,,\label{asympmetric}\\
A^A_{\mathrm{5d}}&=&-p^A\cos\theta d\a+D^{AB}q_Bd\psi+\calo(\frac{1}{\r^2})\,,\label{asympgauge}\\
Y^A&=&\frac{p^A}{U}+\calo(\frac{1}{r^2})\,.
\end{eqnarray}
It is clear that the metric is locally asymptotically
AdS$_3\times$S$^2$ with
$\mathrm{R}_{\mathrm{AdS}}=2\mathrm{R}_{\mathrm{S}^2}=2U$. We have
kept track of some subleading terms as they will be important in
reading off quantum numbers in the next section. Note that we have
in fact a nontrivial $S^2$ fibration over AdS$_3$ described by the
flat connection $\tilde A=\frac{J}{J_{\mathrm{max}}}
(\frac{2dt}{R}-\frac{d\psi}{2})$. As $\tilde A$ depends on the
time coordinate we see that as time progresses the sphere rotates,
implying the solution has angular momentum as expected. In the
same way, going once around the M-theory circle, i.e.
$\psi\rightarrow \psi+4\pi$, induces a rotation of $\frac{2\pi
J}{J_{\mathrm{max}}}$ along the equator\footnote{Remember we chose
the canonical 'z-axis' of our spherical coordinates along the
total angular momentum of the solution.} of the $S^2$.  The
explicit coordinate transformation bringing the above metric in
the form (\ref{asn1}) after dropping the subleading terms will be
given below.

\subsection{CFT quantum numbers}\label{quantumnrs}
In this subsection we will do an analysis of the asymptotic conserved
charges of the decoupled solutions. As we now have an asymptotic AdS geometry we
can use the well developed technology for these spaces. In our case of AdS$_3$ a
nice review can be found in  \cite{Kraus:2006wn}. The asymptotic charges as
determined from the supergravity side can later be compared to various quantum
numbers in the boundary CFT.

To proceed we first rewrite everything asymptotically in terms of a three
dimensional theory on AdS$_3$ by reducing over the asymptotic sphere spanned by
$(\theta,\phi)$.  Reducing five dimensional $\mathcal{N}$=1 supergravity over
the $S^2$ will result in a three dimensional theory with an $SU(2)$ gauge group
in addition to gravity (in an AdS$_3$ background) and the $U(1)$ vector multiplet
fields that descend from five dimensions.  The metric of the reduced theory is
\begin{equation}
ds^2_{\mathrm{3d}}=-\frac{\r^2}{4U^2}dvd\psi+\frac{U^{-4}}{4}\left[-R^2(d^0)^2dv^2+\cald d\psi^2\right]+4U^2\frac{d\r^2}{\r^2}\,.
\end{equation}
This can be put it into a standard form for the asymptotic expansion around
AdS$_3$ by the coordinate transformations
\begin{equation}
\r^2=\frac{e^{\frac{\eta}{U}}4U^2}{R}\,,\qquad dv=-\frac{R}{2}d\bar{w}\,,\qquad d\psi=2dw\,.
\end{equation}
These are related to the coordinates $\tau,\sigma$ we used in
(\ref{asn1})-(\ref{asn3}) by $w=\sigma+\tau$,
$\bar w = \sigma - \tau$. After Wick rotating $\tau \to i \tau$, these become the standard
conjugate holomorphic coordinates on the boundary cylinder, with periodicity $2\pi$. The metric reads
\begin{equation}
ds^2_{\mathrm{3d}}=d\eta^2+e^\frac{\eta}{U}dwd\bar{w}+\frac{1}{U^4}\left(\cald dw^2-\frac{R^4(d^0)^2}{16}d\bar{w}^2\right)\,,\label{expmetric}
\end{equation}
which has the standard form $ds^2_{\mathrm{3d}}=d\eta^2+(e^\frac{2\eta}{R_\mathrm{AdS}}g_{ij}^{(0)}+g_{ij}^{(2)})du^idu^j$. We can now apply the formulas \cite{Kraus:2006wn}:
\begin{eqnarray}
T_{ww}^{\mathrm{grav}}&=&\frac{1}{8\pi G_3 R_\mathrm{AdS}}g^{(2)}_{ww}\,,\label{energymomt}\\
T_{\bar{w}\bar{w}}^{\mathrm{grav}}&=&\frac{1}{8\pi G_3 R_\mathrm{AdS}}g^{(2)}_{\bar{w}\bar{w}}\,.\nonumber
\end{eqnarray}
In our case this becomes\footnote{We used
$G_3=\frac{\ell_5^3}{2R_{\mathrm{S}^2}^2}$. Note furthermore that the
definitions (\ref{energymomt}) are given in unrescaled variables so that both
$R_{\mathrm{AdS}}$ and $R_{\mathrm{S}^2}$ carry a factor $\ell_5$. Thus when
rescaling $g_{ij}\rightarrow\ell_5^2g_{ij}$ all factors of $\ell_5$ drop out of
the energy momentum tensor. This is as expected since we defined our limit in
such a way as to ensure that these energies stay finite as
$\ell_5\rightarrow0$.}
\begin{eqnarray}
T_{ww}^{\mathrm{grav}}&=&\frac{\cald}{8\pi U^3}\,,\label{energymomtmulticent}\\
T_{\bar{w}\bar{w}}^{\mathrm{grav}}&=&\frac{-R^4(d^0)^2}{8\pi\, 16U^3}\,.\nonumber
\end{eqnarray}
Apart from the metric, there are also gauge fields: the SU(2) gauge
field coming from the reduction of the metric on S$^2$ and the U(1) vectors of
the 5d supergravity. These gauge fields do contribute to the asymptotic energy
momentum tensor because the 5-dimensional action contains a Chern-Simons term
involving them. Here we will just present the results of the derivation that is detailed in appendix
\ref{csterm}. The contribution of all the different gauge fields to the energy
momentum is given by

\begin{equation}
   T^{\mathrm{gauge}}_{ww} =\frac{1}{4 \pi} \,  \, \left[ \frac{(p^A \, q_A)^2}{p^3} - (q_A D^{AB} q_B) \right]  \, , \;\;\;\;\; T^{\mathrm{gauge}}_{\bar{w}\bar{w}} =  \frac{1}{4 \pi} \, \frac{(p^A \, q_A)^2}{p^3}+ \frac{R^4}{8\pi} \frac{(d^0)^2}{16U^3}\,.\label{gaugemomentum}
\end{equation}

\noindent So by combining (\ref{energymomtmulticent}) and (\ref{gaugemomentum}), we see that the total energy momentum tensor is:
\begin{eqnarray}
T_{ww}= \frac{1}{4 \pi} \left( \frac{(p^A \, q_A)^2}{p^3} -2q_0\right)\,,  \qquad T_{w\bar{w}} = 0 \, ,\qquad
T_{\bar{w}\bar{w}}= \frac{1}{4 \pi} \,\frac{(p^A \, q_A)^2}{p^3}\,.
\end{eqnarray}
The Virasoro charges $(L_0)_{\rm cyl}$ and $(\tilde{L}_0)_{\rm cyl}$ on the cylinder are obtained from the energy-momentum tensor as
\begin{eqnarray}
(L_0)_{\rm cyl}&=&\oint dw \, T_{ww} = \frac{(p^A \, q_A)^2}{2 p^3} - q_0 \,,  \nonumber \\
(\tilde{L}_0)_{\rm cyl}&=&\oint d\bar{w} \, T_{\bar{w}\bar{w}} = \frac{(p^A \, q_A)^2}{2 p^3} \,,  \label{virasorocharges}
\end{eqnarray}
where the contour integral is taken along a contour wrapped once around the asymptotic
cylinder, i.e.\ $w \to w+2\pi$. These are related to the standard Virasoro charges on the $z=e^{i w}$-plane by the transformations
\begin{equation}
 L_0 = (L_0)_{\rm cyl} + \frac{c}{24}, \qquad \tilde L_0 = (\tilde L_0)_{\rm cyl} + \frac{c}{24},
\end{equation}
with $c$ the Brown-Henneaux central charge:
\begin{equation}
c=\frac{3R_{\mathrm{AdS}}}{2G_3}=p^3.
\end{equation}
These are exactly the quantum numbers of the BPS states of the dual CFT in the Ramond sector as determined in \cite{deBoer:2006vg, Gaiotto:2006wm}, confirming our earlier assertion under (\ref{periodicities}).
Naively one might have thought that the BPS condition would require
$\tilde{L}_0 = c/24$. That this is not so follows from the particular structure of the $(0,4)$ theory under consideration. It has, besides the usual $(0,4)$ superconformal
algebra, several additional $U(1)$ currents, as well as additional
right-moving fermions --- these are superpartners of the center of
mass degrees of freedom of the original wrapped M5-brane description. As was analyzed in \cite{deBoer:2006vg,Gaiotto:2006wm}, the BPS
conditions involve the right-moving fermions in a non-trivial way,
and this modifies the BPS bound into
$\tilde L_0 \geq  \frac{(p^A \, q_A)^2}{2 p^3}
+\frac{p^3}{24}$, consistent with our result above.

Often, it is more convenient to work with different but
closely related quantum numbers, $L_0'$ and
$\tilde{L}_0'$, and similarly $(L_0')_{\rm cyl}$ and
$(\tilde{L}_0')_{\rm cyl}$, which are obtained from the original ones
by subtracting out the contributions of the zero modes of the
additional currents, so only the
oscillator contributions remain. In our case they are given by \cite{deBoer:2006vg}:
\begin{eqnarray}
L_0'-\frac{c}{24} = (L'_0)_{\rm cyl}&=& - \hat{q}_0  := -(q_0-\frac{1}{2} D^{AB} q_A q_B) \nonumber\,, \nonumber \\
\tilde{L}'_0 - \frac{c}{24} = (\tilde L'_0)_{\rm cyl}&=& 0\,. \label{defhatq2}
\end{eqnarray}
These reduced quantum numbers are in many cases more convenient.
They are spectral flow invariant, and when we want to use Cardy's
formula to compute the number of states with given $U(1)$ charges,
we can simply use the standard Cardy formula with
$L_0,\tilde{L}_0$ replaced by $L'_0, \tilde{L}'_0$. The reduced quantum numbers also have a simple
interpretation in the AdS/CFT correspondence. They represent the
contributions to $L_0,\tilde{L}_0$ from the gravitational sector,
ignoring the additional contributions from the gauge fields.

The total energy and momentum, in units of $1/R$, are given by
\begin{equation}
 H = (L_0)_{\rm cyl} + (\tilde L_0)_{\rm cyl} = \frac{(p^A \, q_A)^2}{p^3} - q_0,
 \qquad P = (L_0)_{\rm cyl} - (\tilde L_0)_{\rm cyl} = -q_0 \, , \label{adsangmomt}
\end{equation}
and the reduced energy and momentum by
\begin{equation}
 H' = (L'_0)_{\rm cyl} + (\tilde L'_0)_{\rm cyl} = -\hat{q}_0,
 \qquad P' = (L'_0)_{\rm cyl} - (\tilde L'_0)_{\rm cyl} = -\hat{q}_0 = H' \, .
\end{equation}
The energy $H$ can be seen to equal the BPS energy $E=\frac{|Z|}{\sqrt{G_4}}$ of a D4-D2-D0 particle in a 4d asymptotically flat background with $J^A \to \infty p^A$, $B^A=0$, with the diverging part subtracted off. The reduced energy is the same but now at $B^A = D^{AB} q_B$.

Finally, the $SU(2)_R$ charge can be read off from the sphere reduction connection appearing in the metric (\ref{asn1}). In general it is given by
\begin{equation}
J^I_0=\oint\frac{d\bar{w}}{2\pi}J^I_{\bar{w}}=\frac{c}{12}\oint\frac{d\bar w}{2\pi}A^I_{\bar{w}}\,.
\end{equation}
Details are given in appendix \ref{csterm}. Thus the $SU(2)_R$ charge equals the four dimensional angular momentum:
\begin{equation}
J_0=\frac{R^2d^0}{8}=J\, ,\label{SU2charge}
\end{equation}
where we used (\ref{Jd0rel}). This is as expected, since the $S^2$ descends from the spatial sphere at infinity in four dimensions.

\subsection{Existence and attractor flow trees} \label{existence}

Not all choices of charges $\Gamma_a$ give rise to multicentered solutions in asymptotically flat space at finite $R/\ell_5$. Of those which do, not all survive the decoupling limit $R/\ell_5 \to \infty$. And of those which survive, not all give rise to a single AdS$_3 \times S^2$ throat.

As reviewed in section \ref{afconjecture}, in four dimensional asymptotically flat space, the well supported split attractor flow conjecture states there is a one to one correspondence between attractor flow trees and components of the moduli space of multicentered solutions. In particular, the existence of flow trees implies the existence of corresponding multicentered configurations, which can be assembled or disassembled adiabatically by dialing the asymptotic moduli according to the flow tree diagram. By the uplift procedure we followed, the same correspondence holds for five dimensional solutions asymptotic to $\IR^{1,3} \times S^1$ with a $U(1)$ isometry corresponding to the extra $S^1$.

\FIGURE{
\includegraphics[page=5,scale=0.8]{pictures.pdf}
\caption{This figure is an example of an attractor tree that exists in flat space but that will not survive in the decoupling limit, because the starting point of the split flow will move towards $J = \infty$ and hence cross a wall of marginal stability and decay.}\label{itdies}
}

The 4d K\"ahler moduli scalars $J^A$ are related to the five dimensional normalized K\"ahler scalars $Y^A$ and the radius $R$ of the circle by
\begin{equation}
 J^A = \frac{R}{2 \ell_5} Y^A
\end{equation}
and the four dimensional $B$-field moduli $B^A$ equal the Wilson lines around the $S^1$ of the five dimensional gauge fields. Asymptotically $\IR^{1,3} \times S^1$ solutions surviving the $R/\ell_5 \to \infty$ limit thus correspond to 4d flow trees surviving the $J^A \to \infty \, Y^A$ limit. Figure \ref{itdies} gives an example of a class of flow trees not surviving in this limit. Another example is given by figure \ref{splitflow} in appendix \ref{app:splflnum}, which survives when the limiting direction in the $(J_1,J_2)$-plane in figure \ref{negq0} is taken above the marginal stability line, but not when it is taken below.

Now, not all asymptotically $\IR^{1,3} \times S^1$ configurations
surviving in the limit fit into a single AdS$_3 \times S^2$
throat. For example the D4-D4 bound states studied in appendix
\ref{D4D2D0} have center separations of order $p^3 \, \ell_5$ in
the original coordinates, whereas multicentered configurations
which do fit into an asymptotic AdS$_3 \times S^2$ throat have
separations of order $p^3 \ell_5^3/R^2$. The diverging hierarchy
between these distance scales in the decoupling limit $R/\ell_5
\to \infty$ is manifest in the rescaled consistency condition
(\ref{limitconsistency}) in the decoupling limit: for two D4
centers (or more generally clusters) with nonvanishing mutual
intersection product, the (rescaled) equilibrium separation is
infinite.

Looking at the asymptotics (\ref{asn1})-(\ref{asn3}) of the
decoupled solutions, we see that the value of $Y^A$ at the
boundary of AdS is proportional to $p^A$, and that the
$\theta$-averaged Wilson line $\frac{1}{4 \pi} \oint A_{5d}^A$,
equals $D^{AB} q_B$. This suggest asymptotic AdS$_3 \times S^2$
solutions correspond to 4d attractor flow trees with starting
point at the ``asymptotic AdS$_3$ attractor point''
\begin{equation}
 B^A + i J^A = D^{AB} q_B + i \infty p^A.
\end{equation}
As a test of this suggestion, note that, as pointed out in
\cite{Denef:2007vg}, this eliminates flow trees initially
splitting into two flows carrying only D4-D2-D0 charges, and
therefore configurations of two D4 clusters with nonvanishing
intersection product, which as we just recalled indeed do not fit
in a single AdS$_3 \times S^2$ throat in the decoupling limit. To
see this, it suffices to compute for $\Gamma_a =
(0,p^A_a,q^a_A,q^a_0)$ at $B^A=D^{AB}q_B$, $J^A = \Lambda p^A$,
$\Lambda \to \infty$:
\begin{equation} \label{nostabD4D4}
 \langle \Gamma_1,\Gamma_2 \rangle {\rm Im}(Z_1 \bar{Z_2}) = -\frac{3}{8}(p^A_1q_A^2-p^A_2q_A^1)^2 + {\cal O}(\Lambda^{-1}) < 0.
\end{equation}
This inequality (valid when $\langle \Gamma_1,\Gamma_2\rangle \neq
0$) implies that the initial point can never be on the stable side
of a wall of marginal stability, and hence a flow tree with this
initial split cannot exist. Initial splits involving nonzero
D6-charge on the other hand are not excluded in this way,
consistent with expectations.

Thus, we arrive at the following
\\

{\em \noindent  \textit{ {\bf Conjecture:} There is a one to one correspondence between $(i)$ components of the moduli
space of multicentered asymptotically AdS$_3 \times S^2$ solutions with a $U(1)$ isometry and $(ii)$ attractor flow
trees starting at $J^A=p^A \infty$ and $B^A=D^{AB}q_B$.}}
\\

\noindent In what follows we will refer to this special point in
moduli space as the \emph{AdS point}. It is worth pointing out
that the AdS point may lie on a wall of \emph{threshold}
stability,\footnote{Note that it cannot lie on a wall of
\emph{marginal} stability $\Gamma \to \Gamma_1 + \Gamma_2$: if the
constituents have nonzero D6-charge, these D6-charges have to be
opposite in sign, so in the $J \to \infty$ limit, the central
charges cannot possibly align; if the constituents have zero
D6-charge, (\ref{nostabD4D4}) shows that their central charges
cannot align either if their intersection product is
nonvanishing.} as defined in appendix \ref{margtresh}, for which
the inequality (\ref{nostabD4D4}) may become an equality. As
discussed there, the solution space becomes non-compact in this
case, in the sense that constituents can be moved off to infinity
--- in this case to the boundary of AdS. An example is given by
(\ref{triplecharges}): since the overall D2-charge vanishes, the
AdS point lies on the line $B=0$, which is a line of threshold
stability for splitting off the D0. The flow tree becomes
degenerate as well, as it splits in a trivalent vertex. Keeping
this in mind, the flow tree picture remains valid.

Finally, we should spend a few more words on our choice of $B$-field value for the AdS point. In general, the actual value of $B^A$ at the boundary of AdS depends on the angle $\theta$ with the direction of the total angular momentum:
\begin{equation} \label{bfield}
 B^A|_{\partial AdS} = \frac{1}{4\pi} \oint A_{5d} = D^{AB}q_B - \cos \theta \, \frac{J}{2 J_{\rm max}} \, p^A.
\end{equation}
Hence there is a significant spread of the actual asymptotic value of the $B$-field, proportional to the total angular momentum, which moreover grows with $p$. Although natural, it is therefore not immediately obvious that picking the average value (or equivalently the value at $\theta = \frac{\pi}{2}$) as starting point is the right thing to do, and this is why our conjecture above is not an immediate consequence of the split attractor flow conjecture.

\section{Some examples}\label{examples}

In this section we will briefly describe the decoupling limit for some simple,
but interesting multicentered configurations. The first example is rather
straightforward as we show how the well known case of a single centered black
hole/string fits in our more general story. Afterwards we discuss two 2-center
systems of interest. First, we show that the decoupling limit of a purely
fluxed $\Dp{6}-\antiDp{6}$ bound state is nothing but global
AdS$_3\times$S$^2$ and we discuss the link of this interpretation with spectral
flow in the CFT. Second we analyze configurations leading to the
Entropy Enigma of \cite{Denef:2007vg} in asymptotic AdS space. In the
next section we will show how the Entropy Enigma translated to 5d coincides with a well know instability of small AdS black holes.

Note that from here on we put $R \equiv 1$.

\subsection{One center: BTZ}
\label{1center}
In the case of a single black string we expect to reproduce the
standard BTZ black hole (times $S^2$) as the decoupled geometry \cite{Maldacena:1998bw}. As a
check on our results we show that this is indeed the case and that the entropy
of the BTZ black hole corresponds to the one of the 4d black hole/5d black
string we took the decoupling limit of. Given an M5M2P black string of charge
$(0,p^A,q_A,q_0)$ one can easily calculate that the metric
(\ref{rescaledexplicit}) in the decoupling limit is
\begin{equation}
ds^2=\frac{r}{U}\left[-dt d\psi+\frac{1}{4} \left(1+\frac{1}{rU^3}\left(\frac{S}{\pi}\right)^2 \right)d\psi^2\right]+\frac{U^2dr^2}{r^2}+U^2d\O_2^2\,,\label{formula4}
\end{equation}
where
\begin{equation}
 S = 4 \pi \sqrt{\frac{-\hat{q}_0 p^3}{24}}, \qquad \hat{q}_0 = q_0 - \frac{1}{2} D^{AB}q_A q_B\,,
\end{equation}
is the entropy of the 4d black hole. It is clear that this is indeed of the asymptotically local AdS$_3\times$S$^2$ form as found above. But in this case the full geometry, including the interior, is actually locally AdS$_3\times$S$^2$. To see this perform the coordinate transformation
\begin{equation}
\psi=2(t+\a)\,,\qquad r=U(\rho^2-\rho_*^2)\quad\mbox{and}\quad \rho_*=\frac{S}{\pi U^2}\,,
\end{equation}
to put this metric (\ref{formula4}) into its well known BTZ form:
\begin{equation}
ds^2=-\frac{(\rho^2-\rho_*^2)^2}{\r^2}dt^2+\frac{4\rho^2 U^2}{(\r^2-\r_*^2)^2}d\r^2+\r^2(d\a+\frac{\r_*^2}{\r^2}dt)^2+U^2d\O_2^2\,.
\end{equation}
This is the geometry of a sphere times an extremal rotating BTZ black hole and as is well known \cite{Banados:1992gq}, this can be viewed as a quotient of AdS$_3\times$S$^2$. Calculating the Bekenstein Hawking entropy of this BTZ black hole we find:
\begin{equation}
S_{\mathrm{BH}}=\frac{2\pi\r_*}{4G_3}=S\,,
\end{equation}
in agreement with our expectations.

Note that the horizon topology is $S^1 \times S^2$, so from the 5d
point of view we have a black ring.

\subsection{Two centers: $\mathrm{D6}-\overline{\mathrm{D6}}$ and spinning AdS$_3 \times S^2$}
\label{2center} \label{sec:DbarD}

The first new configurations appear by taking the decoupling limit of 2-center bound states. As follows from the constraint (\ref{limitconsistency}), only 2-centered solutions
where the centers carry (opposite) non-vanishing $\Dp{6}$ charge
exist in asymptotic AdS$_3\times$S$^2$ space. Such centers sit at
a fixed distance completely determined by their
charges:
\begin{equation} \label{eqsep}
r_{12}=\frac{-4\langle\G_1,\G_2\rangle}{p^0_1}\,.
\end{equation}
In general in the bulk the solution is now fully five-dimensional, mixing up the asymptotic sphere and AdS geometries in a complicated way.

The simplest two centered configuration is that of
a bound state of a pure
$\Dp{6}$ and $\antiDp{6}$ carrying only $U(1)$ flux, say $F=\pm \frac{p}{2}$. The
two charges are then:
\begin{eqnarray}
\G_1&=&e^{\frac{p}{2}}=[1,\frac{1}{2},\frac{1}{8},\frac{1}{48}]\,,\nonumber\\
\G_2&=&-e^{-\frac{p}{2}}=[-1,\frac{1}{2},-\frac{1}{8},\frac{1}{48}]\,,
\end{eqnarray}
where we introduced the following notation for (D6,D4,D2,D0)-charges:
\begin{equation} \label{chargenot}
 [a,b,c,d] := (a , b \, p^A , c \, D_{ABC} p^B p^C, d \, D_{ABC} p^A p^B p^C) \, .
\end{equation}

We now show that the lift of such a 2-centered configuration in the decoupling limit yields rotating global
AdS$_3\times$S$^2$. In this limit the harmonic functions are:
\begin{eqnarray}
{H}^0&=&\frac{1}{|{x-x}_1|}-\frac{1}{|{x-x}_2|}\,,\nonumber\\
{H}^A&=&\frac{p^A}{2}\left(\frac{1}{|{x-x}_1|}+\frac{1}{|{x-x}_2|}\right)\,,\\
{H}_A&=&\frac{D_{ABC}p^Bp^C}{8}\left(\frac{1}{|{x-x}_1|}-\frac{1}{|{x-x}_2|}\right)\,,\nonumber\\
{H}_0&=&\frac{p^3}{48}\left(\frac{1}{|{x-x}_1|}+\frac{1}{|{x-x}_2|}\right)-\frac{1}{4}\,.\nonumber
\end{eqnarray}
The equilibrium distance, solution to (\ref{limitconsistency}), is
given by:
\begin{equation}
|{x_1-x}_2|=\frac{2 \, p^3}{3} =: 4 U^3\,.
\end{equation}
After a change of coordinates (see also \cite{Denef:2007yt}):
\begin{eqnarray}
|x-x_1|&=&2U^3(\cosh2\xi+\cos \tilde{\theta})\nonumber\\
|x-x_2|&=&2U^3(\cosh2\xi-\cos \tilde{\theta})\\
t &=& \tau \\
\psi&=&2(\tau+\sigma) \, ,\nonumber
\end{eqnarray}
and letting $\phi$ be the angular coordinate around the axis through the centers (so the coordinates $(2\xi,\tilde{\theta},\phi)$ are standard prolate spheroidal coordinates), the metric takes the simple form:
\begin{equation} \label{globAdSmetric}
ds^2=(2U)^2(-\cosh^2\!\xi\, d\tau^2+\sinh^2\!\xi
\,d\sigma^2+d\xi^2)+U^2(\sin^2\!\tilde{\theta}\, (d\phi+\tilde A)^2+d\tilde{\theta}^2) \, ,
\end{equation}
where
\begin{equation}
 \tilde A = d(\sigma-\tau).
\end{equation}
The general asymptotic form (\ref{asn1}) is obtained from this by
the coordinate transformation $\xi=\frac{\eta}{2U}-\ln U$,
$\tilde{\theta} = \theta$ and taking $\eta \to \infty$.

This metric describes an $S^2$ fibration over \emph{global}
AdS$_3$, with connection $\tilde A$. The connection is flat except
at the origin, where it has a delta function curvature
singularity. Hence this is essentially a particular case of the
geometries considered in
\cite{Balasubramanian:2000rt,Maldacena:2000dr}.\footnote{For the
case of AdS$_3 \times S^3 \times Z$, i.e.\ the (4,4) D1-D5 CFT,
these geometries were further studied in detail in
\cite{Giusto:2004id,Giusto:2004ip}.} The twist of the sphere
around the AdS$_3$ boundary circle $\sigma \to \sigma+2\pi$ is
given by the Wilson line $\oint \tilde A$. In this case the twist
equals a $2\pi$ rotation, in accordance with our general
considerations under (\ref{periodicities}) and the fact that the
angular momentum $J=p^3/12$ is maximal. Translated to the CFT,
this means we have maximal $SU(2)_R$ charge. Moreover, as explained
under (\ref{periodicities}), fermions are periodic around the
AdS$_3$ boundary circle $\sigma \to \sigma+2\pi$, so this geometry
corresponds, in a semi-classical sense, to a maximally charged
R-sector supersymmetric ground state.\footnote{There is of course
a $2J+1$ dimensional space of such ground states in the CFT.
Correspondingly, on the gravity side, a spin $J$ multiplet is
obtained by quantizing the 2-particle $D6-\overline{D6}$ system
\cite{Denef:2002ru,Denef:2007vg}, or equivalently the solution
moduli space. This and related topics are studied in the companion
paper \cite{deBoer:2007}.}

Since the twist amounts to a full $2\pi$ rotation of the sphere, the Wilson line can be removed by a large gauge transformation, that is, a coordinate transformation on the $S^2$,
\begin{equation} \label{coordtransf}
 \phi \to \phi' = \phi + \sigma - \tau \, ,
\end{equation}
which brings the metric to trivial AdS$_3 \times S^2$ direct product form, with $\tilde A'=0$. In general, large gauge transformations of the bulk act as symmetries (or ``spectral flows'') of the boundary theory --- in general they map states to physically different states. Here in particular this large gauge transformation will affect the periodicity of the fermions, since a $2\pi$ rotation of the sphere will flip the sign of the fermion fields. The fermions are then no longer periodic, but antiperiodic around $\sigma \to \sigma + 2\pi$ --- we are now in the NS sector vacuum of the theory, consistent with the symmetries of global AdS$_3$ with $\tilde A=0$.\footnote{Spelled out in more detail, for a fermion field $\psi$, we have in the old coordinates $\psi(\sigma,\phi,\ldots)=\psi(\sigma+2\pi,\phi,\ldots)$. Expressed in the new coordinates, this boundary condition is $\psi(\sigma,\phi,\ldots)=\psi(\sigma+2\pi,\phi'+2\pi,\ldots)=-\psi(\sigma+2\pi,\phi',\ldots)$, where in the last equality we used the fact that $\phi'$ parametrizes rotations of the sphere.}

In the dual $(0,4)$ CFT, this transformation acts as spectral flow generated by the $SU(2)_R$ charge $J_0^3$. The charges discussed in section \ref{quantumnrs} transform under this symmetry as \cite{Schwimmer:1986mf}:
\begin{eqnarray}
L_0&\to&L_0 \,,\nonumber\\
\tilde L_0&\to&\tilde L_0 + 2\epsilon J^3_0 + \frac{c}{6} \epsilon^2\,, \label{specflow} \\
J^3_0&\to&J^3_0+\frac{c}{6} \epsilon\,,\nonumber
\end{eqnarray}
with $\epsilon=1/2$ and $c=p^3$. According to our general results
(\ref{virasorocharges}) and (\ref{SU2charge}), we get for the
original geometry $L_0=0$, $\tilde L_0 = p^3/24$, $J^3_0 =
-p^3/12$. Applying the above spectral flow, we obtain $L_0=0$,
$\tilde L_0=0$, $J_0^3=0$, as expected for the NS vacuum.

More general geometries corresponding to states in the NS sector,
at least in the case of axially symmetric solutions, can be
obtained by applying the spectral flow coordinate transformation
(\ref{coordtransf}) to the R sector solutions we have constructed.

\subsection{Enigmatic configurations} \label{enigma2}

In \cite{Denef:2007vg} it was shown that there are some
regions in charge space where the entropy corresponding to given
total charges (with zero total D6 charge) is dominated not by single centered black holes, but by
multicentered ones. This phenomenon was called the Entropy Enigma. For a short summary see \cite{Denef:2007yn}.


\FIGURE{
\includegraphics[page=6,scale=0.7]{pictures.pdf}
\includegraphics[page=7,scale=0.7]{pictures.pdf}
\includegraphics[page=8,scale=0.7]{pictures.pdf}
\includegraphics[page=9,scale=0.7]{pictures.pdf}
\caption{In the upper left figure, the flow tree for the maximally entropic 2-centered configuration at $h=0$ is shown (i.e.\ $u=1/3$). The other three figures show the total entropy as a function of $h$ for a number of uniformly spaced values of $u$ between $0$ and $1/2$, at three different zoom levels (and different $u$-spacings). The fat red line is the entropy of the BTZ black hole with the same total charge.}\label{entropies}
}

Interestingly, these enigmatic configurations always survive the
decoupling limit, because their walls of marginal stability are
compact, with the stable side on the large type IIA volume side.
This is to be contrasted with the 4d asymptotically flat case at
fixed values of the asymptotic K\"ahler moduli; in this case,
because the unstable region in K\"ahler moduli space grows with
$p$, the enigmatic configurations always disappear when $p \to
\infty$ as the asymptotic moduli will eventually become enclosed by the wall of marginal stability. In this sense, they are most naturally at home in the
decoupled AdS$_3 \times S^2$ setup under consideration, where they
persist for all $p$.

In \cite{Denef:2007vg} section 3.4, a simple class of examples was
given, consisting of 2-centered bound states with centers of equal
entropy. However, this configuration is not the most entropic one
for the given total charge: the total entropy can be increased by
moving charge from one center to the other. The maximal entropic
configuration is obtained when all entropy is carried by one
center only; this can be traced back to the fact that the Hessian of the entropy function of a single black hole has some positive eigenvalues, making multi-black hole configurations generically thermodynamically unstable as soon as charges are allowed to be transported between the centers.

We have not been able to find other, more complicated
configurations, involving more centers, with more entropy. 

Thus we consider two charges $\Gamma_i=(p^0,p^A,q_A,q_0)_i$ of the form
\begin{eqnarray}
\G_1&=&- e^{-u p} = \left[-1,u,-\frac{u^2}{2},\frac{u^3}{6} \right]\\
\G_2&=& p -h \, p^3 - \Gamma_1 = \left[1,(1-u),\frac{u^2}{2},-h - \frac{u^3}{6}\right] \, ,
\label{enigcharges}
\end{eqnarray}
where we used the notation (\ref{chargenot}). The total charge of this system is
\begin{equation} \label{Gamhh}
 \Gamma = (0,p^A,0,-h \, p^3) = [0,1,0,-h] \, .
\end{equation}
If the bound state exists, the angular momentum (\ref{angularmomentum}) and rescaled equilibrium separation (\ref{limitconsistency}) between the centers are, respectively
\begin{equation} \label{eqsepppp}
 J = \frac{1}{4} (u^2 - 2h) p^3 \, , \qquad |\vec x_1-\vec x_2| = 2 (u^2-2h) p^3 \, .
\end{equation}
The entropy is given by
\begin{equation} \label{S12entr}
 S_{2 \rm c}=S_1+S_2, \qquad S_1 = 0, \qquad
 S_2 = \frac{\pi}{3} \, p^3 \, \sqrt{\mbox{$8 (\frac{1}{2}-u)^3-9 (\frac{1}{3}-h-u+\frac{u^2}{2})^2$}}
 \, .
\end{equation}
To get a bound state in the decoupling limit, the equilibrium
separation in (\ref{eqsepppp}) must of course be positive and the
expression under the square root in (\ref{S12entr}) must be
nonnegative. A more detailed analysis using attractor flow trees
shows that if we also require $u \geq 0$, these conditions are
necessary and sufficient. (The latter condition is necessary to
prevent the wall of marginal stability to be enclosed by a wall of
anti-marginal stability.)

The minimal possible value of $h$ is $-\frac{1}{24}$, reached at
$u=\frac{1}{2}$, where $\Gamma_2 = e^{up}$. This corresponds to
the pure fluxed $\mathrm{D6}-\overline{\mathrm{D6}}$ of section
\ref{sec:DbarD}. The maximal value of $h$ attainable by the
configurations under consideration is $9/128 \approx 0.07$.

The entropy for a single center of the same total charge (the BTZ
black hole of section \ref{1center}) is given by
\begin{equation} \label{SBTZz}
 S_{1 \rm c}=4 \pi \sqrt{\frac{-q_0 p^3}{24}} = \frac{\pi \, \sqrt{2h} \, p^3}{\sqrt{3}}.
\end{equation}
One way of phrasing the Entropy Enigma is that in the limit $p \to
\infty$ keeping $q_0$ fixed, the 2-centered entropy is always
parametrically larger than the 1-centered one,\footnote{Note that
if $q_0 > 0$, there is no single centered black hole, so then this
statement is trivially true.} as the former scales as $p^3$, while
the latter scales as $p^{3/2}$. More generally this 2-centered
parametric dominance will occur whenever $h = -q_0/p^3 \to 0$. A
short computation starting from (\ref{S12entr}) shows that in this
limit, the maximal 2-centered entropy is reached at $u=1/3$, with
entropy and angular momentum
\begin{equation} \label{maxentropy}
 S_{2 \rm c} = \frac{\pi \, p^3}{18 \sqrt{3}} \approx 0.100767 \, p^3\, , \qquad J = \frac{p^3}{36} = \frac{J_{\rm max}}{3} \, .
\end{equation}
Indeed this entropy is manifestly parametrically larger than $S_{1
\rm c}$ when $h \to 0$. More precisely the crossover point between
one and two-centered dominance is at $h_c \approx 0.00190622$.
This is illustrated in fig.\ \ref{entropies}. The phase transition
this crossover suggests will be discussed further in section
\ref{sec:phases}.

We should note that we have only analyzed a particular family of
2-centered solutions here. A slight generalization would be to let
both centers have nonzero entropy. However this turns out to give
a lower total entropy for the same total charge --- for example in
the symmetric 2-centered case described in \cite{Denef:2007vg},
the maximal attainable entropy is $S = \frac{\pi \, p^3}{48}$.
Similarly for other generalizations such as sun-earth-moon
systems, we were unable to find configurations with higher
entropy. We cannot exclude however that they exist. If so, this
would affect the precise value of the crossover point $h_c$, but
not its existence.

All of these 2-centered solutions have nonvanishing angular
momentum, except in the degenerate limit of coalescing centers,
when $u^2 = 2h$. In this case the entropy is always less than the
single centered one, as it should not to violate the holographic
principle. One might therefore suspect that the Entropy Enigma
disappears when restricting to configurations with zero angular
momentum. This is not the case though. A simple example of a
multicentered solution with zero angular momentum but entropy $S
\sim p^3$ is obtained as follows. Instead of one particle of
charge $\Gamma_1=-e^{-up}$ orbiting around a black hole of charge
$\Gamma_2 = \Gamma - \Gamma_1$, consider $k>1$ particles of charge
$\Gamma_1(u) = -e^{-up}$ orbiting on a halo around a black hole of charge
$\Gamma_2'(k,u,h) = \Gamma(h) - k \Gamma_1(u)$. Then by positioning the
particles symmetrically on their equilibrium sphere around the
black hole, we get configurations of zero angular momentum, but
with entropy still of order $p^3$ at large $p$. This can be
extended quantum mechanically: quantizing the halos as in
\cite{Denef:2002ru,Denef:2007vg}, we get a number of spin zero
singlets from tensoring $k$ spin $j$ single particle ground
states.

Note that the entropy of the $k$-particle configuration at given $u$ and $h$ can be related to that of our original $k=1$ solution by
\begin{equation}
 S(k,u,h) = S(\Gamma_2'(k,u,h)) = \frac{1}{k} S(\Gamma_2'(1,ku,k^2h)) = \frac{1}{k} S(\Gamma_2(ku,k^2h)) \, .
\end{equation}
The equilibrium separation between a $\Gamma_1(u)$ particle and the $\Gamma_2'$ core, for given $h$ and $u$, does not depend on $k$, so
\begin{equation}
 x_{12}(k,u,h) = x_{12}(1,u,h) = \frac{1}{k^2} x_{12}(1,ku,k^2h) \, . 
\end{equation}
From these relations, we can immediately deduce the existence conditions and maximal entropy configuration for $k>1$ particles using the results for the $k=1$ case derived above. In particular we see that the entropy is maximized at $u=1/3k$, and for e.g.\ $h=0$ equal to
\begin{equation}
 S_{(1+k) \rm c} = \frac{1}{k} \cdot \frac{\pi \, p^3}{18 \sqrt{3}} \, .
\end{equation}
Note that due to the factor $k$ in the denominator, the $k \geq
2$ (possible spin zero) configurations are thermodynamically disfavored
compared to the $k=1$ (necessarily spinning) configurations.

\section{Demystifying the Entropy Enigma}\label{entropyenigma}

\subsection{Interpretation as black hole localization on the sphere}

\FIGURE{
\includegraphics[page=10,scale=0.63]{pictures.pdf}
\includegraphics[page=11,scale=0.585]{pictures.pdf}
\caption{On the left a representation is shown of the single
centered 4d black hole; this lifts to the BTZ black hole (times
$S^2$) at the center of AdS$_3$. Surfaces of constant spherical
coordinate $r$ in $\IR^3$ are indicated --- these become the $S^2$
fibers of AdS$_3 \times S^2$. On the right one of the 2-centered
4d configurations of section \ref{enigma2} is depicted; this lifts
to a BMPV-like black hole roughly localized on the north pole of
the $S^2$ and at the center of AdS$_3$. Surfaces of constant
prolate spheroidal coordinate $\xi$ are indicated. As is clear
from (\ref{globAdSmetric}), these are the $S^2$ fibers of AdS$_3
\times S^2$ in the zero size limit of the black hole at the north
pole, i.e.\ the R vacuum. When the black hole has finite size, the
metric near it will be deformed to that of a BMPV black hole in 5
dimensions. \label{EEvsBTZ}} }

From the discussion in section \ref{enigma2}, it transpires that
the entropy ``enigma'' is in fact nothing but a supersymmetric
version of a well known general instability phenomenon in the
(nonsupersymmetric) microcanonical ensemble on AdS$_p \times S^q$,
first pointed out in \cite{Banks:1998dd}: Schwarzschild-AdS black
holes become thermodynamically unstable once their horizon radius
shrinks below a critical value of the order of the AdS radius --- at this point it becomes entropically
favorable at the given energy to form a Schwarzschild black hole
localized on the $S^q$. Related thermodynamical as well as
dynamical instabilities were studied in
\cite{Gregory:1993vy,Peet:1998cr,Martinec:1999sa,Cvetic:1999ne,Chamblin:1999tk,Chamblin:1999hg,Gubser:2000ec,Gubser:2000mm,Yamada:2006rx,Balasubramanian:2007bs}
and other works.

We see something very similar here: when the BTZ black hole radius
is lowered below a critical value of the order of the AdS radius,
it becomes thermodynamically unstable --- at this point it is
entropically favorable at the given energy and total charge to
form a BMPV-type BPS black hole \cite{Breckenridge:1996is}
localized on the $S^2$, which is precisely the ``enigmatic''
configuration studied in the previous subsection. This is
illustrated in fig.\ \ref{EEvsBTZ}.

In particular, we see now that the statement that multicentered
black holes dominate the entropy in the small $h$ regime is
somewhat misleading. From the 4d point of view, the (presumably)
dominant solution described in section \ref{enigma2} is two
centered, with one zero entropy, pure fluxed D6 center; a naked
timelike singularity. But from the 5d point of view, there is
really only one black hole, since the 4d D6 singularity lifts to
smooth geometry. Thus, the dominant configuration remains a single
black hole --- just one that is localized on the sphere.

To the best of our knowledge, this is the first instance of such
an instability in a supersymmetric setting. As we spelled out in
this article, the presence of supersymmetry makes it possible to
write down completely explicit solutions, which is not possible in
general nonsupersymmetric cases studied before. This might make
explorations of this phenomenon as well as its dual CFT
description more tractable.

\subsection{Phase transitions} \label{sec:phases}

As suggested by figure \ref{entropies} and the discussion in
the previous subsection, the microcanonical ensemble exhibits a
phase transition in the $p \to \infty$ limit. By microcanonical
ensemble we mean more precisely here the statistical ensemble at
fixed total charge $\Gamma=(0,p^A,q_A,q_0)$ and fixed total energy
saturating the BPS bound, but variable $S^2$ angular momentum.  Thus we
introduce a potential $\omega$ dual to say the
$3$-component $J^3$. For concreteness we further specialize to the
situation of section \ref{enigma2}, putting $q_A=0$ and $q_0 = -h
p^3$.

\FIGURE{
\includegraphics[page=12,scale=0.7]{pictures.pdf}
\includegraphics[page=13,scale=0.7]{pictures.pdf}
\caption{Left: Entropy as a function of $h$ in the limit $p \to \infty$. Right: $J^3/J_{\rm max}$ as a function of $h$ (the branch depending on the sign of $\omega$), for $p \to \infty$.}
\label{Jh}}

Let us assume that, as our analysis suggest, the entropy below a
critical value $h=h_c$ is indeed dominated by the black hole
localized on the $S^2$, while for $h>h_c$ it is dominated by the
BTZ black hole. Since the localized black holes have macroscopic
angular momentum, we see that in the limit $p \to \infty$ keeping
$\omega$ fixed, we get
\begin{equation}
 \langle J^3 \rangle = \pm j_*(h)  \quad (h<h_c), \qquad \langle J^3 \rangle = 0  \quad (h>h_c) \, ,
\end{equation}
where $j_*(h)$ is the angular momentum of the most entropic
configuration and the sign is determined by the sign of $\omega$.
This is illustrated in fig.\ \ref{Jh}. If we assume either BTZ or
single sphere localized black holes dominate, the critical value
is $h_c \approx 0.00190622$, and in the large $p$ limit, we have a
sharp first order phase transition, with order parameter given by
the angular momentum. However as we mentioned before, although we
were unable to find any, we cannot exclude the existence of more
complicated, more entropic multi-black hole / particle gas
configurations which would push up $h_c$, and perhaps even
smoothen the entropy and angular momentum as a function of $h$,
changing the order of the phase transition

\FIGURE{
\includegraphics[page=14,scale=0.7]{pictures.pdf}
\caption{Free energy $F$ as a function of $T$ in the limit
$p \to \infty$ for BTZ (fat red line) and sphere localized black
holes at different values of $u$ ranging from $0$ to $1/2$. The
bottom fat blue line corresponds to $u=1/2$, that is, AdS$_3
\times S^2$ without black holes. The end points of the black hole
lines correspond to the 4d equilibrium separation and angular
momentum becoming zero, i.e.\ becoming indistinguishable from BTZ
in the asymptotic region.} \label{canonical}}

We can also consider the ``canonical'' ensemble, trading $-q_0$
for its dual potential $\beta=1/T$ while still keeping the $q_A$
fixed (say $q_A \equiv 0$, which for simplicity of exposition we
assume from now on), and keeping the total energy at BPS
saturation.\footnote{As in the microcanonical ensemble we still
allow the angular momentum $J$ to vary and work at fixed $\omega$,
but we will suppress this in the explicit formulae below --- its
only effect in the end at $p \to \infty$ is to select a low
temperature ground state.} As we will see below, in the dual CFT,
$T$ has an interpretation as the ``left-moving temperature'',
conjugate to $(L_0)_{\rm cyl} = H = h p^3 = -q_0$ (see section
\ref{sec:CFTtransl}), while the constraint of BPS saturation can
be enforced by taking the right-moving temperature $\tilde T \to
0$. Although $T$ is strictly speaking not a real temperature, we
will use terminology as if it were. The relation between $h$ and
$T$ and the free energy are given by the Legendre transform
\begin{equation} \nonumber
 \frac{1}{T} = \frac{\partial S}{\partial H} = -\frac{\partial S}{\partial q_0} \, , \qquad F = H - T S \, .
\end{equation}
For the BTZ black hole, (\ref{SBTZz}) thus gives 
\begin{equation} \label{TFFF}
 h(T) = \frac{(2 \pi T)^2}{24} \, , \qquad F(T)=-\frac{\pi^2 T^2}{6} p^3 \, .
\end{equation}
This means the BTZ black hole charge at thermal equilibrium is $\Gamma(T)=(0,p,0,-h(T) \, p^3)$.
For the localized black holes of section \ref{enigma2} we get more
complicated expressions. The localized black hole charge and
entropy in thermal equilibrium are, using the notation
(\ref{chargenot}):
\begin{equation} \nonumber
 \Gamma_2 = \left[1,(1-u),\frac{u^2}{2},\frac{(1-2 u)^{3/2}}{3\left(\pi ^2 T^2+1\right)^{1/2} }-\frac{u^3}{6}-\frac{u^2}{2}+u-\frac{1}{3} \right], \qquad
   S_2 =  \pi ^2 T \frac{(1-2 u)^{3/2} \, p^3}{3\left(\pi ^2 T^2+1\right)^{1/2}}  \, .
\end{equation}
The resulting free energies as a function of $T$ are shown in
fig.\ \ref{canonical}. Again we see a phase transition in the
large $p$ limit: above a certain temperature $T_c$, the BTZ black
hole minimizes the free energy due to its large entropy; below it
the spinning global AdS$_3 \times S^2$ vacuum
(\ref{globAdSmetric}) (with $J^3 = \pm \frac{p^3}{12}$) takes
over, as dumping energy into the reservoir becomes entropically
favorable. (Both phases will also contain a thermal gas of
particles, since we have coupled the system to a heat bath.) The
free energy of the vacuum ($u=1/2$) is easy to compute as it has
zero entropy: $F_{\rm vac} = H_{\rm vac} = -\frac{p^3}{24}$. By
equating this with the BTZ free energy we get the critical
temperature:
\begin{equation} \label{critT}
 T_c = \frac{1}{2 \pi}
\end{equation}
(in units of $1/R$).

This phase transition is nothing but (a BPS version of) the
Hawking-Page transition \cite{Hawking:1982dh}. Its existence in a
supersymmetric context was observed already in
\cite{Dijkgraaf:2000fq}, by examining the elliptic genus of the
Hilbert scheme of $k$ points on $K3$ and its AdS$_3 \times S^3
\times K3$ dual. Here we see its physical origin more directly.

Note that again the angular momentum jumps: from $\langle J^3
\rangle=0$ at $T>T_c$ to $\langle J \rangle = \pm J_{\rm max}= \pm
\frac{p^3}{12}$ at $T<T_c$. The AdS-CFT correspondence therefore
implies a phase transition in the dual 1+1 dimensional CFT
breaking the continuous $SU(2)_R$ symmetry. This is not in
contradiction with the Coleman-Mermin-Wagner theorem
\cite{Coleman,Mermin-Wagner}, since there is only a true phase
transition in the strict limit $p \to \infty$. At any finite $p$,
the combined free energy is smooth.

In any case, we are led to conclude that BTZ black holes much
smaller than the AdS radius in fact do not provide stable
classical ($p \to \infty$) backgrounds representing macroscopic
(thermodynamic) states in the CFT. This is just as well, as the
opposite situation would lead to various paradoxes. For example,
according to the philosophy of e.g.
\cite{Mathur:2005zp,Balasubramanian:2005mg,Alday:2006nd,Bena:2007kg},
the BTZ black hole, when it exists as a proper classical geometry,
should be obtained by coarse graining over all microstates of
given energy or temperature, consistent with its interpretation as
a purely thermal state \cite{Maldacena:1998bw}. However, when the
BTZ black hole is small, it is hard to see how it could be the
result of coarse graining over the ensembles of multicentered
configurations, which typically extend far beyond the BTZ horizon
size.

We end this subsection by giving an alternative way to arrive at
the critical temperature (\ref{critT}). Let us start from the pure
fluxed $\mathrm{D6}-\overline{\mathrm{D6}}$ system studied in
section \ref{sec:DbarD}. Now add a number $N$ of D0-branes (which
according to (\ref{limitconsistency}) have to lie on the plane
equidistant from the $\mathrm{D6}$ and $\overline{\mathrm{D6}}$).
This is essentially the setup of \cite{Denef:2007yt}. It was shown
there that the D0-branes together with the D6 and anti-D6 can
adiabatically\footnote{By adiabatic we mean here by a evolution
process with energy arbitrarily close to the BPS bound.} collapse
into a scaling solution (or abyss) which approaches the single
centered D4-D0 black hole arbitrarily closely, if and only if
\begin{equation}
  N \geq \frac{p^3}{12} \qquad, \, \mbox{ i.e. } \,\, h = \frac{N-\frac{p^3}{24}}{p^3} \geq \frac{1}{24}\, .\nonumber
\end{equation}
This is in fact a direct consequence of the equilibrium
constraints (\ref{limitconsistency}). In the AdS$_3 \times S^2$
picture, what we have is a gas of gravitons and other massless
modes orbiting at constant radius in AdS$_3$ and at fixed $\phi$
on the equator of $S^2$, which can adiabatically collapse into a
BTZ black hole if $h > \frac{1}{24}$. From the relation
(\ref{TFFF}) between $T$ and $h$, this is equivalent to
$T>\frac{1}{2\pi}$, coinciding with the critical temperature
(\ref{critT}).

Thus, below the critical temperature $T_c$, there is a potential
barrier preventing adiabatic gravitational collapse of the system under consideration into a BTZ
black hole, above $T_c$, this is not the case. We leave the clarification of the 
deeper meaning of this coincidence of critical temperatures, and its implications for the fuzzball proposal (for reviews see \cite{Mathur:2005zp,Bena:2007kg}) to future work.


\section{Interpretation in the $(0,4)$ CFT}\label{cftenigma}

We will now discuss the interpretation in the dual CFT of the Entropy Enigma and other phenomena we observed.

\subsection{Translation to CFT} \label{sec:CFTtransl}

The quantum numbers of the decoupled solutions were given in
section~\ref{quantumnrs}. In particular, $L_0$ and $\bar{L}_0$
were given in (\ref{virasorocharges}), and we also defined reduced quantum numbers $L_0'$ and $\tilde{L}_0'$ in (\ref{defhatq2}). In the regime $L_0' \gg \frac{c}{24}$, the Cardy formula gives the microcanonical entropy of the CFT:
\begin{equation}
 S_{\rm Cardy} = 4\pi\sqrt{\frac{c}{24} (L'_0-\frac{c}{24})} = 4 \pi \sqrt{-\frac{\hat{q}_0 p^3}{24}} = S_{BTZ} \,,
\end{equation}
where $c=p^3$, reproducing precisely the BTZ black hole entropy. Note that the
regime where sphere localized black holes come to dominate is at
$(L_0'-\frac{c}{24})/c \ll 1$; this is the opposite of the Cardy regime.

In both the microcanonical and the canonical ensembles we consider in the previous section, we kept the M2 charge $q_A$ fixed and for simplicity we chose
\begin{equation}
 q_A = 0 \, .
\end{equation}
We will do this here too. In this case the distinction between reduced and original Virasoro charges disappears, and we have the identifications
\begin{equation}
 (L_0)_{\rm cyl} = L_0 - \frac{c}{24} = - q_0 = h c, \qquad (\tilde{L}_0)_{\rm cyl} = \tilde{L}_0 - \frac{c}{24} = 0.
\end{equation}
This implies furthermore $H=hc$, explaining our notation $q_0 = -h c$ used in (\ref{Gamhh}) and in the definition of the canonical ensemble in section \ref{sec:phases}.

The regime of particular interest to us is $h$ small and positive, which is where the phase transitions are expected to occur based on the black hole picture.

\subsection{Entropy for $L_0\sim\frac{c}{24}$} \label{sec:entropynearcrit}

There are not too many tools available to determine the number of
states in a CFT for $h = (L_0 - \frac{c}{24})/c \to 0$. There is certainly no
universal answer to this question, and in addition the answer may
depend on moduli and other parameters --- after all, it is not a
protected quantity. In order for the $N=(0,4)$ CFT, which is dual
to the geometries we have been studying, to accommodate the
sphere localized / multicenter solutions with entropy $S \sim p^3 = c$ near $h \to 0$,
the number of states at small $h$ in the CFT should grow accordingly. One can view this as a prediction of AdS/CFT for the (presumably strongly coupled) $N=(0,4)$ CFT.

The simplest possible model where one could investigate this
question is in the CFT of $c$ free bosons, which has partition function $Z :={\rm Tr}\, q^{L_0-\frac{c}{24}} = Z_1^c$ where
\begin{equation}
 Z_1 = q^{-\frac{1}{24}} \prod_{i>0} \frac{1}{(1-q^i)} = \frac{1}{\eta(q)} \, .
\end{equation}
Then the coefficient of $q^0$ can be estimated at large $c$ by
saddle point approximation. Parametrizing $q=e^{2 \pi i \tau}$:
\begin{equation} \label{saddleint}
 d(0) = \oint e^{c \log Z_1} \, d\tau \approx e^{c \log Z_1(\tau_*)} \, , \qquad \frac{\partial \log Z_1}{\partial \tau}|_{\tau_*} = 0 \, .
\end{equation}
The numerical solution to this is
\begin{equation} \label{solsolsol}
 \tau_* \approx 0.523524, \qquad \log d(0) \approx 0.176491 \, c \, ,
\end{equation}
so this indeed gives an entropy of order $c=p^3$ at $h=0$.
Comparing to (\ref{maxentropy}), we see that the coefficient is
different; of course there was no reason to expect it to be the
same, since the coefficient is model dependent. For example,
replacing $Z_1$ with a more general weight $w$ modular form
\begin{equation}
 Z_1(q) = a_0 q^b + a_1 q^{b+1} + \cdots \, ,
\end{equation}
we can estimate (\ref{saddleint}) by writing $Z_1(\tau) = a_0 (-i \tau)^{-w} e^{- \frac{2 \pi i b}{\tau}} + \cdots$ which leads to
\begin{equation}
 \tau_* \approx \frac{2\pi i b}{w} \, , \qquad \log d(0) \approx (\log a_0 - w(1+\log(2 \pi b/w))) \, c \,.
\end{equation}
For this to be a good approximation we need $e^{\frac{-2 \pi
i}{\tau_*}}=e^{-w/b} \ll 1$. For the free boson, we have $w=-1/2$
and $b=-1/24$, so this is satisfied and indeed plugging in the
numbers gives $\log d(0) \approx \frac{1}{2}(1+\log
\frac{\pi}{6})c$, reproducing (\ref{solsolsol}) to very good
accuracy.

In addition to similar saddle point approximations, a more refined
analysis of the large $c$ growth of $d(0)$ for various modular
forms, using the Fareytail expansion, was done in
\cite{Manschot:2007ha}, and in agreement with the simple estimates
given here.

Of course, since $c$ is a measure for the number of degrees of
freedom, it is hardly a surprise that the entropy for a fixed
nonzero amount of energy per degree of freedom $L_0/c = 1/24$
grows linearly in the number of degrees of freedom $c$. More
interesting would be to compute the actual proportionality
constant. Despite the model dependence of this number,
(\ref{maxentropy}) nevertheless suggests a universal number for all CFTs dual to
AdS$_3\times$S$^2\times$CY$_3$ in the large $c$ limit:
\begin{equation} \label{d0pred}
 \log d(0) = \frac{\pi}{18 \sqrt{3}} \, c \,.
\end{equation}
As mentioned earlier, this universality might however be an artifact of our
lack of imagination in finding more entropic configurations.

In theories in which a ``long string'' picture exists, we can
count the number of states in the long string CFT,
which typically has reduced central charge $\hat{c}=c/k$ and
increased excitation energy $\hat{L}_0=k L_0$. For $k$
sufficiently large, we can then use Cardy even if the original
$L_0$ was of the order of $c/24$, and we find
\begin{equation}
 \log d(0) = \frac{\pi}{6} \, c \,.
\end{equation}
This does not agree with (\ref{d0pred}), but clearly our analyses
on both sides are far from conclusive at this point. 

To make further progress, it is necessary to delve into the
intricacies of the actual dual CFTs. We will initiate this in the
next subsection, improving the analysis of \cite{Maldacena:1997de}
by more carefully identifying entropic modes important at small
$h$.

\subsection{The MSW string} \label{mswstring}

\subsubsection{Content and supersymmetry conditions}

The MSW (0,4) 1+1 dimensional sigma model on $W= \IR \times S^1$
arising from wrapping an M5 brane on $W \times P$ with $P$ a very
ample divisor has the following massless field content
\cite{Maldacena:1997de,Minasian:1999qn,LopesCardoso:1999ur}:
\begin{itemize}
 \item $h^{0,2}(P) \approx p^3/6$ complex non-chiral scalars $z^i$ arising from holomorphic deformations of $P$.\footnote{Consistent with our practice in the rest of the paper, we suppress (large $p$) subleading corrections to various Hodge numbers.}
 \item 3 real scalars $\vec{x}$, the position in $\IR^3$
 \item $b^2(P) \approx p^3$ real scalars from the reduction on $P$ of the self-dual 2-form field $b$ on the M5:
     \begin{equation}
      b = b^\alpha \Sigma_\alpha\,,
     \end{equation}
     where $\{ \Sigma_\alpha \}$ is an integral basis of the space of harmonic 2-forms $H^2(P)$. In such a basis the scalars are periodic: $b^\alpha \simeq b^\alpha + n^\alpha$, $n^\alpha \in {\mathbb Z}$. Furthermore they have to satisfy the self-duality constraint
     \begin{equation} \label{selfduality}
      db^\alpha \wedge \Sigma_\alpha = *_W db^\alpha \wedge *_P \Sigma_\alpha \, ,
     \end{equation}
     which implies there are $b^2_+(P) = 2 h^{2,0}(P) + 1 \approx p^3/3$ right-moving ($*_W = +1$) degrees of freedom and $b^2_- = h^{1,1}(P) - 1 \approx 2p^3/3$ left-moving ($*_W = -1$). The left-right split depends on the deformation moduli $z^i$ and the background complex and K\"ahler moduli.
 \item $4 h^{2,0}(P)+4 \approx 2p^3/3$ real right-moving fermions $\psi^\kappa$. These pair up with the in total $4 h^{0,2}(P)+4$ real right-moving scalars, as required by $(0,4)$ supersymmetry.
\end{itemize}

Motion of the string is supersymmetric if and only if it is
(almost) purely left-moving\footnote{As usual, the the extra
winding term in $b^{\alpha}$ can be written, using
$\tau=\frac{1}{2}(\tau+\sigma) + \frac{1}{2} (\tau-\sigma)$ as the
sum of left-movers and right-movers, and the left-moving
contribution can be absorbed in $b^{\alpha}(\tau+\sigma)$. We
chose for convenience a convention in which the winding term
depends on $\tau$ only.} :
\begin{equation} \label{leftmoving}
 z^i(\tau,\sigma)=z^i(\tau+\sigma), \qquad b^\alpha(\tau,\sigma)=b^\alpha(\tau+\sigma) - 2 (q \cdot \tilde{J}) \tilde{J}^\alpha \, \tau \, .
\end{equation}
Here $q \cdot \tilde J = q_A \tilde{J}^A$ with $q_A$ the M2-charge
and $\tilde{J} = \tilde{J}^A D_A$  proportional to the K\"ahler
form of $X$, normalized such that $\int_P \tilde{J}^2 \equiv 1$.
Furthermore the components $\tilde{J}^\alpha$ are defined by
decomposing $\tilde{J}$ pulled back to $P$: $\tilde{J} =
\tilde{J}^\alpha \Sigma_\alpha$. The reason for the presence of
the $\tau$-dependent term on the right hand side is the fact that
supersymmetry is nonlinearly realized when $q \cdot \tilde{J}$ is
nonvanishing \cite{Maldacena:1997de}, which is related to the fact
that $q \cdot \tilde{J}$ is proportional to the imaginary part of
the central charge $Z$, and therefore that a different subset of
four supercharges out of the original eight is preserved for
different $q \cdot \tilde{J}$. It is also closely related to the
difference between $L_0$ and $L_0'$ as discussed at the end of
section \ref{quantumnrs}.

In addition (\ref{leftmoving}) is a solution to the equations of motion if and only if the selfduality constraint (\ref{selfduality}) is satisfied. On the profile $(z^i(s),b^\alpha(s))$, $s \in S^1$ introduced in (\ref{leftmoving}) this constraint becomes the anti-selfduality condition
\begin{equation}
 \dot{b}^\alpha(s) \, \Sigma_\alpha - (q \cdot \tilde{J}) \tilde{J}  = -* [(\dot{b}^\alpha(s) \, \Sigma_\alpha)
 - (q \cdot \tilde{J}) \tilde{J}] \, .
\end{equation}
The dot denotes derivation with respect to $s$, and we used the
fact that the right-moving contribution in (\ref{leftmoving})
automatically obeys the self-duality constraint
(\ref{selfduality}). Harmonic 2-forms on $P$ are anti-selfdual if
and only if they are of type $(1,1)$ and orthogonal to
$\tilde{J}$. Following appendix G of \cite{Denef:2007vg}, the first
condition can be written as
\begin{equation} \label{susyconstraint1}
 \dot{b}^\alpha(s) \, \partial_i \Pi_\alpha(z(s))  = 0 \, ,
\end{equation}
while the second one is
\begin{equation} \label{susyconstraint2}
 \dot{b}^\alpha(s) \, \tilde J_\alpha = q \cdot \tilde{J} \, .
\end{equation}
Here $\Pi_\alpha(z)$ is the period of the holomorphic 3-form on a
3-chain with one boundary on the 2-cycle in $P(z)$ Poincar\'e dual
to $\Sigma_\alpha$, and $J_\alpha$ is the integral of the K\"ahler
form $J$ over the same 2-cycle.

\subsubsection{Supersymmetric solutions}

One could now try to get the BPS spectrum by quantizing this moduli space of supersymmetric configurations. In general however this is a complicated system of coupled equations.

Things simplify when we consider linearized oscillations around some arbitrary fixed point $(z_*^i,b_*^\alpha)$. Because there are about $p^3$  $b^\alpha$ and $p^3/3$ $z^i$ real degrees of freedom, (\ref{susyconstraint1})-(\ref{susyconstraint2}) will to lowest order just constrain the $b^\alpha$ to lie on a $2p^3/3$-dimensional plane, while $\delta z^i$ can oscillate freely. Hence we can think of this as in total $p^3$ free bosonic modes. At large $L_0$, these oscillator modes will dominate the entropy, approximately reproducing the BTZ entropy.

In addition, since they are periodic, we can allow the scalars
$b^\alpha$ to have nonzero winding number $k^\alpha$ in $H^2(P)$;
this corresponds to turning on worldvolume flux on the M5 (and in particular these modes can therefore carry M2 charge). Still at fixed $z_*$, integrating (\ref{susyconstraint1}) over the $S^1$
then gives the constraint
    \begin{equation} \label{kconstr}
      \partial_i W(z_*) = 0 \, , \qquad W(z) := k^\alpha \Pi_\alpha(z) \, .
     \end{equation}
     For generic $z_*$ and generic integral $k^\alpha$, this will not be satisfied.  Only for $k^\alpha$ in the sublattice $L_X$ of $H^2(P,\IZ)$ pulled back from the ambient Calabi-Yau $X$, this will be automatic (because these forms are always integral (1,1)).

     Based on this and the fact that in the full M-theory, M2 instantons can interpolate between winding numbers
     except those in $L_X$, \cite{Maldacena:1997de} rejected the possibility of turning on winding numbers except for those in $L_X$. However, at \emph{special} points $z_*$, (\ref{kconstr}) \emph{will} have solutions. Indeed these equations can be viewed as a superpotential critical point condition for $z^i$ (formally identical to the one obtained for D4 flux vacua in appendix G of \cite{Denef:2007vg}), and as such it will have isolated critical points for sufficiently generic $k^\alpha$; all $z^i$ have become effectively massive.
     Integrating \ref{susyconstraint2} over $S^1$ gives the constraint $k^\alpha \, J_\alpha = q \cdot J$. This is automatically satisfied,
     because the winding modes are exactly the origin of the M2 charge, as they correspond to M5 worldvolume flux;
     in general one can read off from the WZ terms in the M5-brane effective action that $q_A = \int_P D_A \wedge k^\alpha \Sigma_\alpha$.

     So, once we specify a winding vector $k^\alpha$, the string will still be supersymmetric when located at a critical
     point $z_*(k)$, and some or all of the $z^i$ zeromodes will be lifted. At the semiclassical level, these are definitely
     valid supersymmetric ground states --- and in fact there is a huge number of them, not quite unlike the landscape
     of string flux vacua. Instantons might tunnel between them and mix the states quantum mechanically, but this does
     not mean that they should not be considered; in particular when computing the Witten index, all these semiclassical vacua must be summed over (with signs).

     The contribution of these winding modes to $-q_0 = P = (L_0)_{\rm cyl} - (\tilde L_0)_{\rm cyl}$ is half the topological intersection product:
     \begin{equation}
      \Delta P = - \frac{1}{2} Q_{\alpha \beta} k^\alpha k^\beta \, , \qquad Q_{\alpha \beta}:=\int_P \Sigma_\alpha \wedge \Sigma_\beta \, .
     \end{equation}
     If in addition to (\ref{kconstr}) we also set $q \cdot J = 0$ (for example by restricting to the $q_A=0$ sector),
     then $k^\alpha \Sigma_\alpha$ is anti-selfdual, and therefore $\Delta L_0 =
	 \Delta P \geq 0$. Moreover,
     in the notation of section \ref{quantumnrs}, we have $\Delta L_0' \geq 0$.

There are more complicated solutions to
(\ref{susyconstraint1}) possible, for example when we let the
string loop around a nontrivial closed path $z(s)$ in the divisor
moduli space and at the same time on some loop in the
$b^\alpha$-torus. This can give rise to complicated twisted
sectors. As stressed in \cite{Minasian:1999qn}, there will in
general be monodromies $b^\alpha \to {M^\alpha}_\beta b^\beta$
acting on the $b$-torus when circling around the discriminant
locus in the divisor moduli space. Hence we should think of the
target space of the string as a quotient of the total space of the
$b$-torus fibration over Teichm\"uller space by the monodromy
group. Closed strings can begin and end on different points
identified by this group, leading to twisted sectors and possibly
long strings.

Finally, we can form bound states of the localized winding strings
described above. For example we can form a bound state of a closed
string winding $k_1$ at some $z_*(k_1)$ and one winding $k_2$ at
$z_*(k_2)$, by connecting them with two interpolating pieces of
string. Note though that now the constraint
(\ref{susyconstraint2}) becomes important: indeed generically
$k_1^\alpha J_\alpha \neq k_2^\alpha J_\alpha$, so the string we
just described cannot have constant $\dot{b}^\alpha J_\alpha$ and
we do not get a proper supersymmetric solution. It is conceivable however
that in some cases at least the string will be able to relax down to a BPS configuration for which
$\dot{b}^\alpha J_\alpha$ is constant everywhere.

This is reminiscent of brane recombination. Moreover, note that
the condition of having $k_1^\alpha J_\alpha = k_2^\alpha
J_\alpha$ corresponds to being on a wall of marginal stability for
the two M5-branes represented by the two strings. Hence there is
an obvious candidate for the gravitational interpretation of such
configurations: they should correspond to the M5-M5 2-centered
bound states of section \ref{D4D2D0}. It would be interesting to
make this more precise, but this is beyond the scope of the paper.

\subsubsection{Statistical mechanics}

\FIGURE{
\includegraphics[page=15,scale=0.7]{pictures.pdf}
\includegraphics[page=16,scale=0.7]{pictures.pdf}
\caption{Various entropies as a function of $h$, for $h$ near 0
(left), and for a larger range of $h$ (right). The blue line is
the total entropy derived from (\ref{toyZq}), the yellow line is
the entropy in the winding modes, the green line is the entropy in
the oscillator modes, and the red line is the BTZ entropy.}

\label{comparison}}

In this subsection we will give a rudimentary analysis of the
statistical mechanics of the BPS sector of the MSW string, to see
if we can reproduce some of the features we found on the black
hole side.

We can roughly model the ensemble of winding and oscillator modes
ignoring nonlinearities, say in the $q_A=0$ sector, by the
partition function
\begin{equation} \label{toyZq}
 Z(q) = {\rm Tr} \, q^{L_0-\frac{c}{24}} = \left( \frac{\vartheta_3(q)}{\eta(q)} \right)^c
\end{equation}
with $c=p^3$. Here the theta function models the winding mode
contributions and the eta function the oscillator contributions.\footnote{Note that despite the fact that
turning on winding modes is generically lifting \emph{zeromodes} of $z^i$, it is not true that it also lifts the oscillator modes; in the presence of winding, it remains true that (\ref{susyconstraint1}) reduces the number of \emph{local} fluctuation (oscillator) degrees of freedom by $p^3/3$, so at our level of approximation the oscillator mode counting is essentially unaffected by winding: the number of oscillating degrees of freedom remains $p^3/3 + p^3 - p^3/3 = p^3 = c$.}
By numerical saddle point evaluation, the total entropy and the
(entropy maximizing) distribution of it over the oscillator and
winding modes at given $h=(L_0-\frac{c}{24})/c$ can be
straightforwardly computed. The result is shown in fig.\
\ref{comparison}. The inclusion of winding modes actually improves
the match to the BTZ entropy compared to the most naive model with
only free oscillators; it is almost perfect already slightly above
the threshold. This can also be checked analytically: Because
$Z(q)$ has weight 0, the total entropy computed by saddle point
evaluation is exactly $S = 4 \pi \sqrt{\frac{h}{24}} c = S_{\rm
BTZ}$; for the free oscillator model, there are corrections.

We also see that at $h=0$, there is still an entropy of order
$c=p^3$, and almost all of it is in the winding modes. There are
still no phase transitions in this model of course, since the
system is noninteracting.

Let us turn our attention now to the $SU(2)_R$ $R$-charge $J^3$;
the $S^2$ angular momentum on the gravity side, which appeared as
an order parameter $J^3/p^3$ for the phase transition we
discussed. The only fields transforming nontrivially under
$SU(2)_R$ are $(i)$ the fermions, transforming in the ${\bf 2}$, but
they are all rightmoving so cannot be excited except for their
zeromodes, and $(ii)$ the position $\vec x$ transforming in the ${\bf
3}$, but this represents only three oscillators out of order
$c=p^3$, so one expects their contribution to the total $R$-charge
to be negligible in the thermodynamic limit $p \to \infty$ (in the
sense of their $J^3$ having an expectation value growing slower
than $p^3$).

So, where does the large angular momentum, $J = \frac{p^3}{12}$,
of the $L_0=0$ gravity solution come from then? The answer is from
the center of mass zero modes of the string. Since shifting the
$b^\alpha$ by constants independent of the string coordinate $s$
corresponds to a gauge transformation, the only physical zero mode
space is the deformation moduli space $\CM_P$ of $P$. These
bosonic zero modes together with the fermionic ones (which we can
have since they are independent of $s$) will give ground state
wave functions in one to one correspondence with harmonic
differential forms on $\CM_P$. The form number corresponds to
fermion number and therefore to $R$-charge --- or in other words
the $SU(2)_R$ is identified with Lefschetz $SU(2)_R$ on cohomology
(see for example \cite{Denef:2002ru} for a pedagogical
explanation). This is analogous to how angular momentum is
produced in the D4-brane model \cite{Denef:2007vg}. Since the
moduli space $\CM_P = \ICP^{p^3/6}$ (where as before we are
dropping terms subleading to $p^3$), this means the $L_0=0$ ground
states assemble into a spin $J=\frac{p^3}{12}$ multiplet, exactly
as expected from the gravity side.

Now, when we turn on some small $L_0$, we expect from what we observed on the gravity
side that $J$ will go down somewhat (see fig.\ \ref{Jh}). We propose the following
picture of how this happens on the CFT side. At very small $L_0$,
a small number of winding modes will get turned on. This will
typically freeze a small number of the moduli $z^i$, reducing the
moduli space $\CM_P$ to a lower dimensional space. The maximal
Lefschetz spin always equals half the complex dimension $n$ (this
is the spin of the multiplet created by starting with 1 and
subsequently wedging with the K\"ahler form on the moduli space
till the volume is reached). Therefore the maximal $J$ will go
down. The higher $L_0$, the more winding modes get turned on, the
smaller the dimension of the residual moduli spaces, and the
smaller $J$. Eventually when $L_0$ becomes sufficiently large, so
many winding modes will be turned on that all moduli will
generically be frozen, and the expectation value of $J$ becomes
zero. This is in agreement with what we observe on the gravity
side.

Again, this is only a rudimentary qualitative picture, and in
particular too rough to be able to address how phase transitions
could arise. Perhaps a variant of the toy models
of \cite{liu} would be of help to make further progress. A more in depth analysis is left for future work. 

\subsubsection{The field theory description of the MSW string} \label{commquest}



One puzzle we have encountered several times in the paper has to
do with the nature of the MSW sigma model which describes the
low-energy excitations of the wrapped M5-brane. This sigma model
is obtained from a suitable KK reduction of the M5-brane theory
over the four-cycle over which the M5-brane is wrapped.
Classically, this sigma model is a $(0,4)$ superconformal field
theory, and the target space of the sigma model is the entire
moduli space of supersymmetric four-cycles in the Calabi-Yau
manifold.

The puzzle is that on the one hand, field theory arguments suggest
that this sigma model also describes a quantum $(0,4)$
superconformal field theory which still probes the entire moduli
space of supersymmetric four-cycles, whereas the bulk analysis
shows that not all M5-brane bound states fit into a single
asymptotically AdS$_3\times S^2$ geometry, which strongly suggests
that a quantum SCFT which captures the entire moduli space does
not exist.

The field theory arguments are based on claims in the literature
that, unlike $(2,2)$ sigma models, $(0,4)$ sigma models are always
finite \cite{Howe:1987qv,Howe:1988cj}, in the sense that all
renormalizations can be absorbed in finite field redefinitions, so
that in particular the beta functions vanish and the theory is
conformal also quantum mechanically. However there are potential
caveats \cite{Callan:1991dj}, to which in turn some
counterarguments have been given in \cite{Howe:1992tg}; see also
\cite{Witten:1994tz}. To the best of our knowledge, this issue
remains not fully settled.

Perhaps our results shed some new light on this. As we observed in
section \ref{sec3}, M5-M5 bound states of the type constructed in
appendix \ref{D4D2D0} will not fit in a single asymptotically
AdS$_3 \times S^2$ geometry, but split in two (or more) separated
AdS$_3 \times S^2$ throats. At values of the normalized K\"ahler
moduli $Y^A$ sufficiently far away from the AdS attractor point
$Y^A=p^A/U$, they do exist as supersymmetric states of the MSW
string, and we suggested a possible explicit MSW string
realization of them above. When moving the $Y^A$ to the attractor
point, all of these states decay. Hence they cannot be part of the
CFT which is dual to a single AdS$_3 \times S^2$ geometry.

There are therefore, in our view, two possibilities:
     \begin{enumerate}
       \item The MSW sigma model is a quantum SCFT for all values of the K\"ahler moduli $Y^A$. If so, it is not equivalent to quantum gravity in asymptotic AdS$_3 \times S^2 \times X$, and therefore presents a situation very different from the usual AdS-CFT lore. It is not clear to us what the precise new prescription for a correspondence would be in this case.
       \item The beta function in fact does not vanish for $Y^A$ different from the attractor point and the $Y^A$ undergo RG flow till they
       reach the attractor point, an IR fixed point. Along the flow, the constituents of M5-M5 bound states (whose gravity description
       is of the type studied in appendix \ref{D4D2D0}) decouple from each other; each of them has its own IR fixed point corresponding to an AdS$_3 \times S^2$.
      \end{enumerate}

The second possibility seems much more attractive to us, but would
imply that the MSW $(0,4)$ model does undergo RG flow. This need
not be in contradiction with the finiteness of $(0,4)$ models,
since the relevant non-renormalization theorems assume that the
sigma model is weakly coupled and non-singular, and both
assumptions are almost certainly violated for the MSW $(0,4)$
model. The latter can become strongly coupled whenever two-cycles
in the moduli space shrink to zero volume (similar to what happens
in the D1-D5 CFT), and is most likely singular when the four-cycle
self-intersects: intersecting M5-branes support extra light
degrees of freedom, coming from stretched M2-branes, and these
need to taken into account in a proper low-energy description. The
classical MSW CFT, however, does not take these additional light
degrees of freedom into account, and usually this gives rise to
singularities in the incomplete low-energy theory. Finally, the
nontrivial interaction between the $b^\alpha$ and $z^i$ modes leading to (\ref{susyconstraint1}),
will further complicate the RG flow. 

It would be interesting to study this further.


\section{Conclusions and discussion}\label{conclusions}

In this paper we described a large number of supersymmetric bound
states of black holes and black rings in a space-time which is
asymptotic to AdS$_3 \times S^2 \times {\rm CY}_3$. M-theory on
the latter space is supposedly a well-defined theory of quantum
gravity and is equivalent to a particular $N=(0,4)$ CFT in two
dimensions, the MSW CFT at the attractor point. Therefore, one can
hopefully ask more precise questions (and provide more precise
answers) about these black hole bound states than one can do in
asymptotically flat space. In particular, no states will come in
from or move out to infinity, and the number of states with given
quantum numbers should be unambiguous. Indeed, the moduli of the
CY are completely fixed at the attractor point at the boundary of
AdS. Exactly how and when the low-energy description of wrapped
M5-branes, which exists for all values of the moduli, flows to
this $(0,4)$ SCFT at the attractor point remains puzzling. We
already discussed this in section~\ref{commquest} and here we will
simply take it as a given fact.

So given the $(0,4)$ CFT dual of AdS$_3 \times S^2 \times {\rm
CY}_3$, a first obvious question that arises is to determine the
dual description of the black hole bound states in the CFT. This
was one of the original motivations of this project, but clearly 
more work remains to be done to be able answer this question. Black
hole/ring bound states should be described by suitable density
matrices that include several generalized chemical potentials. The
simplest example where this can be made explicit is the black
ring, which has an extra chemical potential multiplying a dipole
moment operator, both in its thermodynamic description
\cite{Emparan:2004wy,Copsey:2005se} as well as in the dual density
matrix description \cite{Alday:2005xj}. The black hole bound
states we have been considering should clearly involve many more
chemical potentials. Each of the centers is described by
$n=2(b_2+1)$ different charges, and the total entropy depends on $kn$
quantum numbers, with $k$ the number of centers. Therefore, in the
first law of thermodynamics for the black hole bound states, we
expect to see at least $kn$ different chemical potentials
appearing. It would be very interesting to find a nice basis for
these chemical potentials and to determine to which CFT operators
they couple. Though some insight can in principle be obtained by
studying the subleading behavior of the supergravity solutions
near the AdS boundary, lack of detailed knowledge of the $N=(0,4)$
CFT makes it difficult to proceed in this direction. Qualitatively
we expect that as we turn on more and more chemical potentials we
can describe increasingly more complicated black hole bound
states. In the limit where the number of chemical potentials goes
to infinity, we can resolve individual microstates in the CFT and
find the corresponding microstate geometries.

An alternative approach to understanding the CFT duals of the
multi-centered black holes is to use their description in terms of
attractor flows. As we have discussed, we expect to be able to
associate a unique flow tree to a given supergravity solution.
There are several subtleties which may invalidate this statement:
\begin{itemize}
\item
There exist multicentered ``scaling'' or ``abyss'' solutions
\cite{Denef:2007vg,Denef:2002ru,Bena:2007qc,deBoer:2007}, for
which the centers' coordinates can be partitioned in groups in
such a way that within each group the centers can approach each
other arbitrarily closely. These are most naturally viewed as
being continuously connected to a solution where each group is
collapsed into a single center, and as such should be associated
to the attractor flow tree of the latter solution. Unfortunately,
the split attractor flow conjecture only addresses the existence
of the collapsed solution, but does not say anything about the
existence of a particular scaling solution.
\item
It is not completely clear what the right starting point for the
attractor flow tree should be, since the moduli vary on the
boundary of AdS$_3$. More precisely, it is the $B$-field that
varies, see equation (\ref{bfield}).
\item
As discussed in appendix~\ref{margtresh}, for special values of
the moduli charges can lie on a wall of ``threshold'' stability.
If this happens, attractor flow trees can be continuously deformed
into each other.
\end{itemize}
If we nevertheless assume the split attractor flow conjecture to
hold, then it naturally leads to a partitioning of the Hilbert
space of the CFT. The most obvious guess for the CFT dual to a
particular black hole bound state would then be a density matrix
that involves all the states that live in the sector of the CFT
corresponding to the associated flow tree. However, we have not
yet been able to make either this description or the description
in terms of chemical potentials very explicit, nor have we been
able to understand the phase transitions we encountered directly
in the CFT. In would clearly be very interesting to make progress
in any of these directions.

We would also like to understand in more detail the connection
between the multi-centered solutions and the fuzzball proposal
(for reviews see e.g.\  \cite{Mathur:2005zp,Bena:2007kg}).
Roughly, the idea is that the space of smooth BPS solutions of the
supergravity equations of motion with given charges forms a phase
space\footnote{The notion of smoothness is observer dependent, but
for multi-centered solutions a minimal criterion is that each of
the centers should represent a single state and not carry any
entropy. Typically, the centers will therefore have to be single
branes that carry fluxes only.}, and that quantization of this
phase space precisely reproduces the BPS states of the dual CFT.
For 4d multicenter solutions such a proposal was put forward in
\cite{Balasubramanian:2006gi}. If this works, one can establish a
precise connection between density matrices in the CFT and coarse
grained bulk geometries, and in particular black holes are the
result of coarse graining over a large number of underlying
microstate geometries \cite{Balasubramanian:2005mg,Alday:2006nd}.
Before any of these ideas can be tested, we need to be able to
quantize the moduli space of solutions. This is a problem of
independent interest and will be discussed in a companion paper
\cite{deBoer:2007}. As we will show there, quantization of the
moduli space of solutions does lead to results that are in
agreement with the wall-crossing formula of \cite{Denef:2007vg},
at least in the two- and three-center cases, which is suggestive
but certainly not enough to establish the validity of the fuzzball
proposal. An obvious problem is that it is not clear that we can
avoid the use of excitations that non-trivially involve the
Calabi-Yau manifold (for the simpler case of 1/2 BPS states in
AdS${}_3\times $S$^3\times K3/T^4$ one already needs excitations
that involve $K3/T^4$ to account for the total number of states,
but they are under partial technical control, see e.g.
\cite{Kanitscheider:2007wq}) nor is it clear that one can avoid
the use of stringy excitations. In fact, we do not even know
whether we have here the full set of half-BPS solutions of 5d
$N=1$ supergravity. All our solutions have a $U(1)$ isometry, and
there may well exist many more solutions that have no spatial
isometries at all.

The quantization of the moduli space in \cite{deBoer:2007} will
also enable us to discuss issues related to the possible
non-compactness of the space of solutions and to the
aforementioned multicentered ``scaling'' or ``abyss'' solutions.

One of the issues that our work sheds some light on is the issue
of giant gravitons in AdS${}_3$. Originally, these were thought
not to give rise to BPS states \cite{Lunin:2002bj} because, unlike
in other dimensions, their radial position is a free parameter.
However, in \cite{Mandal:2007ug,Raju:2007uj} this issue was
re-analyzed and it was found that there are bound giant gravitons
in global AdS${}_3$ but not in Poincar\'e coordinates. This
analysis was done for AdS${}_3\times $S$^3$, but a similar result
can be seen to hold for AdS${}_3\times $S$^2$: a giant graviton in a
BTZ black hole background is described by a two-center solution,
each of which carries D4D2D0 charge only. If the inner product
of the two charges does not vanish, there are no supersymmetric
solutions, and if it does vanish, the two centers are mutually BPS
and do not form a bound state. Therefore, we never obtain BPS
states in this way. A giant graviton in global AdS, however, is in
our setup described by a three center solution, consisting of a
pure fluxed D6 brane, a pure fluxed $\bar{\rm D6}$ brane, and a
center with D4D2D0 brane charge only. As we reviewed in
section~\ref{2center} following \cite{Denef:2007yt}, the pure
fluxed D6 and $\antiDp{6}$ solution is equivalent to global
AdS${}_3\times $S$^2$. Thus after a suitable change of coordinates
this three center solution does describe a single giant graviton
in global AdS${}_3$. It is indeed a bound state, as the original
three centers do form non-trivial bound states. Thus, we can
understand bound states in global AdS${}_3$ by adding a pure
fluxed D6 brane and a pure fluxed $\bar{\rm D6}$ brane and by
considering this extended configuration in this framework. It
would be interesting to explore this further and to understand in
more detail the precise action of spectral flow on multiple bound
states.

We also found the need to improve the notion of walls of marginal
stability to distinguish walls where the number states jumps
(still called walls of marginal stability) from walls where the
topology of the flow changes but the number of states does not
jump (called walls of threshold stability). This distinction is of
particular importance in AdS${}_3$ as the moduli at infinity are
fixed and can in fact lie on a wall of threshold stability. A detailed
understanding of these two types of walls and their applications
to state counting problems and stability questions of bound states
of branes is clearly desirable but left to future work.

Among the many other open problems that remain we would like to
mention the applications of our results to refine the computations
of the elliptic genus of the $N=(0,4)$ CFT
\cite{Gaiotto:2006ns,deBoer:2006vg}. To leading order this
partition function roughly looks like $|Z_{\rm top}|^2$, but this
contribution is entirely coming from two-center configurations,
and by including three and more centers we should be able to make
a more detailed study of the corrections that arise. These results
would then carry over to the OSV conjecture restricted to infinite
K\"ahler moduli and configurations without D6-brane charge since
in this limit the BPS index reduces to the elliptic genus of the
CFT.

Coming back to the description of the multi-centered black holes
in the dual CFT, we would like to raise a few more points. First,
one may ask the following question: which geometry is dual to the
density matrix that consists of all states with fixed $L_0$? If
this is a single geometry, then it has to be spherically symmetric
(because the density matrix is rotationally invariant), and the
only spherically symmetric solution is the BTZ black hole. For
large values of $L_0$ this seems fine, since the entropy of the
BTZ black hole agrees with the entropy of the CFT computed using
the Cardy formula in this regime. However, for smaller values of
$L_0$ this is no longer true, as the two-centered solutions start
to dominate the entropy. This shows that for small values of
$L_0$, there cannot be a single, semiclassically reliable
geometric dual of the density matrix consisting of all states.
This is quite surprising, and it shows that for small $L_0$ the
appropriate bulk dual description of this very simple density
matrix should be in terms of a sum over geometries. The small BTZ
black hole is then dual to a density matrix which contains only a
small subset of the total number of states. In \cite{Denef:2007vg}
various arguments for and against were presented that the number
of states in this small subset is equal to the index of the total
number of states. This would require a large amount of
cancellation in the index (some preliminary numerical evidence for
this was presented in \cite{Huang:2007sb}) and it would be
interesting to explore this further. We have not found any obvious
mechanism for this cancellation in the space of multi-centered
solutions.

It is also worth pointing out that a better understanding of the
CFT description of the multi-centered solutions would probably
allow us to give a CFT explanation of the entropy of the BMPV
black hole: we would simply count the number of CFT states dual to
the two-centered solution described in section~\ref{enigma2},
consisting of a BPMV black hole and an (entropyless) purely fluxed
$\overline{\rm D6}$.

Finally, it is well-known that moving into the interior of
AdS$_3$, and ignoring the outside region, is like RG flow in the
dual field theory. The moduli in the interior of AdS$_3$ will
approximately follow the attractor flow tree. Therefore, we should
be able to understand the attractor flow tree, and the
corresponding rearrangement of the degrees of freedom in seemingly
disconnected pieces, from the point of view of the RG flow. In
other words, as we lower the scale, we should encounter phase
transitions whenever we cross a wall of marginal stability in
which the degrees of freedom of the CFT split up into a tensor
product of decoupled sectors. The mechanism responsible for this
could be quite similar to tachyon condensation, as it is a
tachyonic degree of freedom which is responsible for the decay of
BPS states \cite{Denef:2002ru}. If we could make such a "split RG
flow" picture more precise, we would in particular be able to
explain the entropy of all 4d black holes in terms of a 2d CFT. We
hope to come back to this in the near future.

\section*{Acknowledgements}
The authors gratefully acknowledge useful discussions with D.~Anninos, I.~Bena, M.~Cheng,
R.~Emparan, E.~Gimon, M.~Guica, I.~Kanitscheider, A.~Kashani-Poor, F.~Larsen,
H.~Liu, G.~Moore, K.~Papadodimas, J.~Raeymaekers, S.~Raju, B.~van Rees, A.~Strominger and X.~Yin.

DVdB is an Aspirant of the FWO Vlaanderen and thanks the High Energy Theory
Group at Harvard University for its hospitality. FD and DVdB are partially
supported by the European Community's Human Potential Programme under contract
MRTN-CT-2004-005104 `Constituents, fundamental forces and symmetries of the
universe', by the FWO - Vlaanderen, project G.0235.05 and by the Federal Office
for Scientific, Technical and Cultural Affairs through the `Interuniversity
Attraction Poles Programme - Belgian Science Policy' P6/11-P.  The work of JdB,
SES and IM is supported financially by the Foundation of Fundamental Research on
Matter (FOM).

\appendix
\section{Conventions and notation}\label{conventions}
For various common definitions we refer to appendix A of \cite{Denef:2007vg}, whose notation we follow.

In this appendix we give some more details on the conventions we take for various physical quantities.
We work in units in which $c=\hbar=k_B=1$ but we will keep
dimensions of length explicitly in most part of the paper. The coordinates $x,t$ we take to
have dimension of length. Angular coordinates, most of the time
denoted by Greek letters as $\alpha\,, \theta\,, \psi$ etc will be
dimensionless however. Furthermore we will take forms to be
dimensionless. As e.g.\  $\omega=\omega_i dx^i$ is dimensionless this
implies the components of forms have dimensions of inverse length,
i.e. $[dx^i]=L,\ [\omega_i]=L^{-1}$ and $[\omega]=1$. This convention
implies that the Hodge star is dimensionful: $[\star]=L^{d-2p}$
when acting on a $p$-form.

In each dimension we define a natural Planck length $l_d$ ($[l_d]=L$
of course) by normalising the Einstein-Hilbert action as
\begin{equation}
S^{\mathrm{EH}}_d=\frac{2\pi}{(l_d)^{d-2}}\int
\sqrt{-g_{d}}\,\calr_{d}\,,
\end{equation}
and a reduced planck length by
\begin{equation}
\ell_d=\frac{l_d}{4\p}\,.
\end{equation}

\subsection{M-theory vs IIA conventions}
We start from the following $11d$ M-theory metric:
\begin{equation}
ds^2_{11d}=R^2\,e^{4\phi/{3}}\,\Theta^2+e^{-2\phi/{3}}\,ds^2_{10d}\,,
\end{equation}
where $ds^2_{10d}$ is a ten-dimensional metric and $R$ is a constant
with dimensions of length. The one form $\Theta=d\theta+2\pi\,A$,
with $\theta=\theta+2\pi$ and $A$ is a one form on the ten
dimensional space. Furthermore $\phi$ is normalized in such a way
that $\phi(\infty)=0$. The M2-branes of this theory have a tension
\begin{equation}
T_{M2}=\frac{2\pi}{l_M^3}\,,
\end{equation}
with $l_M=l_{11}$ is the 11 dimensional Planck length.

We can relate these to IIA quantities by reduction on the $\theta$
circle. As an M2 wrapped around this circle is a fundamental string
we find:
\begin{equation}
T_{F1}=\frac{2\pi}{l_s^2}=2\pi R\,T_{M2}=\frac{4\pi^2}{l_M^3}R\quad\Rightarrow\quad
l_{M}^3=2\pi R\,l_s^2\,.
\end{equation}
From the relation between M2 and D2 one easily infers
\begin{equation}
l_M^3=g_s\,l^3_s\,,
\end{equation}
where in our conventions $T_{\Dp{}p}=\frac{2\pi}{g_s\,l_s^{p+1}}$.
The constants $g_s$ and $l_s$ are respectively the string coupling
constant (at infinity) and the string length. They are arbitrary
constants related to the 10 dimensional Planck length $l_{10}$ by
\begin{equation}
l_{10}^4=g_s\,l_s^4\,.
\end{equation}
We can now reduce both the 11d and 10d theory on the same Calabi-Yau
manifold $X$. Since
\begin{equation}
ds^2_{11d}=R^2\,e^{4\phi/{3}}\,\Theta^2+e^{-2\phi/{3}}\left(\,ds^2_{4d}+ds^2_{CY\,\IIA}\right)\,,
\end{equation}
we can relate the effective 5d and 4d metrics:
\begin{equation}
ds^2_{5d}=R^2\,\left(\frac{V_{\IIA}}{V_M}\right)^{2/3}
\Theta^2+\left(\frac{V_{\IIA}}{V_M}\right)^{-1/3}\,ds^2_{4d}\,,
\end{equation}
where we used that
\begin{equation}
e^{2\phi}=\frac{V_{\IIA}}{V_M}\label{dilaton}\,.
\end{equation}
In a slightly more transparent form this is
\begin{equation}
ds^2_{5d}=\tilde{V}_{\IIA}^{2/3}\,\ell_5^2\,(2\Theta)^2+\tilde{V}_{\IIA}^{-1/3}\,\frac{\hat{R}}{2}\,ds^2_{4d}\,,\label{final5d}
\end{equation}
with $\ell_5$ the reduced 5 dimensional Planck length, we will also use the
notation $2\Theta=d\psi+A^0_{\mathrm{4d}}$. We use
various dimensionless objects:
\begin{equation}
\tilde{V}_M=\frac{V_M}{l_M^6},\qquad
\tilde{V}_\IIA=\frac{V_\IIA}{l_s^6},\qquad \hat{R}=\frac{R}{\ell_5}\,.
\end{equation}
The Calabi-Yau reduction relates all the different parameters at
infinity. We will give the relations that will be of most importance
to us. The relation between the 4d Planck length $l_4$ and the
string length is
\begin{eqnarray}
l_4=g_{4d}\,l_s\,,
\end{eqnarray}
where
\begin{equation}
g_{4d}^2=\frac{g_s^2}{\tilde{V}^{\infty}_{\IIA}}=\frac{1}{\tilde{V}_M}\,.
\end{equation}
The effective $4d$ string coupling $g_{4d}$ is in a hypermultiplet
and thus constant in the solutions we will consider. Note that the
same is true for $\tilde{V}_M$. The 4d and 5d Planck lengths are
related by the size of the M-theory circle:
\begin{equation}
\ell_5=\sqrt{\frac{\hat{R}}{2}}\,\ell_4\,.\label{plancklengths}
\end{equation}
Furthermore this size of the circle is immediately related to the
size of the Calabi-Yau at infinity and thus to the value of the
K\"ahler moduli at infinity, i.e.
\begin{equation}
\frac{\hat{R}^3}{8}=\tilde{V}^{\infty}_{\IIA}=\frac{1}{6}(J_{\infty})^3\,.\label{asympmoduli}
\end{equation}
Finally let us relate the reduced 4d plank length $\ell_4$ to Newton's constant
that appears in front of the 4d Einstein-Hilbert action as
\begin{equation}
S^{\mathrm{EH}}_{4}=\frac{1}{16\pi\,G_4}\int\sqrt{-g_4}\calr_4\,.
\end{equation}
This gives the relation
\begin{equation}
\ell_4=\sqrt{2\,G_4}\,,
\end{equation}
and by (\ref{plancklengths}) this implies
\begin{equation}
\sqrt{G_4}=\frac{\ell_5^{3/2}}{\sqrt{R}}\,.
\end{equation}

\section{Marginal vs threshold stability}\label{margtresh}

In this appendix we refine
the commonly used notion of \textit{marginal stability}. This refinement is, in
our view, useful as there are two different physical situations that both go
under the name of marginal stability in the current literature. Distinguishing
between them is useful in analyzing the decoupling limit. A somewhat similar distinction was already proposed in \cite{Denef:2001ix}.

The notion that we want to refine and that is commonly referred to as marginal
stability is that of two BPS states having aligned central charges for certain
values of the moduli. In our case of interest, multicentered black holes in
$\mathcal{N}$=2 supergravity, the BPS states are characterised by their charge $\G$ and
their central charge is determined in terms of this charge and the scalar moduli
$t$ by $Z(\G,t)=\langle \G,\Omega(t) \rangle$. The length of the central charge
vector, $|Z|$, corresponds to the mass, as we are considering BPS states, and
its phase, $\a$, characterises the supersymmetries left unbroken by this state. In case the
moduli are such that for two BPS states $\G_1$ and $\G_2$ the phase aligns, the
two BPS particles preserve the same supersymmetries and the binding energy of a BPS bounds state of them
vanishes (if it exists),
as $|Z_{1+2}|=|Z_1|+|Z_2|$. This is equivalent to the condition
\begin{equation}
\mathrm{Im}(\bar Z_1Z_2)=0\quad\mbox{and}\quad\mathrm{Re}(\bar Z_1Z_2)>0\,.\label{marggen}
\end{equation}
The second inequality is needed to ensure that the central charges not only
align but also point in the same direction. As the condition (\ref{marggen}) is
a single real equation it will, in general, be satisfied on a codimension one
surface in moduli space. Crossing such a surface or 'wall' may correspond to the
decay of the bound state formed by the two charges, but it does not have to. Whether a
bound state decays or not depends on the intersection product of these charges.
In the case $\langle\G_1,\G_2\rangle=0$, i.e. the charges are mutually local,
there will be no decay whereas if charges are mutually non-local,
$\langle\G_1,\G_2\rangle\neq0$, there will be a decay.  This follows because in
the constraint equation, (\ref{consistency2}), the RHS depends on
$\mathrm{Im}(\bar Z_1Z_2)$ so the inter-center separation is given by
\begin{equation}\label{2centersep}
    r_{12} =
    \frac{\langle\G_1, \G_2\rangle}{\langle h , \G_1 \rangle } =
    \frac{\langle\G_1, \G_2\rangle\, |Z_1 + Z_2|}{2 \,\mathrm{Im}(\bar Z_2
    Z_1)}\biggr|_\infty\,.
\end{equation}

This qualitative difference when approaching or crossing such a hypersurface in
moduli space prompts us to name them differently so we can easily refer to the
appropriate picture.  Therefor we define
\begin{eqnarray*}
\mbox{\textbf{Marginal stability:}}\quad& \mathrm{Im}(\bar Z_1Z_2)=0\,,\quad\mathrm{Re}(\bar Z_1Z_2)>0\quad \mbox{ and } \langle\Gamma_1,\Gamma_2\rangle\neq0\\
\mbox{\textbf{Threshold stability:}}\quad& \mathrm{Im}(\bar Z_1Z_2)=0\,,\quad\mathrm{Re}(\bar Z_1Z_2)>0\quad \mbox{ and } \langle\Gamma_1,\Gamma_2\rangle=0
\end{eqnarray*}
Thus we will refer to the codimension one hypersurfaces on which this condition
is satisfied as \textbf{walls of marginal/threshold stability}, respectively. As
mentioned above the physics of bound states is rather different when crossing a
wall of marginal stability or one of threshold stability.  So let us shortly
review this physics to make things clear. The discussion can be most easily
understood when illustrated by an example although the story is general and
holds for all multicenter black holes.

We take as our example a simple three center solution consisting of the charges
\begin{equation}
\G_1=(1,\frac{p}{2},\frac{p^2}{8},\frac{p^3}{48})\,,\quad \G_2=(-1,\frac{p}{2},-\frac{p^2}{8},\frac{p^3}{48})\quad\mbox{and}\quad\G_3=(0,0,0,-n)\,.\label{triplecharges}
\end{equation}
This configuration is discussed in some detail in \cite{Denef:2007vg} and an attractor flow tree is given in figure \ref{triplesplit}.
\FIGURE{
\includegraphics[page=2,scale=0.7]{pictures.pdf}
\caption{Attractor flow for the charges $\G_1=(1,1,1/2,1/6)$, $\G_2=(-1,1,-1/2,1/6)$ and $\G_3=(0,0,0,-1/100)$. The attractor point for $\G_1$ is the box on the $B$-axis on the left, that for $\G_2$ the one on the right. The attractor point for $\G_3$ lies at infinite $J$.}\label{triplesplit}
}
In this figure \ref{triplesplit} the green line is a wall of marginal stability for the charges $\G_2$ and $\G_1+\G_3$. More precisely on this line $\mathrm{Im}(\bar Z_2Z_{1+3})=0$. As the intersection product $\langle \G_2, \G_1+\G_3\rangle=\frac{p^3}{6}-n$ is non-vanishing this is thus a wall of marginal stability in our refined sense. In this same example the $J$-axis, i.e. $B=0$, is a wall of threshold stability for the charges $\G_1+\G_2$ and $\G_3$, i.e. $\mathrm{Im}(\bar Z_{1+2}Z_{3})=0$ at $B=0$ and $\langle\G_1+\G_2,\G_3\rangle=0$. We will now look at the behavior of the split flow and the solution space in approaching this wall of marginal or threshold stability respectively. In both cases we start from the attractor flow depicted in figure \ref{triplesplit}, which has its starting point at $B=-1$ and $J=9/4$. First we will discuss what happens while we keep $B$ fixed and lower $J$ thus approaching the wall of marginal stability discussed above. Secondly we will see what happens when one keeps $J$ fixed but moves $B$ towards positive values thus crossing the wall of threshold stability at $B=0$ pointed to above.

Starting at a negative value for the $B$-field modulus and a large enough K\"ahler modulus a split flow $(\G_2,(\G_1,\G_3))$ exists and in spacetime this corresponds to a supergravity solution corresponding to the $\antiDp{0}$ ``orbiting'' the $\Dp{6}$ which then together bind to the $\antiDp{6}$, see figure \ref{modmarg} A. We can now see what happens in case we start moving the starting point of the attractor flow tree. We keep the value of the B-field fixed and lower the K\"ahler modulus towards zero. In this way we will approach the wall of marginal stability for the charges $\G_2$ and $\G_1+\G_3$, the green line in figure \ref{triplesplit}. Approaching this wall corresponds to the $(\G_1,\G_3)$ cluster being forced further and further away from the $\G_2$ center. A plot of the solution space for values of the moduli closer and closer to marginal stability is given in figure \ref{modmarg} A through C. The centers are forced infinitely far apart and decay the moment the starting point coincides with the wall of marginal stability and the solution ceases to exist once the wall has been crossed. This is the familiar decay of multicenter bound states when crossing a wall of marginal stability. Also microscopically the bound state disappears out of the spectrum and the BPS index makes a jump. The way this is manifested in the split flow picture is by the fact that the split flow tree only exists on one side of the wall of marginal stability.

\FIGURE{
\includegraphics[page=3,scale=0.7]{pictures.pdf}
\caption{On the left the attractor flows for the charges of fig. \ref{triplesplit}
 are shown for different values of the starting moduli. On the right the corresponding solution moduli space is plotted. The red points are the positions of $\G_1$ (right) and $\G_2$ (left). In blue are the possible positions of $\G_3$. Note the difference in the scale in the last plot, this as once we approach marginal stability the relative position of the centers diverges.}\label{modmarg}
}

In case of crossing a wall of threshold stability the physics is different.  We can start from the same initial configuration but now deform it in a different way. We now move the starting point in moduli space towards the $J$-axis along a trajectory of constant $J$. We have plotted the solution space along this trajectory in figure \ref{modthresh} A through E. Approaching the wall of threshold stability $B=0$, the orbit of the $\Dp{0}$ around the $\Dp{6}$ becomes more and more deformed and it expands. Once we reach threshold stability the $\Dp{0}$ is equally bound to the $\Dp{6}$ as to the $\antiDp{6}$ and can sit anywhere on the equidistant plane between $\Dp{6}$ and $\antiDp{6}$. Note that this plane is non-compact, i.e. the $\Dp{0}$ can move arbitrarily far away along this plane, while the orbits before were large but always compact. Continuing further to positive values for $B$ the orbit of the $\Dp{0}$ becomes compact again but has now become an orbit around the $\antiDp{6}$. This corresponds to the fact that the split flow has now changed topology from $(\G_2,(\G_1,\G_3))$ to $(\G_1,(\G_2,\G_3))$. In this process no states have decayed and no solutions have been lost.

\FIGURE[h!]{
\includegraphics[page=4,scale=0.7]{pictures.pdf}
\caption{Here we show the same type of plots as in fig. \ref{modmarg}, but now taking the starting point through a wall of threshold stability, in this case the $J$-axis.}\label{modthresh}
}

This example illustrates the general behavior that we can summarize as follows:
\begin{itemize}
\item A wall of marginal stability (in the refined sense) corresponds to a boundary between a region in moduli space where a certain multicenter solution exists and a region where it no longer exists. In the supergravity picture the disappearance of the bound state happens as a number of centers separate further and further towards infinite separation at marginal stability. Crossing a wall of marginal stability corresponds to a decay of states and a jump in the index counting these states.

\item A wall of threshold stability corresponds to a boundary between two regions of different 'topology'. This holds both on the level of the flow tree that changes topology, i.e. the type and order of splits changes, as on the level of the solution space that changes topology as a manifold. This change of topology of the solution manifold can happen as exactly at threshold stability the solution space becomes non compact. Note that when crossing a wall of threshold stability no states decay, they only change character.
\end{itemize}

So at threshold stability some centers are allowed to move of to infinity but it is also possible for them to sit at finite distance to the other centers; they are not forced to infinite separation as is the case for marginal stability. Although the solution space is non-compact, it turns out to have finite symplectic volume when considered as a phase space \cite{deBoer:2007}. One can check explicitly that this number of states equals that on both sides of the wall of threshold stability and so crossing a wall of threshold stability does \textit{not} correspond to a decay of states --- rather, at the wall, the BPS states exist as bound states \emph{at threshold} (hence the name), similar to D0-branes in type IIA string theory in flat space. As will be discussed in more detail in \cite{deBoer:2007} non-compactness of the solution space can only appear at threshold. Furthermore we will discuss there how the definitions generalize to the case where more than two charges have aligned central charges at a single point in moduli space.

\section{Rescaled solutions}\label{app_rescaled}

In this appendix we provide the explicit form of the multicentered solutions in
rescaled coordinates $\mathsf{x}_i$ and in terms of the rescaled harmonics
$\mathsf{H}$,
\begin{equation}
\mathsf{H}=\sum_a\frac{\G_a}{\sqrt{R}\,|\mathsf{x}-\mathsf{x}_a|}-2\ell_5^{3/2}\mathrm{Im}(e^{-i\alpha}\O)|_\infty\,.\label{rescaledharmonicsa}
\end{equation}
The rescaled solution is given by
\begin{eqnarray}
 ds^2_{\mathrm{4d}} & = & -\frac{1}{\S}(dt+\frac{\o}{\sqrt{R}})^2+\S\, d\mathsf{x}^id\mathsf{x}^i\,, \nonumber\\
 \cala^0 & = & \frac{-L}{\S^2}\left(\sqrt{R}dt+\o\right)+\omega_0\,,\label{rescaledmulticentera}\\
 \cala^A & = &\frac{\mathsf{H}^AL-Q^{3/2}y^A}{\mathsf{H}^0\S^2}\left(\sqrt{R}dt+\o\right)+\cala_d\,,\nonumber\\
 t^A&=&B^A+i\,J^A=\frac{\mathsf{H}^A}{\mathsf{H}^0}+\frac{y^A}{Q^{\frac{3}{2}}}\left(i\S-\frac{L}{\mathsf{H}^0}\right).\nonumber
 \end{eqnarray}
These relate to the other rescaled functions appearing in
(\ref{rescaledmulticentera}) through:
\begin{eqnarray}
 d\omega_0 & = &\sqrt{R}\, \star d\mathsf{H}^0 \,,\nonumber\\
 d\cala_d^A & = & \sqrt{R}\,\star d\mathsf{H}^A \,,\nonumber\\
 \star d\o & = & \sqrt{R}\,\langle d\mathsf{H},,\mathsf{H}\rangle \,\nonumber\\
 \S&=&\sqrt{\frac{Q^3-L^2}{(\mathsf{H}^0)^2}}\,,\label{rescaledconditionsa}\\
 L&=&\mathsf{H}_0(\mathsf{H}^0)^2+\frac{1}{3}D_{ABC}\mathsf{H}^A\mathsf{H}^B\mathsf{H}^C-\mathsf{H}^A\mathsf{H}_A\mathsf{H}^0\,,\nonumber\\
 Q^3&=&(\frac{1}{3}D_{ABC}y^Ay^By^C)^2\,,\nonumber\\
 D_{ABC}y^Ay^B&=&-2\mathsf{H}_C\mathsf{H}^0+D_{ABC}\mathsf{H}^A\mathsf{H}^B\,.\nonumber
 \end{eqnarray}
Of course the form of the rescaled consistency condition doesn't
change:
\begin{equation}
  \langle \mathsf{H},\G_s\rangle|_{\mathsf{x}=\mathsf{x}_s}=0\,.
 \label{rescaledconsistencya}
\end{equation}

The rescaled 5d lift is
\begin{eqnarray}
\frac{1}{\ell_5^2}\,ds^2_{5d}&=&\tilde{V}_{\IIA}^{2/3}\,\left(d\psi+\cala^0\right)^2+\frac{R}{2}\tilde{V}_{\IIA}^{-1/3}\,ds^2_{\mathrm{4D}}\,,\nonumber\\
A_{\mathrm{5d}}^A&=&\cala^A+B^A\left(d\psi+\cala^0\right)\,,\label{rescaled5Dsolutiona}\\
Y^A&=&\tilde{V}_{\IIA}^{-1/3}\,J^A\,,\qquad\tilde{V}_{\IIA}=\frac{D_{ABC}}{6}J^AJ^BJ^C=\frac{1}{2}\left(\frac{\Sigma}{Q}\right)^3\,.\nonumber
\end{eqnarray}

The more explicit form of the five dimensional metric becomes in terms of the rescaled
variables
\begin{eqnarray}
\frac{1}{\ell_5^2}\,ds^2_{5d}&=&2^{-2/3}\,Q^{-2}\left[-(\mathsf{H}^0)^2(\sqrt{R}dt+\omega)^2-2L(\sqrt{R}dt
+\omega)(d\psi+\omega_0)
+\Sigma^2(d\psi+\omega_0)^2\right]\nonumber\\&&+2^{-2/3}R\,Q\,d\mathsf{x}^id\mathsf{x}^i\,.\label{rescaledexplicita}
\end{eqnarray}

\section{$\Dp{4}\Dp{2}\Dp{0}$ bound
states}\label{D4D2D0}

In this appendix we show that bound states of two
$\Dp{4}\Dp{2}\Dp{0}$ centers exist for arbitrarily large asymptotic
K\"ahler moduli (i.e. $J^3_\infty >> 1$). It was noticed in \cite{Denef:2007vg} that such a bound state of charges $\G_1=(0,p^A_1,q_A^1,q_0^1)$ and $\G_2=(0,p^A_2,q_A^2,q_0^2)$ doesn't exist in case $J^A_\infty= p^A\lambda$ with $\lambda>>1$ and $B^A|_\infty=D^{AB}q_B$. This can be understood by computing the following quantity:
\begin{equation}
\langle \G_1,\G_2\rangle\mathrm{Im}(Z_1\bar{Z}_2)=-\frac{3}{8}(p^A_1q_A^2-p^A_2q_A^1)^2+\calo(\l^{-1})<0\,.
\end{equation}
As the distance between the two centers is given by $\frac{\langle \G_1,\G_2\rangle}{2\mathrm{Im}(Z_1\bar{Z}_2)}$ this implies no such solutions can exist for these asymptotic moduli. So bound states of $\Dp{4}\Dp{2}\Dp{0}$ centers don't exist in the large volume region of
moduli space if the asymptotic K\"ahler moduli are proportional to the
$\Dp{4}$- charge, $J^A_\infty\sim p^A$, and the asymptotic B-field moduli are $B^A|_\infty=D^{AB}q_B$. But a priory nothing forbids
to consider asymptotic moduli of a more general form. Indeed, as we will show in this appendix,
$\Dp{4}\Dp{2}\Dp{0}$ bound states turn out to exist for more general
moduli. Although we expect such bound states to exist quite
generically, a thorough analysis of these type of multicenters is
outside the scope of this paper. We will only present a simple class
of such solutions for a given compactification to show that they do
indeed exist. Related examples where discussed recently in \cite{Diaconescu:2007bf}.

Note that in this appendix we will work in {\bf unrescaled} variables.

\subsection{A class of solutions}
The simplest setting one can consider to find these $\Dp{4}\Dp{2\Dp{0}}$ bound states is
in case of a two dimensional moduli space. As an example we take the
resolution of the hypersurface $x_1^8+x_2^8+x_3^4+x_4^4+x_5^4=0$ in
$\mathbb{P}^{(1,1,2,2,2)}$ as our Calabi-Yau manifold (see e.g.\
\cite{Candelas:1993dm}). This Calabi-Yau has two K\"ahler moduli and
its intersection numbers are $D_{11A}=0, D_{122}=4, D_{222}=8$ (and
permutations). We will often parametrize the two K\"ahler moduli as
\begin{equation}
J^A=\tilde{V}_\IIA^{1/3} Y^A\,,
\end{equation}
with $\tilde{V}_\IIA^{1/3}=(\frac{1}{6}D_{ABC}J^AJ^BJ^C)^{1/3}$ and $2\tilde{V}_\IIA^{1/3}|_\infty=\hat{R}$.
In this specific case this implies the constraint
\begin{equation}
\frac{4(Y^1)^3}{3}+2Y^1(Y^2)^2=1\,.
\end{equation}
Note furthermore that the $Y^A$ are related to the M-theory K\"ahler moduli as $J^A_M=\tilde{V}_M^{1/3}Y^A$.

We will now show that there exist bound states of charges
\begin{eqnarray}
\G_1&=&(0,\begin{pmatrix}p^1\\p^2\end{pmatrix},\begin{pmatrix}q\\-\l q\end{pmatrix},q_0)\,,\nonumber\\
\G_2&=&(0,\begin{pmatrix}p^1\\p^2\end{pmatrix},\begin{pmatrix}-q\\\l
q\end{pmatrix},q_0)\,,\label{charges}
\end{eqnarray}
in case $q_0<0$. The total charge is thus
\begin{eqnarray}
\G&=&(0,\begin{pmatrix}2p^1\\2p^2\end{pmatrix},\begin{pmatrix}0\\0\end{pmatrix},2q_0)\,.\label{totalcharge}
\end{eqnarray}
Note that as $q_0<0$ this total charge can also exists as a single center
BPS black hole \cite{Denef:2007vg}. We show the existence of the two center bound state
by proving that an attractor flow exists if certain conditions on
$q$ and $\l$ are met. Furthermore we check the existence of the
solution explicitly for a numerical example.

To show that a well defined split attractor flow exists it
is enough to show the following:
\begin{itemize}
\item There exist moduli at infinity for which the stability
constraint
\begin{equation}
\langle\G_1,\G_2\rangle\,\mathrm{Im}(\bar{Z}\,Z_1)|_\infty>0\,\label{stability}
\end{equation} is satisfied.
\item The single center attractor flow for the total charge,
starting at these moduli at infinity, crosses a wall of marginal
stability for the split.
\item The two centers exist separately as single centers.
\end{itemize}

We will show that these three conditions can all be fulfilled for a
split of any given total $\Dp{4}\Dp{0}$ charge with negative
$\Dp{0}$ charge, in charges of the form (\ref{charges}), if one
chooses $q,\l$ and the asymptotic moduli $t^A|_\infty$
appropriately.

\paragraph{Stability condition} Take the moduli at infinity to be
\begin{eqnarray}
\hat{R}&>>&1\,,\\
Y^1_\infty&=&Y^2_\infty=(\frac{3}{10})^{1/3}\,,\\
B^A_\infty&=&0\,.
\end{eqnarray}
It is not difficult to verify that for the charges (\ref{charges})
and on the Calabi-Yau we consider
\begin{eqnarray}
\langle\G_1,\G_2\rangle&=&-2 (p^1-\l p^2) q\,,\nonumber\\
\mathrm{Im}(\bar{Z}\,Z_1)|_\infty&=&-\frac{3}{20} (\l-1) (p^1+4 p^2)
q+\calo(\hat{R}^{-1})\,. \label{stability2}
\end{eqnarray}
The stability constraint (\ref{stability}) is thus satisfied if
$(\l-1)(p^1-\l p^2)>0$ i.e.
\begin{equation}
1<\l<\frac{p^1}{p^2}\qquad\mbox{or}\qquad
\frac{p^1}{p^2}<\l<1\,.\label{condition}
\end{equation}

\paragraph{Flow and marginal stability} One can easily calculate that
the attractor moduli for the total charge (\ref{totalcharge}) are
given by
\begin{eqnarray}
B^A_{*}&=&0\,,\\
J^1_{*}&=&-\frac{8p^1q_0}{S(\G_1+\G_2)}\,,\\
J^2_{*}&=&-\frac{8p^2q_0}{S(\G_1+\G_2)}\,,
\end{eqnarray}
where
\begin{equation}
S(\G_1+\G_2)=8\sqrt{-\frac{2(p^2)^2}{3}q_0(3p^1+2p^2)}
\end{equation}
is the entropy corresponding to the total charge. We will now show
that the single center attractor flow for this charge has to cross a
wall of marginal stability for a split of the form (\ref{charges}).
For this note first that the B-field moduli are $B^A=0$, both at
infinity and at the attractor point.  One can check that indeed the
B-field stays constant under the attractor flow and that the moduli
only flow in the $J^1, J^2$ plane. Secondly, note that the moduli at
infinity lie far away from the origin on the line $J^1=J^2$ and that
at the attractor point (for the total charge) they lie at some finite distance from the origin on
the line $J^1=\frac{p^1}{p^2}J^2$. One can verify however that there
exists a line of marginal stability $J^1=\l J^2$ for a split in
charges (\ref{charges}).  In the case that $p^1<p^2$ we can choose
$\l$ such that $\frac{p^1}{p^2}<\l<1$ and in this case the single
center flow has to cross the line of marginal stability somewhere on
its flow from infinity to the attractor point. The situation is
illustrated in figure \ref{negq0}.
\FIGURE{
\includegraphics[page=17,scale=0.6]{pictures.pdf}\caption{In this figure the $(J^2,J^1)$-plane is shown. The red dot
is the single center attractor point that lies on the (red) line
$J^1=\frac{p^1}{p^2}J^2$. The solid blue line is a line of marginal
stability $J^1=\l J^2$, $1>\l>\frac{p^1}{p^2}$. It is clear that if
one takes the boundary moduli in the shaded blue area, the flow to
the attractor point always has to cross the wall of marginal
stability. This is thus e.g.\  the case if we choose our moduli at
infinity to lie on the orange line $J^1=J^2$. The numerical values have
been taken from the example below.}\label{negq0}
} The case
$p^1>p^2$ is analogous, with now $1<\l<\frac{p^1}{p^2}$. So we can always find charges of the form (\ref{charges}) such that the single center atractor flow for the total charge has to cross a wall of marginal stability.

\paragraph{Existence of the separate centers} As shown in e.g.\  \cite{Denef:2007vg}, to check if a single
center $\Dp{4}\Dp{2}\Dp{0}$ exists it is enough to verify that
\begin{equation}
\hat{q}_0=-\frac{1}{2}D^{AB}q_Aq_B+q_0<0\,.
\end{equation}
Evaluating this constraint for two charges of the form
(\ref{charges}) gives a single constraint:
\begin{eqnarray}
\frac{p^1+2 \l p^2 +2 p^2 }{8 (p^2)^2}q^2+q_0&<&0\,,
\end{eqnarray}
this thus gives a constraint on the size of $q$ i.e.
\begin{eqnarray}
q^2&<&\frac{-8 q_0(p^2)^2}{p^1+2 \l p^2 +2 p^2 }\,.
\end{eqnarray}
\vspace{1cm}
The three discussions above show that a valid split attractor flow exists
for two-centers of the form
\begin{eqnarray}
\G_1&=&(0,\begin{pmatrix}p^1\\p^2\end{pmatrix},\begin{pmatrix}q\\-\l q\end{pmatrix},q_0)\,,\nonumber\\
\G_2&=&(0,\begin{pmatrix}p^1\\p^2\end{pmatrix},\begin{pmatrix}-q\\\\l
q\end{pmatrix},q0)\,,
\end{eqnarray}
in case
\begin{eqnarray}
p^A&>&0\,,\\
q_0&<&0\,,\\
1&<&\l<\frac{p^1}{p^2}\qquad\mbox{or}\qquad
\frac{p^1}{p^2}<\l<1\,,\\
q^2&<&\frac{-8 q_0(p^2)^2}{p^1+2 \l p^2 +2 p^2 }\,.
\end{eqnarray}
It is clear that any total $\Dp{4}\Dp{0}$ state with negative
$\Dp{0}$ can be split in such a way.

Note that the equilibrium distance between the centers is
\begin{equation}
|x_1-x_2|=\frac{2(p^1-\l p^2)
}{(\frac{3}{10})^{\frac{1}{3}}(\l-1)}\ell_5+\calo(\ell_5^2)\,,
\end{equation}
where we used the relations between $R, \ell_5$ and $G_4$, see appendix \ref{conventions}. Note that this distance scales as $\ell_5$ while the distance between bound centers carrying D6 charge scales as $\ell_5^3/R^2$. This has some important consequences when considering the decoupling limit as discussed in section \ref{existence}.

\subsection{Numerical example} \label{app:splflnum}
Here we will numerically compute the split attractor flow for a specific example of the class of solutions presented in the previous subsection. Take the following charges\footnote{In principle the charges need to be very large
 to satisfy all kind of assumptions silently made. This can be easily obtained by using a scaling symmetry \cite{Denef:2007vg} to scale the charges uniformly to some big value.}
\begin{eqnarray}
\G_1&=&(0,\begin{pmatrix}1\\3\end{pmatrix},\begin{pmatrix}2\\-1 \end{pmatrix},-3)\,,\nonumber\\
\G_2&=&(0,\begin{pmatrix}1\\3\end{pmatrix},\begin{pmatrix}-2\\1\end{pmatrix},-3)\,.\label{chargesex}
\end{eqnarray}
The total charge is thus
\begin{eqnarray}
\G&=&(0,\begin{pmatrix}2\\6\end{pmatrix},\begin{pmatrix}0\\0\end{pmatrix},-6)\,.\label{totalchargeex}
\end{eqnarray}
\FIGURE[left]{
\includegraphics[page=18,scale=0.5]{pictures.pdf}
\caption{In this figure the $(50 B^1,J^1,J^2)$ subspace in moduli space is shown.
The blue surface is a wall of marginal stability for the split
(\ref{chargesex}). The central red line is the attractor flow for a
single center solution with the same total charge as the two center.
The pink surface shows the values the moduli take in the two center
solution.}\label{splitflow}
}
There is a wall of marginal stability (at $B^A=0$))
\begin{equation}
J^1=\frac{1}{2}J^2\,.
\end{equation}
The attractor point for the total charge is
\begin{equation}
B^A_{*t}=0\,,\qquad J^1_{*t}=\frac{1}{3 \sqrt{2}}\,,\qquad
J^2_{*t}=\frac{1}{\sqrt{2}}\,.
\end{equation}
Note that indeed in the $(J^1,J^2)$-plane, the wall of marginal
stability $(1,2)s$ separates this attractor point
$\frac{1}{3\sqrt{2}}(1,3)$ from the starting point at infinity:
$\hat{R}(\frac{3}{10})^{1/3}(1,1)$. Figure \ref{negq0} shows the
attractor point and the line of marginal stability in the
$(J^1,J^2)$-plane for this example. As discussed in the previous subsection, if
we take the moduli at infinity to be
\begin{equation}
\hat{R}=200,\qquad
Y^1_\infty=Y^2_\infty=(\frac{3}{10})^{1/3},\qquad
B^A_\infty=0\,,
\end{equation}
the single center flow crosses the wall of marginal stability and the split flow corresponding to the charges (\ref{chargesex}) exists.
The two centers of the split
have the following attractor points:
\begin{equation}
B^1_{*1}=-\frac{17}{36}\,,\qquad B^2_{*1}=\frac{1}{6}\,,\qquad
J^1_{*1}=\frac{\sqrt{\frac{11}{3}}}{9}\,,\qquad
J^2_{*1}=\frac{\sqrt{\frac{11}{3}}}{3}\,,
\end{equation}
and
\begin{equation}
B^1_{*1}=\frac{17}{36}\,,\qquad B^2_{*1}=-\frac{1}{6}\,,\qquad
J^1_{*1}=\frac{\sqrt{\frac{11}{3}}}{9}\,,\qquad
J^2_{*1}=\frac{\sqrt{\frac{11}{3}}}{3}\,,
\end{equation}
note that they both exist as a single center and the central charges
at the attractor points are $Z_{*1}=Z_{*2}=\frac{88}{9}>0$. A
numerical computation of the split attractor flow is shown in figure
\ref{splitflow}.


\section{Gauge field contribution to conserved charges} \label{csterm}
In this appendix we give some more detail concerning the contribution of the
various gauge fields to the conserved boundary charges. The five dimensional
$\mathcal{N}$=1 supergravity theory of which our asymptotic AdS$_3\times$S$^2$ configurations are solutions has $b_2$ U(1) vectorfields, where $b_2$ is the second Betti number of the Calabi-Yau we compactified on. After reduction over the asymptotic two-sphere we end up with an additional SU(2) gauge field as will be explained in some detail below. Analyzing how the action on the boundary of AdS varies with respect to these gauge fields and the metric gives the conserved currents of the boundary theory that can be identified with a 2d CFT.

Before doing this analysis explicitly one can save some work by considering the behavior of the theory near that asymptotic boundary. As in (\ref{expmetric}) we can in general write an asymptotic AdS$_3$ metric as
\begin{equation}
ds^2_{\mathrm{3d}}=d\eta^2+(e^\frac{2\eta}{R_\mathrm{AdS}}g_{ij}^{(0)}+g_{ij}^{(2)})du^idu^j\,,
\end{equation}
where the boundary is at $\eta=\infty$ and $g_{ij}^{(0)}$ is the metric on the boundary. A generic action for a gauge field in 3 dimensions has the following form
\begin{equation}
S=a\int \mathrm{Tr}(F\wedge\star F)+b\int \mathrm{Tr}(A\wedge F+\frac{2}{3}A\wedge A\wedge A)\,,
\end{equation}
with $a$ and $b$ some coupling constants. Now one should remark that due to the appearance of the metric in the first term, this term decreases as $e^{-\eta}$ near the boundary while the second term is purely topological and will thus dominate near the boundary. This implies that to calculate the boundary charges we only need to keep track of the topological Chern-Simons part of the gauge field action. In the following subsection \ref{redux} we calculate these 3d Chern-Simons terms explicitly for the case of our concern. In subsection \ref{lightning} we shortly review the general idea behind the calculation of the boundary charges from Chern-Simons theory and in \ref{u1} and \ref{su2} we calculate these for our solutions.

\subsection{Reduction of the Chern-Simons term}\label{redux}
The three dimensional Chern-Simons term is a reduction over the sphere of the
Chern-Simons term of five dimensional $\mathcal{N}$=1 supergravity, which itself has its origin in such a topological term in the M-theory action. Starting from the Chern-Simons part of the 11-dimensional supergravity action and reducing over a CY$_3$, one gets the following action in 5-dimensions (where we went to Euclidean signature)

\begin{equation}\label{5dcs}
    I_{CS} = \frac{i}{192 \pi^2} \int D_{ABC} A^A \wedge F^B \wedge F^C\,.
\end{equation}

The ansatz for the gauge field $A^A$ to further reduce to 3 dimensions is given by the asymptotic form found in (\ref{asympgauge}). So we propose as our general reduction ansatz a field strength of the form
\begin{eqnarray}
    F^A = \frac{1}{2} p^A  e_2 + d C^A\,,
\end{eqnarray}
where $C^A$ is a one-form on AdS$_3$. The two-from $e_2$ is known in the literature as the global angular 2-form \cite{Freed:1998tg}, \cite{Harvey:1998bx}, \cite{Hansen:2006wu}, it is the generalisation of the standard volume of the sphere to an $S^2$ fibration and is defined as follows:
\begin{eqnarray}
   e_2 &=& \epsilon_{ijk} (D y^i \wedge Dy^j - \tilde{F}^{ij}) y^k\,,\nonumber\\
   ds^2& =& ds^2_{\text{AdS}_3} + \frac{1}{l^2} (dy^i - \tilde{A}^{ij} y^j) (dy^i - \tilde{A}^{ik} y^k)\,, \\
      Dy^i& =& dy^i - \tilde{A}^{ij} y^j\,,\nonumber \\
   \tilde{F}^{ij} &= &d \tilde{A}^{ij} - \tilde{A}^{ik} \wedge \tilde{A}^{kj}\,.\nonumber
\end{eqnarray}
Summation over repeated indices is assumed and the $y^i$ are the embedding coordinates of S$^2$ in flat $\mathbb{R}^3$, i.e. $y^iy^i=1$. The $\tilde{A}$ are the SU(2) gauge fields coming from the reduction of the metric over the S$^2$. Keep in mind that $\tilde{A}$ depends only on the AdS$_3$ coordinates.

To make the reduction a bit more tractable we introduce the following quantities
\begin{equation}
   \tilde{A}^{ij} = \epsilon_{ijk} A^k, \;\;\;\;\; F^{ij} = \epsilon_{ijk} F^k\,.
\end{equation}
To get compact expressions, we will associate to every quantity with an SU(2) index $i$, $j$, ... the following notation
\begin{equation}
    \mathcal{O} = \frac{i}{2} O^j \sigma_j\,,
\end{equation}
where $\sigma_j$ are the Pauli matrices which satisfy
\begin{eqnarray}
   [\sigma_i, \sigma_j] = 2 i \epsilon_{ijk} \sigma_k, \;\;\;\;\; \text{Tr} (\sigma_i \sigma_j) = 2 \delta_{ij}, \;\;\;\;\; \text{Tr} (\sigma_i \sigma_j \sigma_k ) = 2 i \epsilon_{ijk} \,.\\
\end{eqnarray}
For example, one has
\begin{eqnarray}
    D \mathcal{Y} &= &d \mathcal{Y} + [ \mathcal{Y}\,, \mathcal{A}], \;\;\;
    \mathcal{F} = d \mathcal{A} - \mathcal{A} \wedge \mathcal{A}\,, \nonumber \\
     e_2 &= &4 \text{Tr} \left[\mathcal{Y} d\mathcal{Y} \wedge d\mathcal{Y} + d (\mathcal{Y} \mathcal{A}) \right] = 2 [\sin \theta d\theta \wedge d\phi - d (y^i A^i)]\,,
\end{eqnarray}
where in the last equation, we used spherical coordinates.

Plugging in eqn (\ref{5dcs}) and reducing over S$^2$ keeping in mind that the only dependence on S$^2$ resides in $e_2$, one ends up with the following Chern-Simons term on AdS$^3$:
\begin{equation}
    I_{gauge} = -\frac{i}{4 \pi} \frac{p^3}{6} \int \text{Tr} \left(\mathcal{A} \wedge d \mathcal{A} - \frac{2}{3} \mathcal{A} \wedge \mathcal{A} \wedge \mathcal{A} \right) + \frac{i}{16 \pi} D_{AB} \int C^A \wedge dC^B\,.\label{CHERNSIMONS}
\end{equation}
As one sees the $\mathcal{A}$ and $C$ fields don't interact with each other, this is as expected from SU(2) gauge invariance. The SU(2) gauge field $\mathcal{A}$ does change under such a gauge transformation but $C$ does not. So one needs two $\mathcal{A}$ and one $C$ for a consistent interaction term. But $\text{Tr} \mathcal{A} \wedge \mathcal{A} = 0$. So there is no interaction term between $\mathcal{A}$ and $C$.

\subsection{Boundary charges: lightning review}\label{lightning}
How in general the presence of Chern-Simons terms leads to contributions to both the boundary SU(2) and U(1) currents and the boundary energy momentum tensor is very nicely reviewed in \cite{Kraus:2006wn} and so we will restrict ourselves to a short recapitulation here. Essential in the derivation is the addition of extra boundary terms to the bulk Chern-Simons action. Let's take the simple example of single U(1) field:
\begin{equation}
S=ik\int_{AdS} A\wedge dA\,.
\end{equation}
We can make the gaugechoice $A_\eta=0$ and furthermore the equations of motion imply that $A$ is a flat connection. As argued in \cite{Kraus:2006wn} there are two reasons to include an additional boundary term to this bulk action. The first is that imposing Dirichlet conditions for both components of the gauge fields, i.e. $\d A|_{\partial AdS}=0$, is too strict. Second is that one wants the current associated to the gauge field to be purely left or rightmoving. This last argument is natural from the canonical quantization of Chern-Simons theory \cite{Elitzur:1989nr}. Without the boundary term one has $\d S\sim \int_{\partial AdS} p\d q+q\d p$, where $p$ and $q$ are both a component of the boundary gauge field. Adding the correct boundary term cancels the second term and gives the natural interpretation to $p$ as the momentum conjugate to $q$. The boundary term that does this is
\begin{equation}
S_{bd}=-\frac{|k|}{2}\int_{\partial AdS}A\wedge\star A\,.
\end{equation}
The absolute value of $k$ is needed to have positive energy as we will see shortly. Introducing the standard complex coordinates $w, \bar{w}$ on the boundary cilinder and noting that $\star dw=idw,\ \star d\bar w=-id\bar w$ one can verify that that once one adds the boundary term indeed
\begin{equation}
\d S =
\begin{cases}2i\int_{\partial AdS}(\d A_{w})A_{\bar w}\qquad\mbox{if }k>0\,,\\
2i\int_{\partial AdS}(\d A_{\bar w})A_{w}\qquad\mbox{if }k<0\,,
\end{cases}
\end{equation}
where we have assumed the bulk fields to be on shell. Now we can impose the Dirichlet boundary conditions $\d A_{w}=0$ and leave $\d A_{\bar w}$ arbitrary in case $k>0$ and vice versa if $k<0$. The addition of this boundary term influences the boundary currents, these are defined as
\begin{equation}
\d S=\int_{\partial AdS}\sqrt{g^{(0)}}(\frac{i}{2\pi} J^i\d A_i + \frac{1}{2}T_{ij}\d g^{ij}_{(0)})\,,
\end{equation}
so one sees that e.g.\  the contribution to the energy momentum tensor comes completely from the boundary term as the bulk term is purely topological.
It is now easy to calculate these currents:
\begin{eqnarray}
T_{ww}&=&\frac{|k|}{2}A_wA_w\,,\quad T_{w\bar w}=0\,,\quad T_{\bar w\bar w}=\frac{|k|}{2}A_{\bar w}A_{\bar w}\,,\nonumber\\
J_w&=&\begin{cases} \quad 0\qquad \mbox{if }k>0\,,\\ 2\pi A_w\quad \mbox{if }k<0\,,\end{cases}\\
J_{\bar w}&=&\begin{cases}2\pi A_{\bar w}\quad \mbox{if }k>0\,,\\ \quad 0\qquad \mbox{if }k<0\,. \end{cases}\nonumber
\end{eqnarray}
Having reviewed the general philosophy we can now calculate the charges in our case of interest, note that the story generalizes straightforward to the non-abelian case \cite{Kraus:2006wn}.

\subsection{The U(1) part}\label{u1}
The U(1) part of the Chern-Simons term (\ref{CHERNSIMONS}) is given by
\begin{equation}
\frac{i}{16 \pi} D_{AB} \int C^A \wedge dC^B\,.
\end{equation}
Due to the fact that $D_{AB}$ as a metric on $\mathrm{H}^2(X)$ has a single positive eigenvector and $b_2-1$ negative ones (see e.g.\  \cite{Maldacena:1997de}) we have to treat the two cases slightly differently, see the discussion above.
The projectors to the positive and negative eigenspaces are
\begin{equation}
     (P^+)^A_B = \frac{1}{p^3} \, p^A \, D_{BC} \, p^C\,,  \;\;\;\;\;\;\; (P^-)^A_B = \delta^A_B - (P^+)^A_B\,,
\end{equation}
which gives
\begin{eqnarray}
    C^{A+} = \frac{1}{p^3} \, p^A \, D_{BC} \, p^B \, C^C =  \frac{1}{p^3} \, (p^B q_B) \, p^A \, d\psi =  \frac{2}{p^3} \, (p^B q_B) \, p^A \, dw, \label{Cplus}\\
    C^{A-} =  \left( D^{AB} - \frac{1}{p^3} \, p^B \, p^A \right) \, q_B \, d\psi = 2 \left( D^{AB} - \frac{1}{p^3} \, p^B \, p^A \right) \, q_B \, dw\,,  \nonumber
\end{eqnarray}
where we used the asymptotic from of our gauge field, eqn. (\ref{asympgauge}).

As explained in the previous subsection, once we add the correct boundary term we have the following boundary conditions left $\d C^{+A}_w=0$ and $\d C^{-A}_{\bar w}=0$. So we have to choose a fixed value for those gauge fields at the boundary. It turns out that the correct choice is $C^{+A}_w=0$ and
$C^{-A}_{\bar w}=0$.

However our asymptotic gauge field (\ref{asympgauge}) doesn't satisfy this boundary condition as one can see from (\ref{Cplus}). This is however easily cured by the following gauge transformation

     $$ C^A \longrightarrow C^A - \frac{4}{R}\frac{p^Bq_B}{p^3} p^A\, dt  \,,$$
which gives
\begin{equation}
    C^{A+} =  2  \frac{1}{p^3} \, p^B\, q_B\, p^A\, d\bar{w}\,.
\end{equation}
Given this split into positive and negative modes one can now apply the general procedure as reviewed in the previous subsection to find
\begin{eqnarray}
    T_{\bar{w}\bar{w}}&=&\frac{1}{4 \pi} \, \frac{1}{p^3} \, (p^A \, q_A)^2 \,,\nonumber\\
     T_{ww} &=& \frac{1}{4\pi} \, \frac{1}{p^3} \, \left[ (p^A \, q_A)^2 - p^3 (q_A D^{AB} q_B) \right]\,,\\
     J^+_{\bar{w}}&=& \frac{1}{4}\frac{1}{p^3} \, p^B \,q_B p^A\nonumber\,,\\
     J^-_{w}&=&\frac{1}{4} \left( D^{AB} - \frac{1}{p^3} \, p^B \, p^A \right) \, q_B\,.\nonumber
\end{eqnarray}

\subsection{The SU(2) part}\label{su2}
The SU(2) part of (\ref{CHERNSIMONS}) is
\begin{equation}
-\frac{i}{4 \pi} \frac{p^3}{6} \int \text{Tr} \left(\mathcal{A} \wedge d \mathcal{A} - \frac{2}{3} \mathcal{A} \wedge \mathcal{A} \wedge \mathcal{A} \right)
\end{equation}
here $k = \frac{1}{4 \pi} U^3 > 0$. Let's look at the value the SU(2) gauge field $\mathcal{A}$ takes in our solution. The general sphere reduction Ansatz has the form
\begin{equation}
ds^2=g_{\mu\nu}dx^\mu dx^\nu+g_{\a\beta}\left(dx^\a+A^I_\mu X_I^\a dx^\mu\right)\left(dx^\beta+A^I_\nu X_I^\beta dx^\nu\right)\,,
\end{equation}
where the $x^\mu$ are in our case the coordinates on AdS$_3$ and the $x^\a$ coordinates on the S$^2$, the $X^\a_I\partial_\a$ are the killing vectors of the sphere. So using the form of the asymptotic metric (\ref{asympmetric}), we have
\begin{equation}
d\theta^2+\sin^2\theta(d\phi+\frac{Rd^0}{2U^3}dv)^2=(d\theta+A^I_\mu X_I^\theta dx^\mu)^2+\sin^2\theta(d\phi+A_\mu^IX_I^\phi dx^\m)^2
\end{equation}
which then implies that the only non vanishing component of the gauge field is
\begin{equation}
A^3_v=\frac{Rd^0}{2U^3}\,,
\end{equation}
or in the complex coordinates we introduced
\begin{equation}
A^3_{\bar{w}}=-\frac{R^2d^0}{4U^3}\,.
\end{equation}
And following the by now standard procedure its contribution to the energy momentum tensor is
\begin{equation}
    T_{\bar{w}\bar{w}} = \frac{R^4}{8\pi} \frac{(d^0)^2}{16U^3}\,.
\end{equation}

\bibliographystyle{jhep}
\bibliography{refslast}

\end{document}